\RequirePackage{lineno}
 \documentclass
 [prd,superscriptaddress,twocolumn,showpacs,preprintnumbers,amsmath,amssymb]{revtex4}
 \RequirePackage[colorlinks,breaklinks,pdftitle={ATLAS draft},pdfauthor={The ATLAS Collaboration}]{hyperref}  
\hypersetup{linkcolor=blue,citecolor=blue,filecolor=black,urlcolor=blue}
\usepackage{subfig}
\usepackage{epsfig}
\usepackage{dcolumn}
\usepackage{bm}
\usepackage{graphicx}
\usepackage{endnotes}
\usepackage[sort&compress]{natbib}
\usepackage{multirow}
\usepackage{xspace}
\usepackage{caption}
\usepackage{atlasphysics}
\usepackage{draftcopy}

\newcommand{\twojTitle}{Search for supersymmetry in $pp$ collisions at $\sqrt{s}=7$~TeV in final
 states with missing transverse momentum and $b$-jets with the ATLAS detector}

\newcommand{\sqR}   {\tilde{q}^{}_{R}}
\newcommand{\sqL}   {\tilde{q}^{}_{L}}

\newcommand{\bone}   {\tilde{b}^{}_{1}}
\newcommand{\tone}   {\tilde{t}^{}_{1}}
\newcommand{\neut}  {\tilde{\chi}^{0}_{1}}  
\newcommand{\chipm}  {\tilde{\chi}^{\pm}_{1}} 
\def\etmiss{\ensuremath{E_{\mathrm{T}}^{\mathrm{miss}}}\xspace}

\def\dphimin{\ensuremath{\Delta\phi_{\mathrm{min}}}}
\def\trmass{\ensuremath{m_\mathrm{T}}}
\def\bjets{$b$-jets\xspace}
\def\bjet{$b$-jet\xspace}

\def\btagging{$b$-tagging\xspace}
\def\btag{\ensuremath{b}-tag\xspace}
\def\btagged{$b$-tagged\xspace}

\def\meff{\ensuremath{m_{\mathrm{eff}}}\xspace}
\def\mt{\ensuremath{m_{\mathrm{T}}}\xspace}

\newcommand{\gl}   {\tilde{g}}
\newcommand{\sq}   {\tilde{q}}
\def\Wjets{$W$+jets\xspace}
\def\Zjets{$Z$+jets\xspace}

\renewcommand{\ttbar} {\ensuremath{t\bar{t}}\xspace}
\newcommand{\ttbb}{\ensuremath{t\bar{t}b\bar{b}}\xspace}
\renewcommand{\met} {\ensuremath{E_{\mathrm{T}}^{\mathrm{miss}}}\xspace}
\renewcommand{\pt} {\ensuremath{p_\mathrm{T}}\xspace}

\newcommand{\lumi}{2.05}
\newcommand{\mywidth}{1.03}
\newcommand\totallumi{2.05~${\rm fb}^{-1}$}

\def\met{\ensuremath{E_{\mathrm{T}}^{\mathrm{miss}}}}

\newcommand{\btheta}{\bf \theta}

\usepackage{preprintcover} 
\PreprintCoverPaperTitle{Search for supersymmetry in $pp$ collisions at $\sqrt{s}=7$~TeV in final
 states with missing transverse momentum and $b$-jets \\ with the ATLAS detector}  
\PreprintIdNumber{CERN-PH-EP-2012-052}
\PreprintCoverAbstract{The results of a search for supersymmetry in events with large missing transverse 
momentum and heavy flavour jets using an integrated luminosity corresponding to \totallumi~of $pp$ collisions at $\sqrt{s}=7$~TeV 
recorded with the ATLAS detector at the Large Hadron Collider are reported. No significant excess is observed with respect 
to the prediction for Standard Model processes. Results are interpreted in a variety 
of $R$-parity conserving models in which scalar bottoms and tops are the only scalar quarks 
to appear in the gluino decay cascade, and in an SO(10) model framework. Gluino masses up to  600--900 GeV are excluded, depending on the model considered.} 
\PreprintJournalName{Physical Review D}  

\begin{document}
\title{\twojTitle}
\author{The ATLAS Collaboration}
\affiliation{}

\begin{abstract} 
The results of a search for supersymmetry in events with large missing transverse 
momentum and heavy flavour jets using an integrated luminosity corresponding to \totallumi~of $pp$ collisions at $\sqrt{s}=7$~TeV 
recorded with the ATLAS detector at the Large Hadron Collider are reported. No significant excess is observed with respect 
to the prediction for Standard Model processes. Results are interpreted in a variety 
of $R$-parity conserving models in which scalar bottoms and tops are the only scalar quarks 
to appear in the gluino decay cascade, and in an SO(10) model framework. Gluino masses up to  600--900 GeV are excluded, depending on the model considered.
\end{abstract}

\maketitle
\normalsize

\section{Introduction}

Supersymmetry (SUSY)~\cite{Miyazawa:1966,Ramond:1971gb,Golfand:1971iw,Neveu:1971rx,Neveu:1971iv,Gervais:1971ji,Volkov:1973ix,Wess:1973kz,Wess:1974tw} is a framework that provides an extension of the Standard Model (SM) and naturally resolves the hierarchy problem \cite{Weinberg:1975gm,Gildener:1976ai,Weinberg:1979bn,Susskind:1978ms} by introducing supersymmetric partners of the known bosons and fermions. In the MSSM~\cite{Fayet:1976et,Fayet:1977yc,Farrar:1978xj,Fayet:1979sa,Dimopoulos:1981zb}, which is an $R$-parity conserving minimal supersymmetric extension of the SM, SUSY particles are produced in pairs and the lightest supersymmetric particle (LSP) is stable, providing a possible candidate for dark matter. In a large variety of models, the LSP is the lightest neutralino, $\neut$. The coloured superpartners of quarks and gluons, the squarks ($\sq$) and gluinos ($\gl$), are expected to be produced in strong interaction processes at the centre-of-mass energy of the Large Hadron Collider (LHC). Their decays 
via cascades ending with the LSP would produce striking experimental signatures. The undetected LSP results in missing transverse momentum (its magnitude is referred to as $\etmiss$ in the following). The final states also  contain  multiple jets and possibly leptons. In the MSSM, the scalar partners of right-handed and left-handed quarks, 
$\sqR$ and $\sqL$, can mix to form two mass eigenstates. The mixing effect is proportional to the corresponding SM fermion masses and therefore becomes important for the third generation. Large mixing can yield scalar bottom (sbottom, $\bone$) and scalar top (stop, $\tone$) mass eigenstates which are significantly lighter than other squarks. Consequently, $\bone$ and $\tone$ 
could be produced with large cross sections at the LHC, either directly in pairs, 
or through $\gl\gl$ production with subsequent $\gl \rightarrow \bone b$ or 
$\gl \to \tone t$ decays (gluino-mediated production).

In this paper, a search for scalar top and bottom quarks using an integrated luminosity corresponding to
2.05 \ifb\ of $\sqrt{s}=7$~TeV proton-proton collisions at the LHC, is presented. 
Events are selected by requiring large $\etmiss$, several jets, including $b$-quark jets ($b$-jets), and either vetoing (0-lepton channel) 
or requiring (1-lepton channel) charged leptons. The search is mostly sensitive to the gluino-mediated  
production of third generation squarks. 
Results are interpreted in the framework of various simplified models in which 
scalar bottoms and tops are the only squarks that appear in the gluino decay 
cascade, and in specific Grand Unification Theories (GUTs) based on the gauge 
group SO(10)~\cite{SO10_2, SO10}. The GUT group SO(10) is especially compelling 
since it allows for gauge and matter unification. In the two SO(10) models 
considered in this paper, we also expect $t$ - $b$ - $\tau$ 
third generation Yukawa coupling unification at $Q = M_{\mathrm{GUT}}$.

The paper is an update of a search presented by the ATLAS collaboration 
using 35~$\ipb$ of data collected in 2010~\cite{Aad:2011ks}, with a number of improvements. 
The analysis has been extended by including more signal regions which profit 
from the increased available integrated luminosity and maximise the 
sensitivity to a large variety of SUSY scenarios. Data-driven methods   
are employed to estimate the contributions of SM background processes.  
Searches for scalar bottom quarks via $\gl\gl$ production have been also 
reported by the CMS~\cite{cms_bjet} collaboration. 
Searches sensitive to direct scalar bottom production irrespective of gluino 
mass have been published by the ATLAS collaboration~\cite{sbottom} using 
the same data-set employed in this paper. 

\section{The ATLAS detector}

The ATLAS detector~\cite{DetectorPaper:2008} comprises an inner detector surrounded by a thin superconducting solenoid and a calorimeter system. Outside the calorimeters is an
extensive muon spectrometer in a toroidal magnetic field.

  The inner detector system is immersed in a $2$\,T axial magnetic field and provides 
tracking information for charged particles in a pseudorapidity 
range $|\eta|< 2.5$ \cite{atlascoordinates}. The highest granularity is achieved around the vertex region using silicon 
pixel and microstrip (SCT)
detectors. These detectors allow for an efficient tagging of jets originating from $b$-quark decays 
using impact parameter measurements and the reconstruction of secondary decay vertices.
The transition radiation tracker (TRT), which surrounds the silicon detectors, contributes to track reconstruction 
up to $|\eta| = 2.0$ and improves electron identification by the detection of transition radiation.

  The calorimeter system covers the pseudorapidity range $|\eta| < 4.9$. 
The highly-segmented electromagnetic calorimeter consists of lead absorbers with liquid argon (LAr)
as the active material and covers the pseudorapidity range $|\eta| < 3.2$. In the region  $|\eta| < 1.8$, a pre-sampler detector using a thin 
layer of liquid argon is used to correct for the energy lost by electrons and photons upstream of the calorimeter.
The hadronic tile calorimeter is a steel/scintillating-tile detector and is situated directly 
outside the envelope of the electromagnetic calorimeter. The two hadronic end-cap calorimeters have liquid argon as the active material, and copper absorbers. The calorimeter coverage is completed by forward calorimeters with liquid argon and copper and tungsten absorber material.

Muon detection is based on the deflection of muon tracks in the large superconducting air-core toroid magnets. Three eight-coil toroids, a barrel and two end-caps, generate the field for the muon spectrometer in the range $|\eta|<2.7$. The toroids are instrumented with separate trigger and high-precision chambers.

\section{Monte Carlo simulation}

Simulated event samples are used to aid in the description of the background and
to model the SUSY signal. 
Top quark pair and single top quark production are simulated with {\tt MC@NLO}~\cite{mcatnlo},
fixing the top quark mass at 172.5~GeV, and using the next-to-leading-order (NLO) parton density function (PDF) set
{\tt CTEQ6.6}~\cite{cteq66}. 
Additional Monte Carlo (MC) samples generated with {\tt POWHEG}~\cite{powheg} and {\tt ACERMC}~\cite{Kersevan:2004yg} 
are used to estimate the event generator systematic uncertainties. 
Samples of \Wjets, \Zjets with light
and heavy flavour jets, and \ttbar with additional \bjets, \ttbb, are generated with
{\tt ALPGEN}~\cite{alpgen} and the PDF set {\tt CTEQ6L1}~\cite{cteq6l1}.
The fragmentation and hadronisation for the {\tt ALPGEN} and {\tt MC@NLO} samples are
performed with {\tt HERWIG}~\cite{herwig}, using {\tt JIMMY}~\cite{jimmy} for the
underlying event.
Samples of Z\ttbar and W\ttbar are generated with {\tt MADGRAPH}~\cite{madgraph}
interfaced to {\tt PYTHIA}~\cite{pythia}. Di-boson ($WW$, $WZ$, $ZZ$) samples
are generated with {\tt HERWIG}.
The signal samples are generated  using the
{\tt HERWIG++}~\cite{Bahr:2008pv} v2.4.2 Monte Carlo program.
The SUSY sample yields are normalised to the results of NLO calculations, as obtained using the {\tt PROSPINO}~\cite{Beenakker:1997ut}
v2.1 program and the parameterisation of the PDFs is done with {\tt CTEQ6.6M}~\cite{cteq6m}. 
The MC samples are produced using parameters 
tuned as described in Ref. \cite{atlastune1,atlastune2} and are
processed through a detector simulation~\cite{atlassimulation} based on {\tt GEANT4}~\cite{geant4}. The collision events considered in this search contain on average five proton-proton interactions per bunch crossing. This effect is included in the simulation, and MC events are reweighted to reproduce the mean expected number of collisions per bunch crossing estimated for data.

The background predictions, normalised to theoretical cross sections,
including higher-order QCD corrections when available, are compared to
data in control regions. The cross sections times branching ratio in the relevant final states used for each Standard Model background process are listed in Table~\ref{table:MCSample}. 
The $W$ and $Z/\gamma^*$ production processes are normalised to the next-to-next-to-leading order (NNLO) cross sections while the $\ttbar$ and single top production 
are normalised to the NLO+NNLL (next-to-next-to-leading logarithms) cross sections. 
The normalisation of the di-boson production is based on cross sections determined at NLO using MCFM~\cite{mcfm,mcfm2}.  
The $\ttbar$ production in association with $W/Z$ or $b\bar{b}$ is normalised to LO. 

For background from jet production from parton scattering processes (multi-jet in the following), no reliable
 prediction can be obtained from a leading-order
Monte Carlo simulation and data-driven methods are used to determine the residual contributions
of this background to the selected event samples, as discussed in Section~\ref{sect:bkg}.

\begin{table}[t]
\begin{small}
\begin{center}
\begin{tabular}{ l|l l}
\hline
\hline
\raisebox{-0.4ex}{Physics process} &  {\raisebox{-0.4ex}{$\sigma \cdot$ BR [nb] (perturbative order) }} &  \\
\hline
\hline
$W \rightarrow \ell \nu$  & 31.4 (NNLO) &\cite{Hamberg:1990np,Melnikov:2006kv,Melnikov:2006di}  \\
$Z/\gamma^* \to \ell^+ \ell^-$  & 3.20 (NNLO) &\cite{Hamberg:1990np,Melnikov:2006kv,Melnikov:2006di} \\
$Z \to \nu \bar \nu$  & 5.82 (NNLO) &\cite{Hamberg:1990np,Melnikov:2006kv,Melnikov:2006di} \\
$\ttbar$          & 0.165 (NLO+NNLL) &\cite{Bonciani:1998vc,Moch:2008qy,Beneke:2009ye}\\
Single top         & 0.085 (NLO+NNLL) &\cite{Bonciani:1998vc,Moch:2008qy,Beneke:2009ye}\\
\ttbb & 0.9 $\times$ 10$^{-3}$   (LO) & \cite{alpgen} \\
$\ttbar$+$W/Z$ & 0.4 $\times$ 10$^{-3}$   (LO) & \cite{madgraph}\\
$WW$, $WZ$, $ZZ$  & 0.071 (NLO) & \cite{mcfm,mcfm2} \\
\hline
\hline
\end{tabular}
\caption{The most important background processes and their production cross sections,
multiplied by the relevant branching ratios (BR).
The $\ell$ indicates all three types of leptons ($e$, $\mu$, $\tau$) summed together.
 Contributions from higher order QCD corrections are included for 
$W$ and $Z$ boson production, for \ttbar\ production and for di-boson production. The $Z/\gamma^* \to \ell^+ \ell^-$ cross section is given for events with a di-lepton invariant mass of at least 40~GeV.
The cross sections for  \ttbb and $\ttbar$+$W/Z$ production 
are  given at leading order.}
\label{table:MCSample}
\end{center}
\end{small}
\end{table}

\section{Object reconstruction}

A pre-selection of electron and muon candidates is used to estimate the contribution from non-isolated leptons and misidentified electrons, to veto on additional leptons in the event when required, and to calculate the value of $\etmiss$. More stringent identification criteria are then applied for the final selections.

Electrons are reconstructed from energy clusters in the electromagnetic calorimeter matched to a track in the inner 
detector.  Candidates for the electron pre-selection must satisfy the  ``medium''~\cite{2011mk} selection based on calorimeter 
shower shape, inner-detector track quality, and track-to-calorimeter cluster matching.  Electrons used in the final 
selection are required to pass the ``tight''~\cite{2011mk} electron definition, which adds  requirements on the 
ratio $E/p$ between the calorimeter cluster energy $E$ and the track momentum $p$, on the detection of transition radiation in the TRT, and on the isolation of the candidate. 
The scalar transverse momentum (\pt) sum of tracks within a cone in the $\eta, \ \phi$ plane of 
radius $\Delta R = 0.2$ around the electron candidate (excluding the electron track \pt itself), $\Sigma_{\pt}$, must be less than 10\% of the electron \pt. 
Medium electrons are required to pass kinematic requirements of $\pt>20$~GeV and $|\eta|<2.47$, while the \pt threshold is raised to 25~GeV for tight electrons. In addition, electrons with a 
distance to the closest jet of $0.2 < \Delta R < 0.4$ are discarded. 

Muons are identified as a match between an extrapolated inner detector track and one or more track segments in the muon 
spectrometer. A requirement on the minimum number of hits in each tracking device ensures the quality of the inner detector track reconstruction.
Muons with a 
distance to the closest jet of $\Delta R<0.4$ are discarded. In order to reject muons resulting from cosmic rays, tight 
criteria are applied on the proximity of the muon trajectories to the primary vertex (PV)~\cite{ATLASPV}: $|z_\mu-z_{\mathrm{PV}}|<1$~mm and 
$|d_0|<0.2$ mm, where $z_\mu$ is the $z$ coordinate of the extrapolated muon track at the point of closest approach to the 
PV, $z_{\mathrm{PV}}$ is the coordinate of the PV, and $|d_0|$ is the magnitude of the impact parameter of the muon in the 
transverse plane. Pre-selected muons are required to satisfy all these requirements, and in addition to have $\pt>10$~GeV and $\eta<2.4$. For muons in the final selection, the \pt requirement is raised to 20~GeV and the muon 
is required to be isolated with $\Sigma_{\pt}<1.8$~GeV.

Jets are reconstructed from three-dimensional calorimeter 
energy clusters by using the anti-$k_t$ jet algorithm~\cite{Cacciari:2008gp, Cacciari200657} 
with a radius parameter of 0.4. The measured jet energy is corrected 
for inhomogeneities and for the non-compensating nature of the calorimeter 
by using $\pt$- and $\eta$-dependent correction factors \cite{JES}. Jets are required to have $\pt > 20$~GeV and 
 $|\eta| < 2.8$. Events with jets failing jet quality criteria against noise and non-collision 
backgrounds are rejected. The quality criteria used are the same as in Ref.~\cite{JES}. Additionally, in the 0-lepton channel the three leading jets, if central ($|\eta| < 2$), are required to have a jet charged fraction (defined as the scalar sum of the transverse momenta of the tracks associated with the jet divided by the jet \pt) of at least 5\%. Jets within a distance of $\Delta R=0.2$ of a pre-selected electron are 
rejected, since these jets are likely to be electrons also reconstructed as jets. For jets in the signal regions, the 
\pt requirement is tightened to 50~GeV to remove jets that are not associated with the hard scattering of interest.

A \btagging algorithm exploiting both impact parameter and secondary vertex information~\cite{btag} is used to identify jets containing a $b$-hadron decay. This 
algorithm has a 60\% efficiency for tagging \bjets in a MC sample of \ttbar events, with a mis-tag rate for light quarks 
and gluons of less than 1\% and for $c$ quarks of less than 10\%. These $b$-jets are identified within the nominal acceptance of the inner detector ($|\eta| < 2.5$) and they are required to have $\pt > 50$~GeV. 

The value of $\etmiss$~\cite{Aad:2012re} is the magnitude of the vector $\vec{\bf E}_{\mathrm{T}}^{\mathrm{miss}}$, which is calculated as the vector sum of the transverse momenta of all reconstructed jets with $\pt>20$~GeV and $|\eta|<4.5$, all preselected electrons and muons, and calorimeter energy clusters which do not belong to other reconstructed objects.

During a fraction of the data-taking period (about 40\% of the total integrated luminosity), a localised electronics failure in the 
LAr barrel calorimeter created a dead region in the second and third calorimeter 
layers ($\Delta \eta \times \Delta \phi \simeq 1.4 \times 0.2$) in which on average 
30\% of the incident jet energy is not measured. Negligible impact is found on the 
reconstruction efficiency for jets with $\pt > 20$~GeV. For events selected during this data period, if any jet with $\pt > 50$~GeV falls in the aforementioned region, the event is rejected. 
The loss in signal acceptance is smaller than 10\% in the affected period for the models considered. 

In the event selection, a number of variables derived from the reconstructed objects are used. The transverse mass \trmass\ formed by \etmiss and the \pt of the lepton is defined as:

\begin{equation}
\trmass = \sqrt{2 \pt^{\mathrm{lep}} \etmiss - 2 {\vec{\bf p}_\mathrm{T}}^{\mathrm{lep}} \cdot \vec{\bf E}_{\mathrm{T}}^{\mathrm{miss}}}
\end{equation}

The effective mass \meff is obtained as the scalar \pt sum of all selected objects in the event:

\begin{equation}
\meff = \sum_{i} (\pt^{\mathrm{jet}})_i + \met + \sum_j (\pt^{\mathrm{lep}})_j
\end{equation}

\noindent where the sums are over the number of jets, $i$, and the zero or one leptons, $j$, in a given signal region.

Finally,  \dphimin\ is defined as the minimum azimuthal separation between the selected jets in a given signal region and the $\vec{\bf E}_{\mathrm{T}}^{\mathrm{miss}}$ direction.

\section{Event Selection}

This search uses proton-proton collisions recorded from March to August 2011 at a centre-of-mass energy of 7~\TeV. 
After the application of beam, detector and data quality requirements, the data 
set consists of a total integrated luminosity of 2.05$\pm$0.08~fb$^{-1}$~\cite{Aad:2011dr,lumiconf}. Two groups of signal regions are defined based on the presence, or otherwise, of a charged lepton ($\ell = e, \mu$) in the final 
state and are further referred to as 0-lepton and 1-lepton channels. In the 0-lepton channel, a veto on pre-selected leptons is applied, while exactly one lepton is required in the 1-lepton channel. Events containing two or more leptons are the subject of a different study~\cite{Tapas}. 

The data  are selected with a three-level trigger system. A trigger requiring a high transverse momentum jet and missing transverse momentum is used to select events for the 0-lepton channel. The plateau efficiency is reached for jets with $\pt > 130$~GeV and $\etmiss > 130$~GeV.  A single electron trigger, reaching the plateau efficiency for offline electrons with $\pt \geq 25$~GeV, and a combined muon-jet trigger, reaching the plateau efficiency for muons with $\pt \geq  20$~GeV and jets with $\pt  \geq 60$~GeV are used for the 1-lepton channel.

Events are required to have a reconstructed primary vertex associated with 
five or more tracks with $\pt > 0.4$~GeV, and must pass basic quality criteria against 
detector noise and non-collision backgrounds. 

For the 0-lepton selection, at least one jet with \pt$>130$~GeV, at least two 
additional jets with $\pt > 50$~GeV and \etmiss$>130$~GeV are required. At least one of the selected jets is required to be $b$-tagged. To reduce the 
amount of multi-jet background, where \etmiss results from mis-reconstructed jets or from neutrinos emitted close to 
the direction of the jet axis, additional requirements of $\dphimin>0.4$  and $\met/\meff > 0.25$ are applied.

\begin{table*}[htp]
 \centering
 \begin{tabular}{l |c|c}
   \hline
   \hline
     Pre-selection& Signal Region name & Selection   \\
   \hline
   \hline
   & SR0-A1 &   at least one $b$-tag, $\meff > 500$~GeV\\
  no leptons, at least three jets,  & SR0-B1 &   at least one $b$-tag, $\meff > 700$~GeV\\	
  $\pt(j1) > 130$~GeV, $\pt(j2, j3) > 50$~GeV,& SR0-C1 & at least one $b$-tag, $\meff > 900$~GeV\\
  $\etmiss > 130$ GeV, $\etmiss/\meff > 0.25$, & SR0-A2 &  at least two $b$-tags, $\meff > 500$ GeV\\
  $\dphimin > 0.4$ & SR0-B2 & at least two $b$-tags, $\meff > 700$ GeV\\
   & SR0-C2 & at least two $b$-tags, $\meff > 900$ GeV\\
   \hline
    one lepton, at least four jets &  & \\
      $\pt(j1) > 60$ GeV, $\pt(j2, j3, j4) > 50$ GeV, & SR1-D & $\meff > 700$ GeV\\
    $\etmiss > 80$ GeV, $\mt > 100$ GeV,  & SR1-E & $\meff > 700$ GeV, $\etmiss > 200$ GeV\\
    at least one \btag & & \\
   \hline
   \hline
   \end{tabular}
   \caption{Signal regions definition for the 0-lepton and 1-lepton channels. The first column summarises the common pre-selection applied, while the last column specifies the selection defining the different signal regions.}
\label{tab:SR}
\end{table*}

Six signal regions are defined in order to obtain good signal sensitivity for the various models and parameter values studied. The regions are chosen by optimising the expected significance in models in which pair-produced gluinos decay with 100\% branching ratio to on- and/or off-shell scalar bottom quarks.  The signal regions are characterised by the minimum number of \bjets in the final state and by different thresholds on \meff. The regions are labelled with the prefix SR0, and are listed in the upper section of Table~\ref{tab:SR}, together with a summary of the full selection applied.

For the 1-lepton channel, events are required to have exactly one lepton, a leading jet with $\pt > 60$~GeV, three further jets with $\pt>50$~GeV, and $\etmiss>80$~GeV. At least one jet is required to be \btagged. SM background processes that lead to the production of a $W$ boson in the final state are rejected by requiring  $\trmass>100$~GeV. Two signal regions, labelled with the prefix SR1 and summarised in Table~\ref{tab:SR}, are defined, based on different thresholds applied on the effective mass and the missing transverse momentum.

\section{Background Estimation}\label{sect:bkg}

   \begin{table*}[htp]
 \centering
 \begin{tabular}{l|c|c}
   \hline
   \hline
     Pre-selection & Control region name & Selection   \\
\hline
\hline
one lepton, at least three jets & & \\
 $\pt(j1) > 130$ GeV, $\pt(j2, j3) > 50$ GeV, & CR0-1 & at least one \btag\\
  $\etmiss > 130$ GeV, $40\ \textrm{GeV} < \mt < 100$ GeV, & CR0-2 & at least two $b$-tags \\
 $\meff > 600$ GeV & & \\
   \hline
  one lepton, at least four jets & & \\
 $\pt(j1) > 60$ GeV, $\pt(j2, j3, j4) > 50$ GeV, & CR1 & at least one \btag\\
  $\etmiss > 80$ GeV, $40\ \textrm{GeV} <\mt < 100$ GeV, $\meff > 500$ GeV & & \\
   \hline
   \hline
   \end{tabular}
   \caption{Control regions definition for the 0-lepton and 1-lepton channels. The first column summarises the common pre-selection applied, while the last column specifies the selection defining the control regions.}
\label{tab:CR}
\end{table*}

\begin{figure*}
  \centering

  \includegraphics[width=\mywidth\columnwidth]{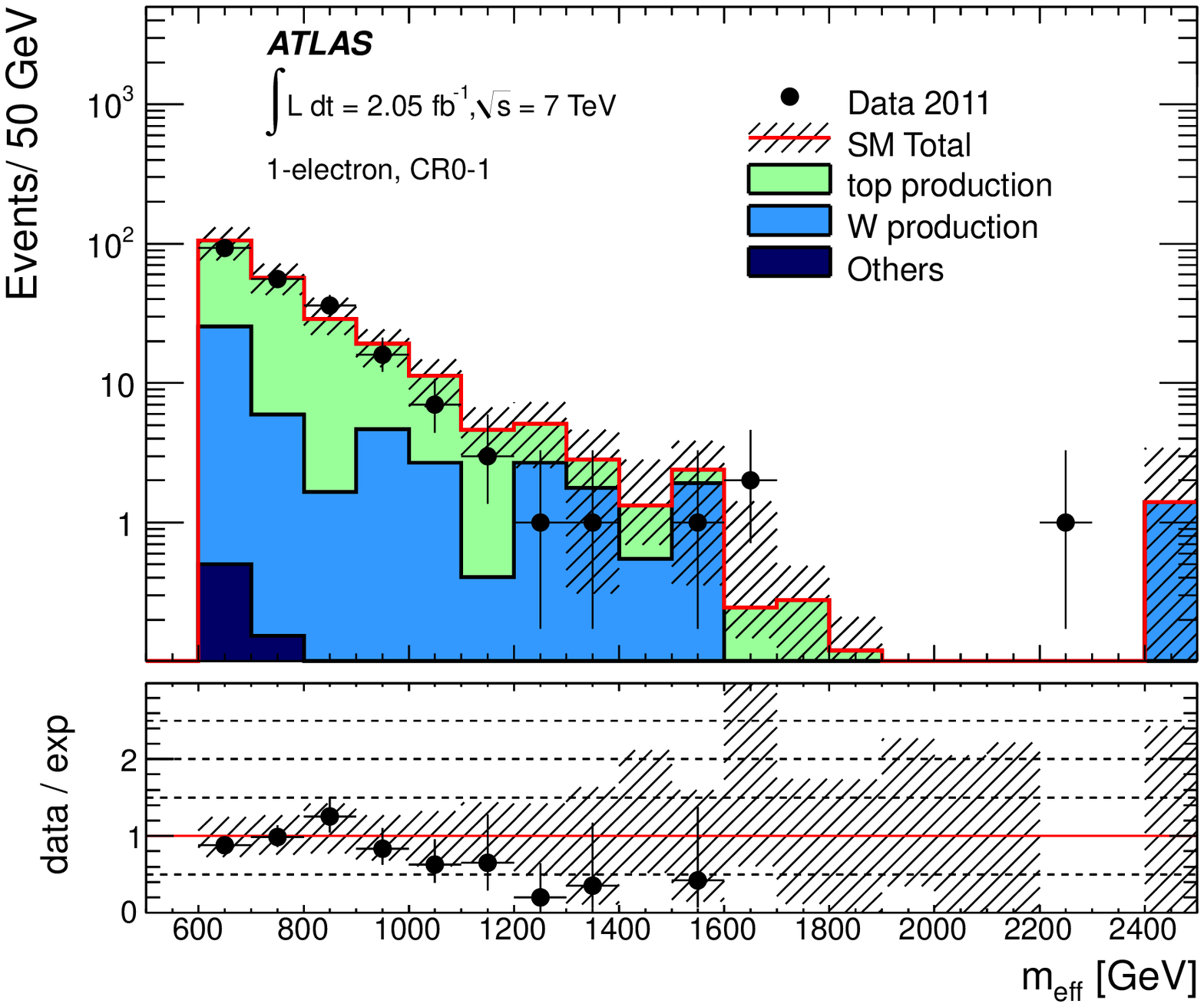}
  \includegraphics[width=\mywidth\columnwidth]{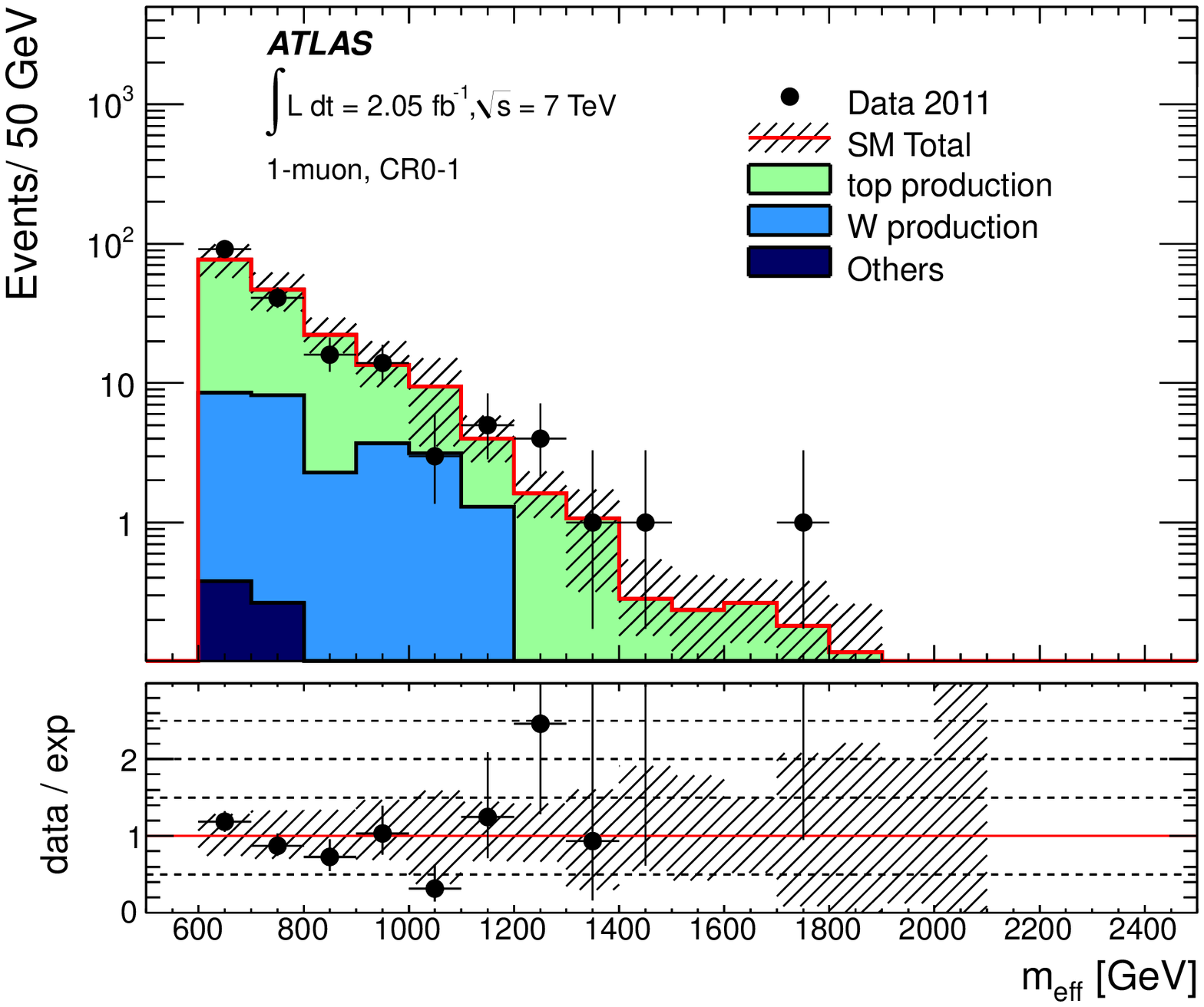} \\	    
  \includegraphics[width=\mywidth\columnwidth]{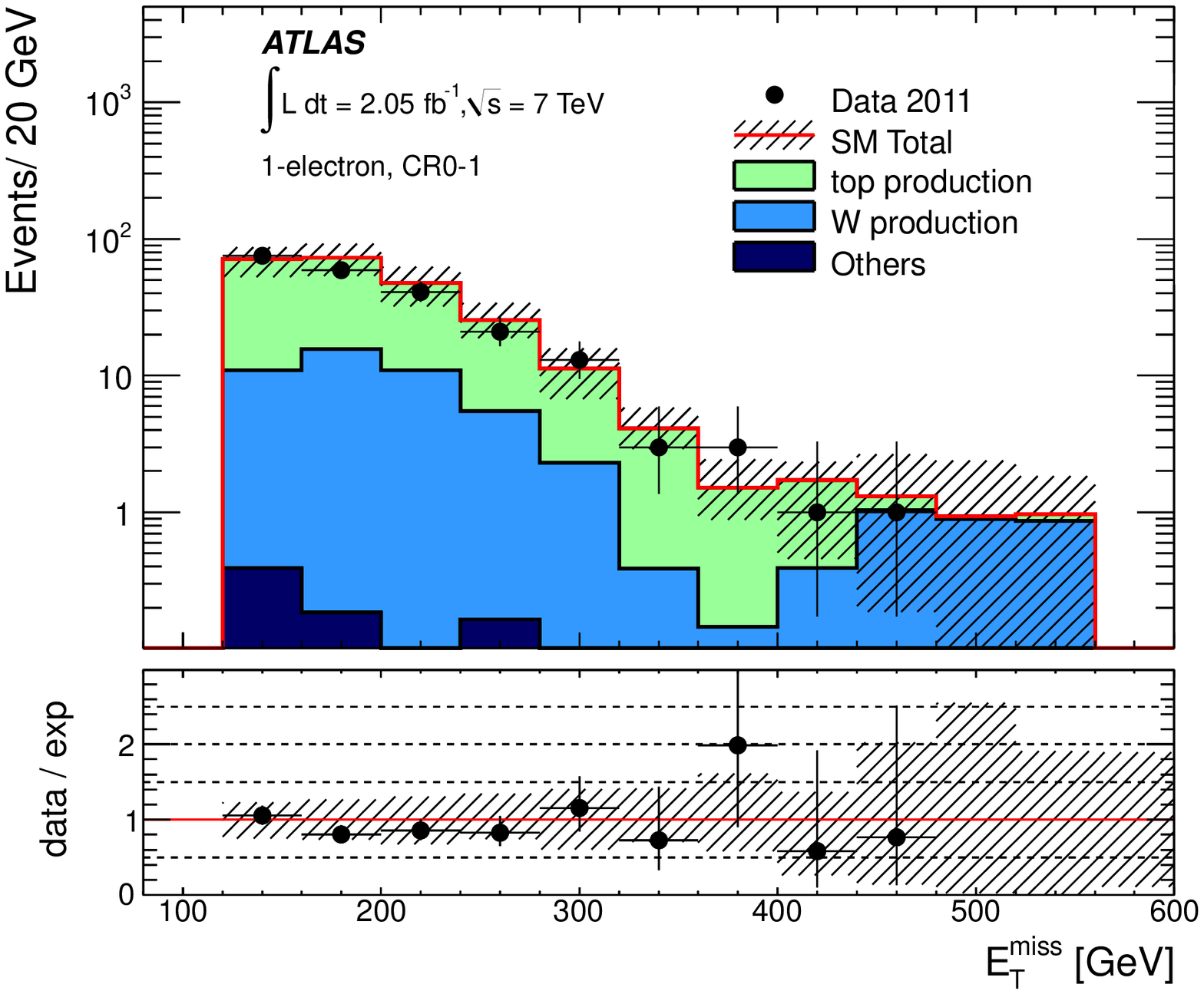}
  \includegraphics[width=\mywidth\columnwidth]{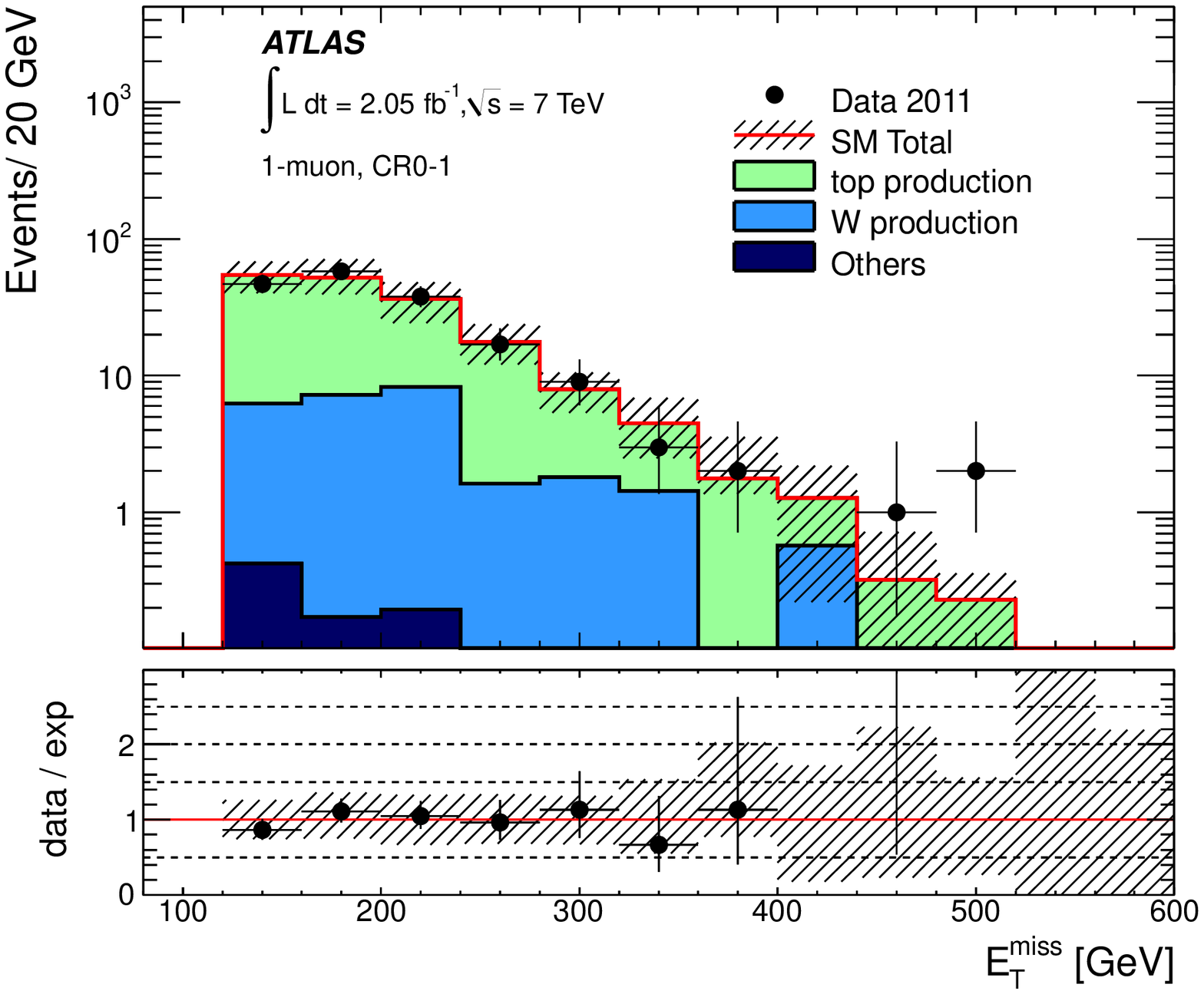} \\	    

\caption{Distribution of the effective mass (top) and \etmiss (bottom) in the CR0-1 control region for the 1-electron (left) and 1-muon (right) channels. The colour labelled ``Others'' includes contributions from $Z$, di-boson and  multi-jet production processes. The hatched band shows the systematic uncertainty, which includes both experimental uncertainties (among which JES and $b$-tagging uncertainties are dominant) and theoretical uncertainties on the background normalisation and shape. The small insets show the ratio between the observed distribution and that predicted for the Standard Model background. Although the distributions are presented separately for $e$ and $\mu$, the background estimation uses the sum of the $e$ and $\mu$ yields in the CR0.}
\label{fig:0lep_nonqcd}
\end{figure*}

\begin{figure*}
  \centering

  \includegraphics[width=\mywidth\columnwidth]{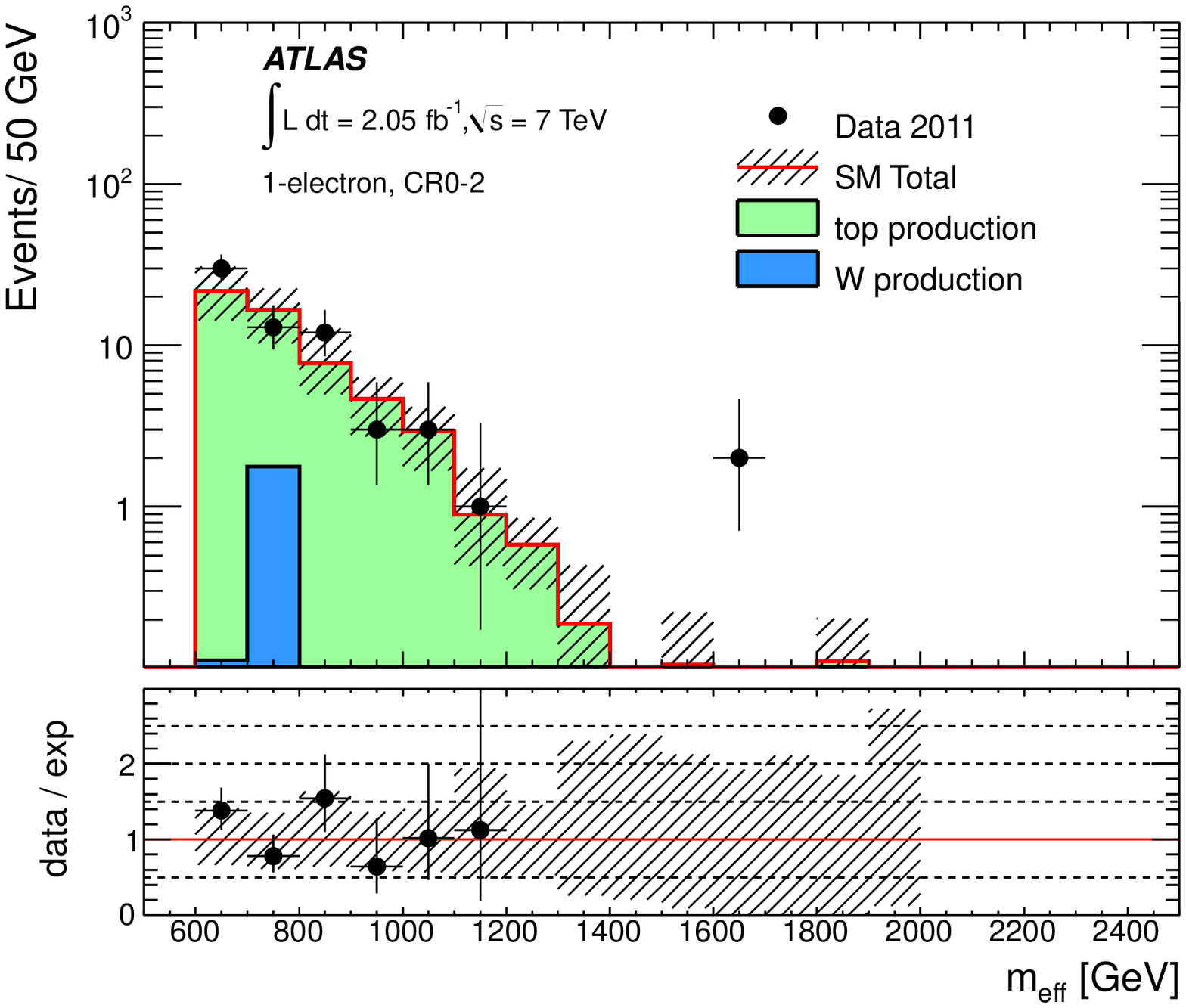}
   \includegraphics[width=\mywidth\columnwidth]{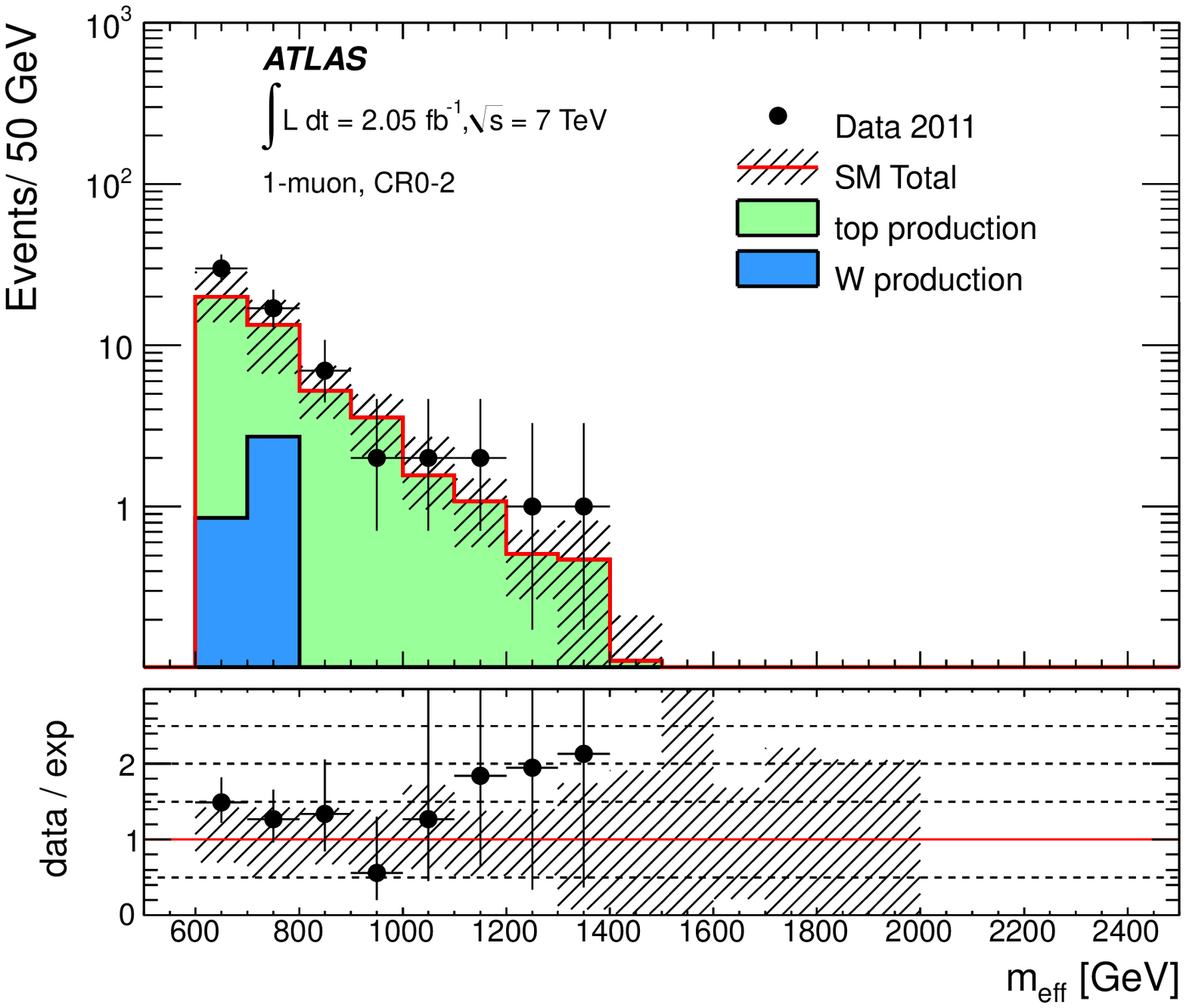}\\
  \includegraphics[width=\mywidth\columnwidth]{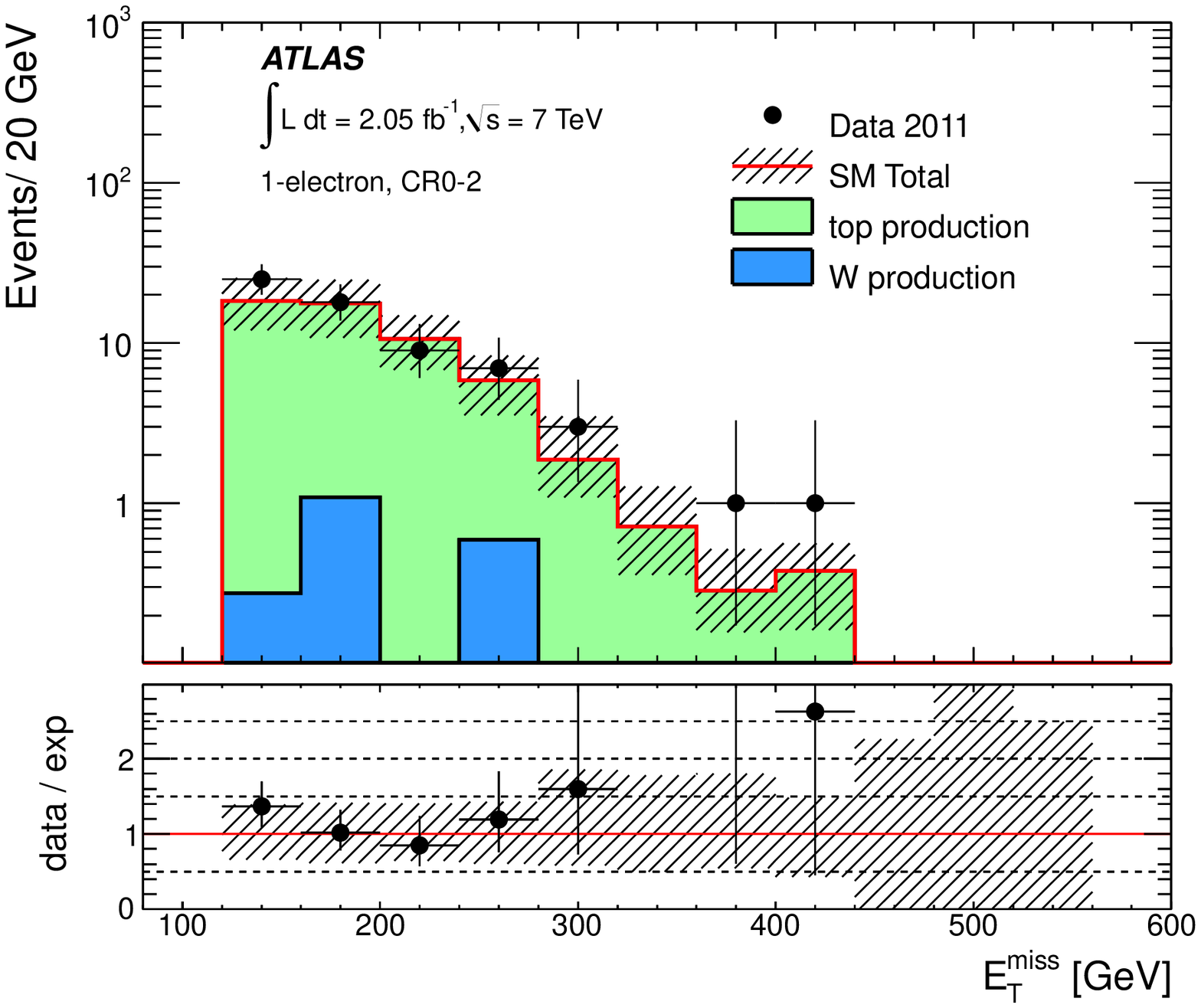}
   \includegraphics[width=\mywidth\columnwidth]{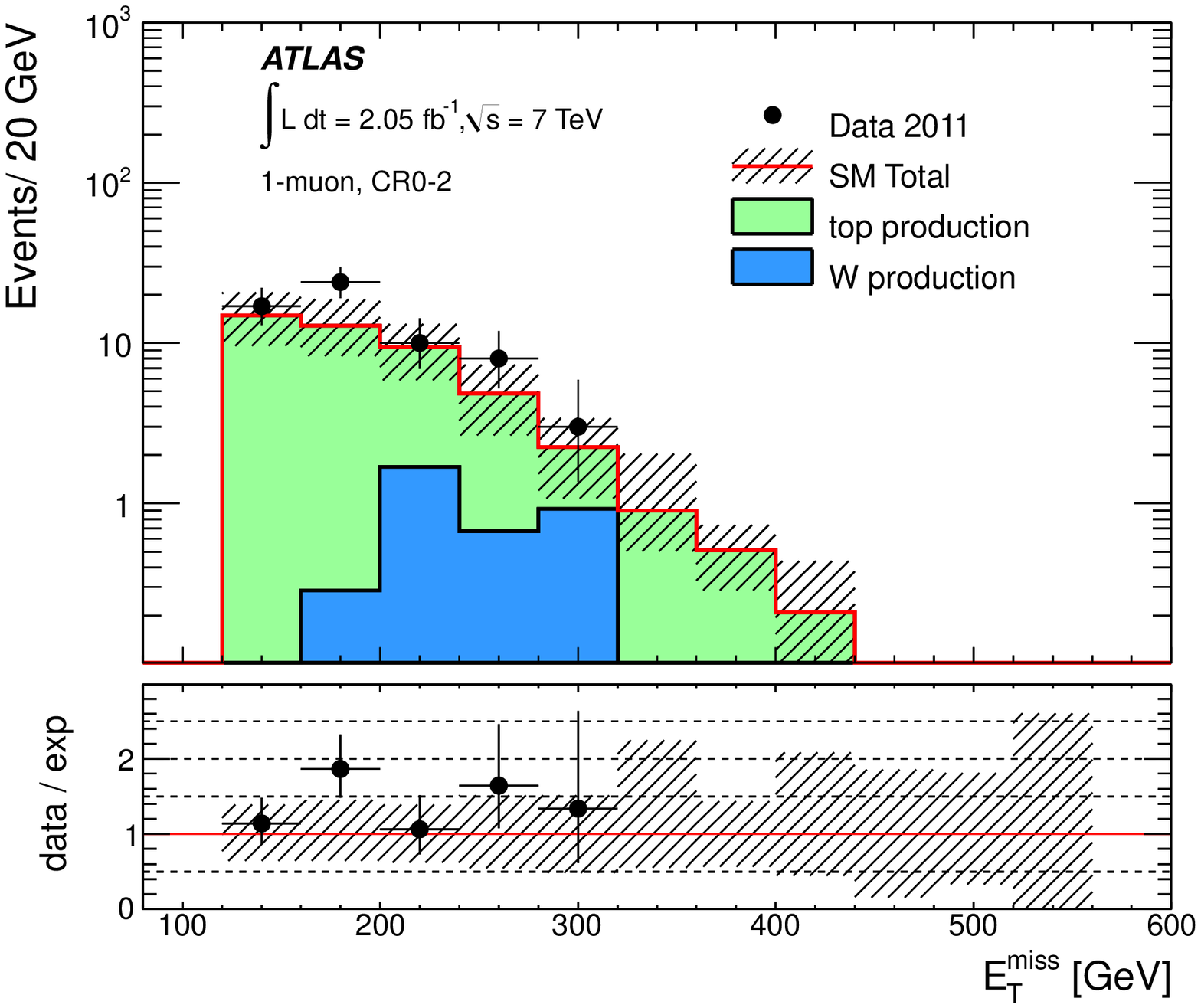}\\

\caption{Distribution of the effective mass (top) and \etmiss (bottom) in the CR0-2  control region for the 1-electron (left) and 1-muon (right) channels. The hatched band shows the systematic uncertainty, which includes both experimental uncertainties (among which JES and $b$-tagging uncertainties are dominant) and theoretical uncertainties on the background normalisation and shape. The small insets show the ratio between the observed distribution and that predicted for the Standard Model background. Although the distributions are presented separately for $e$ and $\mu$, the background estimation uses the sum of the $e$ and $\mu$ yields in the CR0.}
\label{fig:0lep_nonqcd_CR2}
\end{figure*}

Standard Model processes contributing to the total background in the signal regions are top quark production (single and in pairs), the production of a $W$  or a $Z$ boson in association with heavy-flavour quarks (mostly $b$, but also $c$), and multi-jet production. The last enters in the signal regions if missing transverse momentum is produced in the final state, either because of the mis-measurement of one or more of the jets in the event, or because of the semileptonic decay of a heavy-flavour hadron.

\begin{table}[htp]
 \centering
 \begin{tabular}{c|c|c|c||c|c}
   \hline
   \hline
   Control & top & $W/Z$ & multi-jet/ & SM & data \\
   Region &  & &di-boson & & (\lumi\ \ifb) \\
   \hline
   \hline
  CR0-1 (1 $e$)  & 187  & 48 & 1 & $235 \pm 45$ & 217\\ 
  CR0-1 (1 $\mu$)  & 146 & 22 & 1 & $169 \pm 45$ & 177\\ 
  \hline
   CR0-2 (1 $e$)  & 53 & 2 & 0.1 & $55 \pm 20$ & 64\\ 
  CR0-2 (1 $\mu$)  & 42 & 3 & 0.1 & $45 \pm 17$ & 62\\ 
   \hline
   CR1 (1 $e$) & 414 & 40 & 3.6 &$460 \pm 100$ & 465 \\
   CR1 (1 $\mu$) & 377 & 25 & 5.2 & $410 \pm 110$ & 420 \\
\hline
\hline

   \end{tabular}
   \caption{Expected background composition and comparison of the predicted total SM event yield to the measured event yield for \lumi\ \ifb\ for each of the control regions defined in the text. The column ``Top'' includes contributions from the single top, $\ttbar$, \ttbb and $\ttbar + W/Z$ production processes. The quoted uncertainty on the SM prediction includes only experimental systematic uncertainties (among which jet energy scale and $b$-tagging uncertainties are dominant).}
\label{tab:0lep_nonqcd_1}
\end{table}
\vspace{0.5cm}

\paragraph*{\bf Top and $W/Z$ background estimation:} The dominant  SM background contributions to the signal regions are evaluated using control regions with low expected yields from the targeted SUSY signals. They are defined by selecting events containing exactly one lepton, large \meff and low \mt.  The background estimation in each signal region is obtained by multiplying the number of events observed in the corresponding control region by a transfer factor, defined as the ratio of the MC predicted yield in the signal region to that in the control region:

\begin{equation}
N_{\mathrm{SR}} = \frac{N_{\mathrm{SR}}^{\mathrm{MC}}}{N_{\mathrm{CR}}^{\mathrm{MC}}} (N_{\mathrm{CR}}^{\mathrm{obs}}-N_{\mathrm{CR}}^{\mathrm{res}}) = T_f (N_{\mathrm{CR}}^{\mathrm{obs}}-N_{\mathrm{CR}}^{\mathrm{res}})
\label{eq:tf}
\end{equation}

\noindent where $N_{\mathrm{CR}}^{\mathrm{obs}}$ denotes the observed yield in the control region and $N_{\mathrm{CR}}^{\mathrm{res}}$ includes contributions from multi-jet production and, in the 0-lepton case, $W$ and $Z$ production. The advantage of this approach is that systematic uncertainties that are correlated between the numerator and the denominator of $T_f$ largely cancel out, provided that the event kinematics in the corresponding signal and control region are similar. 

Two control regions are defined for the 0-lepton channel, differing only in the number of $b$-tags required. These are used to determine the top background in the six signal regions. They are obtained by applying the same thresholds on the three jets and \etmiss as for the SR0, but requiring exactly one signal electron or muon.  The transverse mass must be in the range $40$~GeV$ < \mt < 100$~GeV and the effective mass $\meff$ should be larger than 600~GeV. The region CR0-1 is required to have at least one $b$-tag, and CR0-2 is required to have at least two $b$-tags. The definition of the control regions for the 0-lepton channel is summarised in the upper part of Table~\ref{tab:CR}. Figures~\ref{fig:0lep_nonqcd} and \ref{fig:0lep_nonqcd_CR2} show the \etmiss and \meff distributions obtained in CR0-1  and CR0-2 respectively, for the 1-electron and 1-muon case.

\begin{figure*}
 \centering

\includegraphics[width=\mywidth\columnwidth]{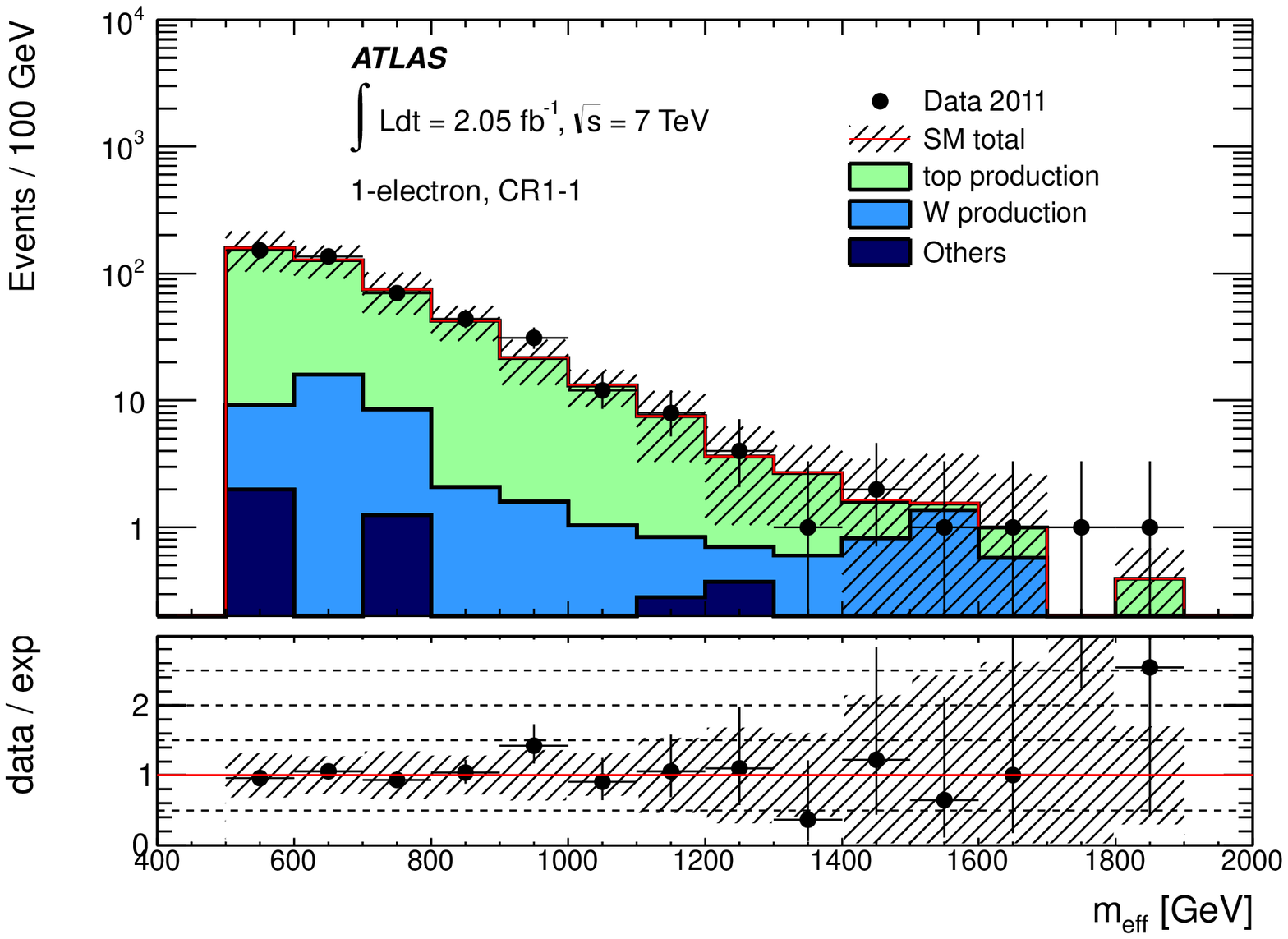}
\includegraphics[width=\mywidth\columnwidth]{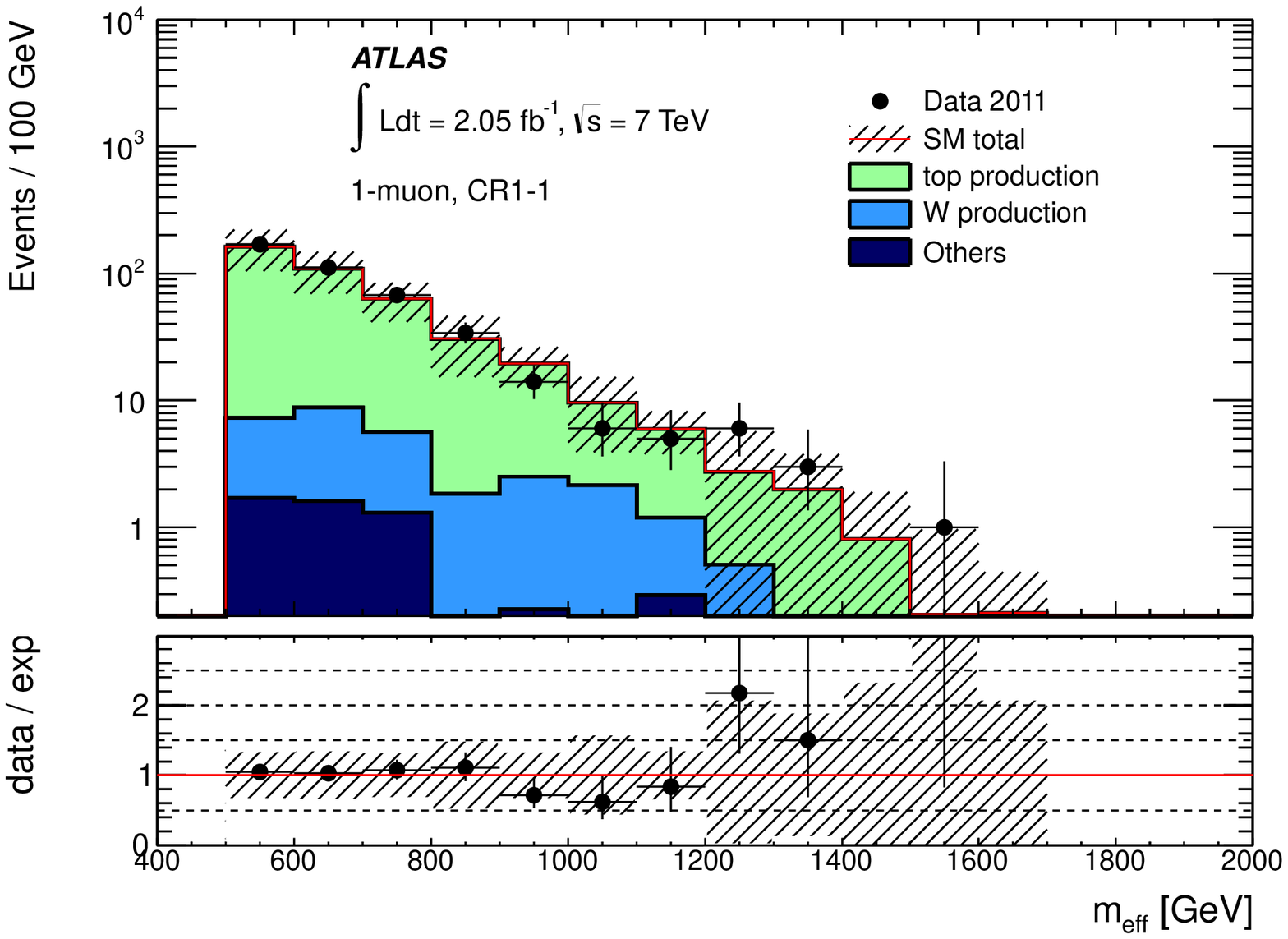}\\

 \includegraphics[width=\mywidth\columnwidth]{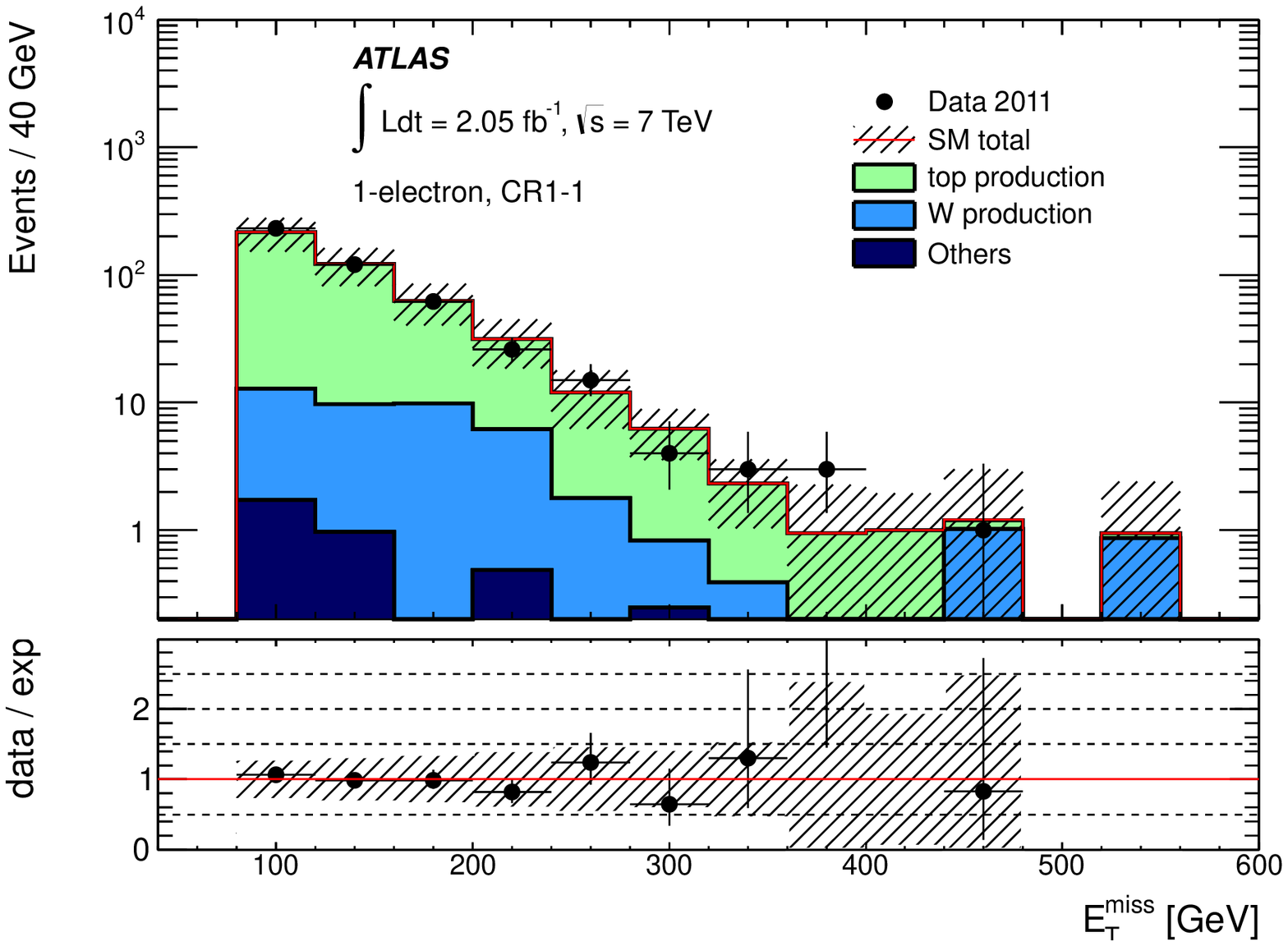}
 \includegraphics[width=\mywidth\columnwidth]{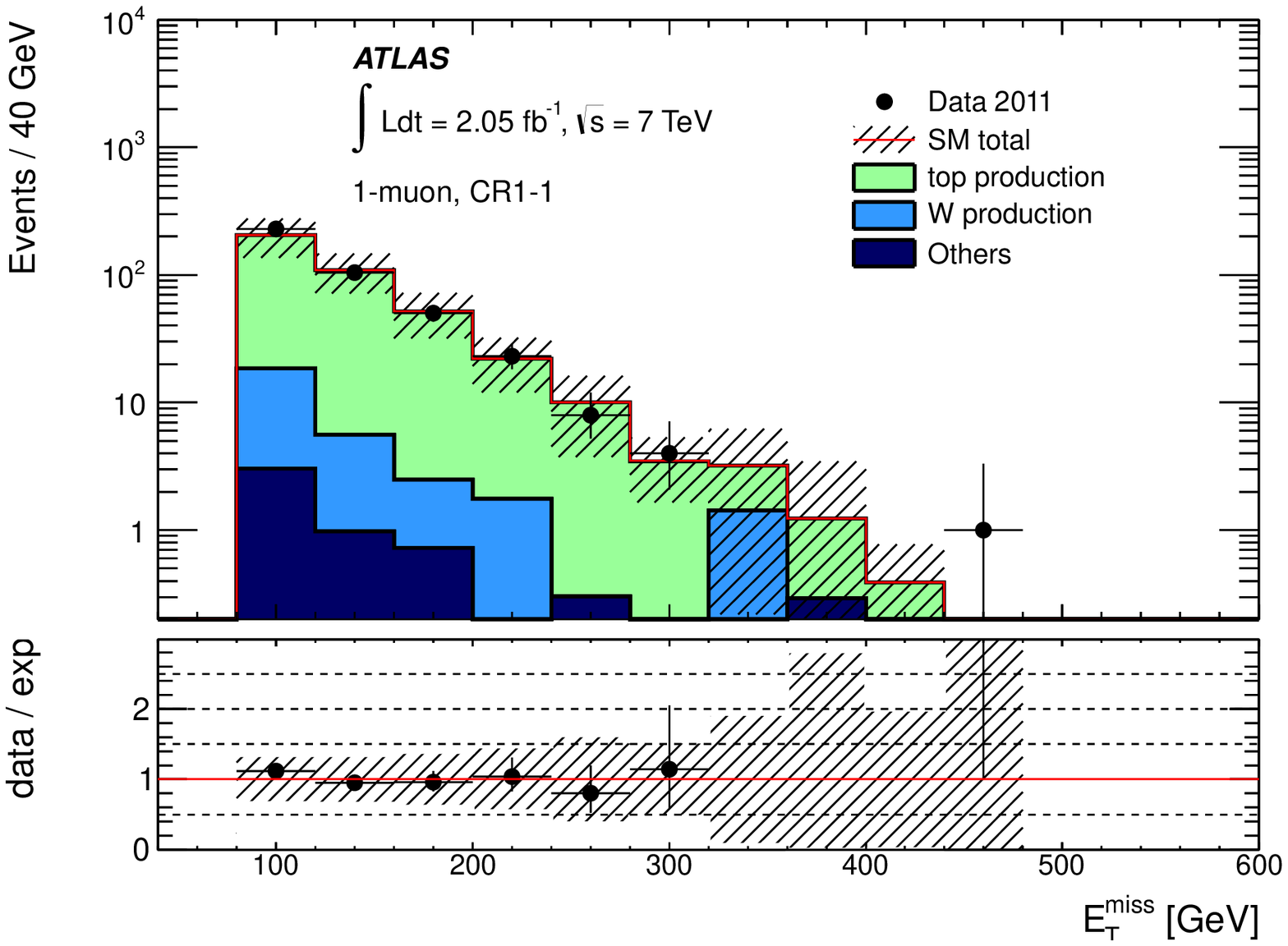}\\

	  \caption{Distribution of the effective mass (top) and \etmiss (bottom) in the CR1 control region for the 1-electron (left) and 1-muon (right) channels. The colour labelled ``Others'' includes contributions from $Z$, di-boson and  multi-jet production processes. The hatched band shows the systematic uncertainty, which includes both experimental uncertainties (among which JES and $b$-tagging uncertainties are dominant) and theoretical uncertainties on the background normalisation and shape. The small insets show the ratio between the observed distribution and that predicted for the Standard Model background.}
 \label{fig:1lep_nonqcd}
\end{figure*}

The formula used to obtain the top background prediction in each of the six signal regions is:

\begin{eqnarray}
N_{\mathrm{SR0\textrm{-}}\alpha j} = T_f^{\alpha j} (N_{\mathrm{CR0\textrm{-}}j}^{\mathrm{obs}} - N_{\mathrm{CR0\textrm{-}}j}^{\mathrm{non\textrm{-}top}}) \\
T_f^{\alpha j} = \frac{N_{\mathrm{SR0\textrm{-}}\alpha j}^{\mathrm{MC,top}}}{N_{\mathrm{CR0\textrm{-}}j}^{\mathrm{MC,top}}}
\end{eqnarray}

\noindent where $\alpha = A,B,C$, $j=1,2$ denote the six signal regions, $N_{\mathrm{CR0\textrm{-}}j}^{\mathrm{non\textrm{-}top}}$ includes the estimate for $W,Z$ and multi-jet production in the control region $j$, and all numbers are the sum of the corresponding electron and muon channel yields.  
	
The remaining SM contributions to the SR0 are mainly from $W$ and $Z$ production in association with heavy-flavour quarks. This corresponds to  about 30\% (10\%) of the total background in the signal regions defined with one \btag (two $b$-tags), and it is estimated from MC simulation.

For the 1-lepton channel signal regions, the total SM background (more than 90\% of which consists of top quark production) is determined using a similar technique, but using one single transfer factor for top, $W/Z$ and di-boson production processes. In this case, only one control region (CR1) is defined, requiring the same kinematic cuts applied in SR1-D, with the exception that the transverse mass should be in the range 40~GeV$<\mt<100$~GeV and that  $\meff > 500$~GeV. The last row of Table~\ref{tab:CR} summarises the event selection for the 1-lepton control region. Figure~\ref{fig:1lep_nonqcd} shows the \etmiss and \meff distributions in CR1.

\paragraph*{\bf Multi-jet background estimation:}
The small contribution of  multi-jet background in the SR0 signal region is estimated with the use of a jet response smearing 
technique~\cite{aad:2011ib}. Multi-jet events with possibly large $\etmiss$ are obtained by smearing jet energies in low $\etmiss$ ``seed'' events according to jet response functions obtained with the MC simulation. The Gaussian core of the response function is tuned to data by considering the jet balance in di-jet events, while its non-Gaussian tail is adapted to reproduce the  response in three-jet events where the $\etmiss$ can be unambiguously associated to a single jet.

The number of  multi-jet events in the CR0, CR1 control regions and SR1 signal regions is estimated using a matrix method similar to the one described in Ref.~\cite{w-paper}. The probability of misidentifying a tight lepton is estimated by 
computing the probability that pre-selected leptons are identified as signal leptons in low-\etmiss control regions dominated by  multi-jet events. 

\paragraph*{\bf Total Background:}
The number of expected events for \lumi\ \ifb\ of integrated luminosity as predicted by the MC and by the data-driven multi-jet estimate for all control regions is compared to that obtained in data in Table~\ref{tab:0lep_nonqcd_1}. The uncertainty quoted on the Standard Model prediction includes experimental systematic uncertainties (jet energy scale and resolution, $b$-tagging efficiency, lepton identification and energy scale, and luminosity determination).  

Further selection regions are used to validate the MC prediction in different kinematic regimes (in particular for small and large values of \mt at low value of \meff, for both the 0-lepton and 1-lepton channels). In all cases, a good agreement between the data and MC predictions is found.
\\
   
\section{Systematic uncertainties on background estimation}

Various systematic uncertainties affecting the background rates in the signal regions have been considered. Their treatment is discussed in the following paragraphs, and their impact on the absolute predicted event yield in the control and signal regions is evaluated. Such uncertainties are used either directly ($W,Z$ for the 0-lepton channel) in the evaluation of the predicted background in the signal regions, or to compute the $T_f$. In the latter case, the uncertainties on the absolute predicted event yield in the control regions and signal regions are propagated using Eq.~\ref{eq:tf} to obtain the signal region uncertainties.

Experimental systematic uncertainties arise from several sources:

\paragraph*{\bf Jet energy scale and resolution uncertainty:} The uncertainty on the jet energy scale (JES), derived using single particle response and test beam data, varies as a function of the jet \pt and pseudorapidity and it is about 2\% at $\pt = 50$~GeV in the central detector region. Additional systematic uncertainties arise from the dependence of the jet response on the number of expected interactions per bunch crossing and on the jet flavour. The total jet energy scale uncertainty at $\pt = 50$~GeV in the central detector region is about 5\% \cite{JES}. The jet energy scale uncertainty is propagated to obtain an uncertainty on the event yield by varying it by $\pm 1 \sigma$ in the MC simulation. Uncertainties related to the jet energy resolution (JER) are obtained with an in-situ measurement of the jet response asymmetry in di-jet events~\cite{JER}. Their impact on the event yield is estimated by applying an additional smearing to the jet transverse momenta. The JES and JER relative uncertainties 
on the event yield amount to a total of 20-40\% (depending on the signal region) and are completely dominated by the JES uncertainty. 

\paragraph*{\bf $b$-tagging efficiency and mis-tagging uncertainties:} The uncertainty associated with the tagging procedure used to identify $b$-jets 
is evaluated by varying the $b$-tagging efficiency and mis-tagging rates within 
the uncertainties evaluated on the central values measured in-situ \cite{btag}. The resulting relative uncertainty on the event
yield is about 20\% (35\%) in the one $b$-tag (two $b$-tags) signal region.

\paragraph*{\bf Further experimental uncertainties:} Other systematic uncertainties arise from the imperfect knowledge of the lepton identification efficiency and energy scale, the rate of  lepton misidentification and from the luminosity determination. Their contribution to the final uncertainty is found to be negligibly small.
\\

All the experimental systematic uncertainties are included, together with process-specific uncertainties, in the evaluation of the background uncertainty:

\paragraph*{\bf Multi-jet background:} The systematic uncertainty on the estimation of the multi-jet background in the SR0 is determined by taking into account statistical uncertainties and possible biases in the selection of the seed events, as well as uncertainties in the tuning of the tail of the jet response function in the three-jet events. The relative uncertainty varies between 50\% and 70\% depending on the SR0 considered.  

The estimated multi-jet background in the SR1 is affected by systematic uncertainties related to the determination of the lepton misidentification  rate and to the subtraction of non-multi-jet contributions to the event yield in the multi-jet enhanced region. The estimated relative uncertainty is 90\% in SR1-D and 100\%  and SR1-E.

\paragraph*{\bf $W$ and $Z$ production processes:} Systematic uncertainties on $W$ and $Z$ production are evaluated by varying the relative cross sections of the samples generated with the {\tt ALPGEN} MC with different numbers of outgoing partons \cite{toppaper}, resulting in an uncertainty of about 30\%. Additional uncertainties of about 70\% on the production cross section of $W$ and $Z$ bosons in association with $b$-quarks are considered. They are derived from direct measurements \cite{Aad:2011kp,sbottom}, and extrapolated using the MC simulation to include differences in the phase space regions probed by this analysis. Uncertainties related to the parton density function choice have been evaluated and found to be small compared to the large uncertainty already considered.

\begin{table}[htp]
 \centering
 \begin{tabular}{c|c|c|c|c|c||c}
   \hline
   \hline
	SR & JES/ & $b$-tag & lepton ID & theory & others & total\\
              & JER & & &   & & \\
\hline
   SR0-A1 & 4 & 3 & 2 & 11 & 10 & 15 \\
   SR0-B1 & 3 & 3 & 2 & 20 & 10 & 22 \\	
   SR0-C1 & 3 & 4 & 2 & 35 & 11 & 37 \\
   \hline
   SR0-A2 & 3 & 3 & 2 & 15 & 11 & 19 \\
   SR0-B2 & 3 & 4 & 2 & 20 & 10 & 22 \\
   SR0-C2 & 3 & 2 & 2 & 30 & 12 & 32 \\
   \hline
SR1-D  & 6 & 1 & 1 & 34 & 7 & 35  \\  
SR1-E  & 7 & 1 & 1 & 53  &  10 & 55  \\

\hline
   \end{tabular}
   \caption{Relative systematic uncertainties (in percent) associated with the background estimated by using transfer factors for all the signal regions considered. The column ``others'' includes statistical uncertainties on the event yield in the control regions, and, in the case of the 0-lepton channel, systematic uncertainties on the non-top production contributions subtracted from the control regions. The column ``theory'' contains theoretical uncertainties on the top production process addressed as discussed in the text.}
\label{tab:tf_summary}
\end{table}

\paragraph*{\bf Top production processes:} Theoretical uncertainties on the shape of \ttbar and single top kinematic distributions are evaluated by comparing different LO and NLO generators ({\tt ALPGEN} or {\tt POWHEG}, the latter using both {\tt PYTHIA} and {\tt HERWIG} as parton shower), and using different parton shower tunes, still consistent with data from previous experiments \cite{toppaper}. An additional uncertainty of 100\% is considered for \ttbar production in association with \bbbar\ or $W/Z$.
\\

The $T_f$,  used for the top and total SM background determination in the SR0 and SR1, respectively, are computed using MC predictions. Their values span from 1.8 to 0.05 depending on the signal region considered. Their associated uncertainty arises from both experimental (JES and JER, $b$-tagging efficiency and fake rate, lepton identification and energy scale) and event-generator level uncertainties. The use of control regions with similar kinematical properties to those of the signal regions strongly suppresses experimental uncertainties. Theoretical uncertainties typically dominate the total uncertainty on the $T_f$, which varies between 15\% and 35\%. 

A summary of the systematic uncertainties for the background estimates with the use of transfer factors is shown in Table~\ref{tab:tf_summary}.

\section{Results}

The \meff and \etmiss distributions are shown in Figure~\ref{fig:0lep_res} for SR0-A1 and SR0-A2, 
and in Figure~\ref{fig:1lep_res} for SR1-D. 
Tables~\ref{tab:results} and \ref{tab:results_1lep} show the Standard Model background predictions and the observed number of events corresponding to \lumi\ \ifb\ in all signal regions. The top background in the 0-lepton signal regions is estimated making use of the transfer factors, and its uncertainty corresponds to the total systematic uncertainty of Table~\ref{tab:tf_summary}. In parentheses, the MC prediction is reported for comparison. The $W/Z$ background and uncertainty in the SR0 are estimated directly with the MC simulation. The  multi-jet background contribution in the SR0, obtained with a data-driven estimate, is summed together with that of di-boson background. The SM background uncertainty in the SR1 corresponds to the total systematic uncertainty of Table~\ref{tab:tf_summary}, plus a small contribution arising from the data-driven estimate of the multi-jet background.

\begin{figure*}
  \centering
  \includegraphics[width=\mywidth\columnwidth]{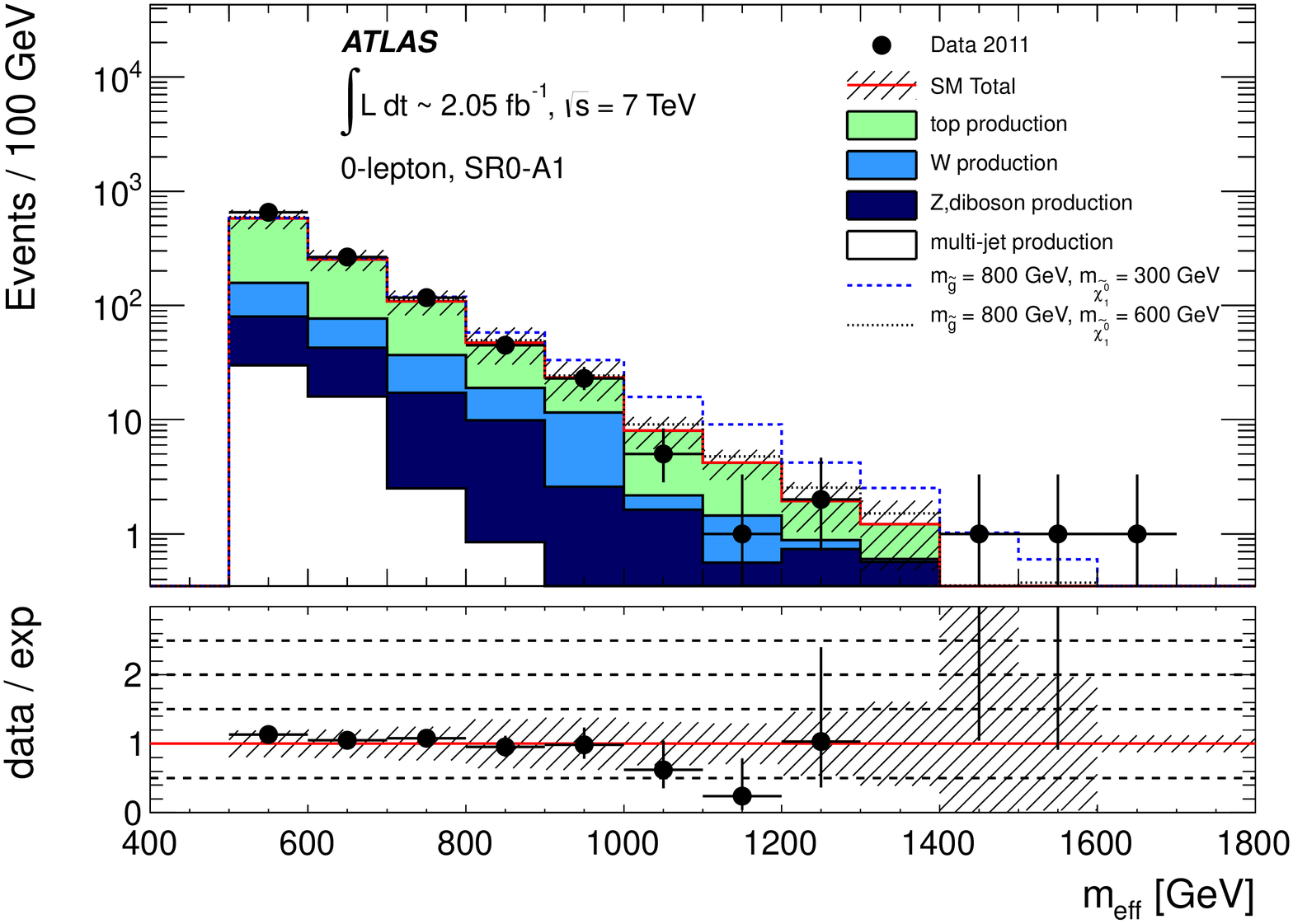}\includegraphics[width=\mywidth\columnwidth]{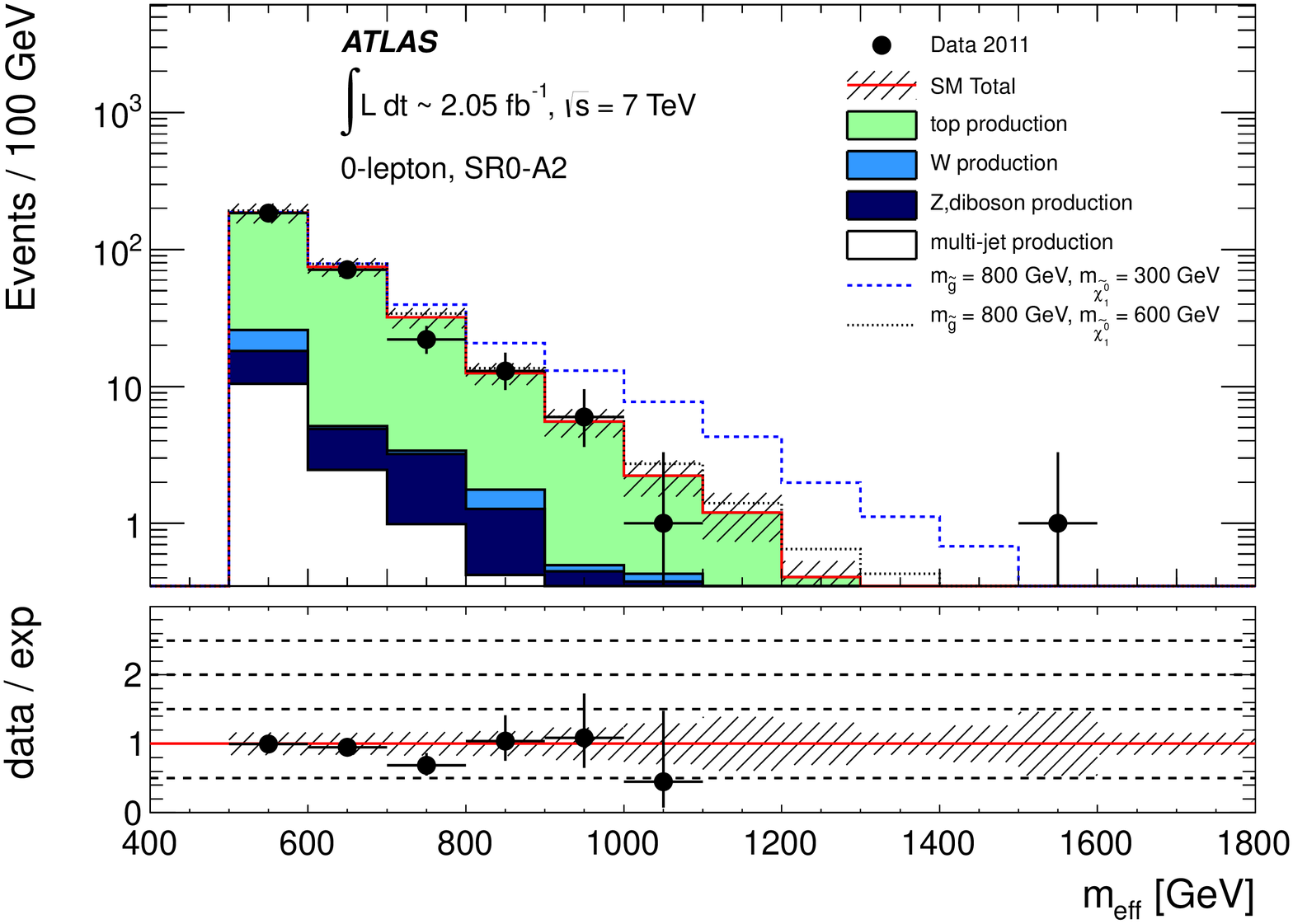} \\ 	    
  \includegraphics[width=\mywidth\columnwidth]{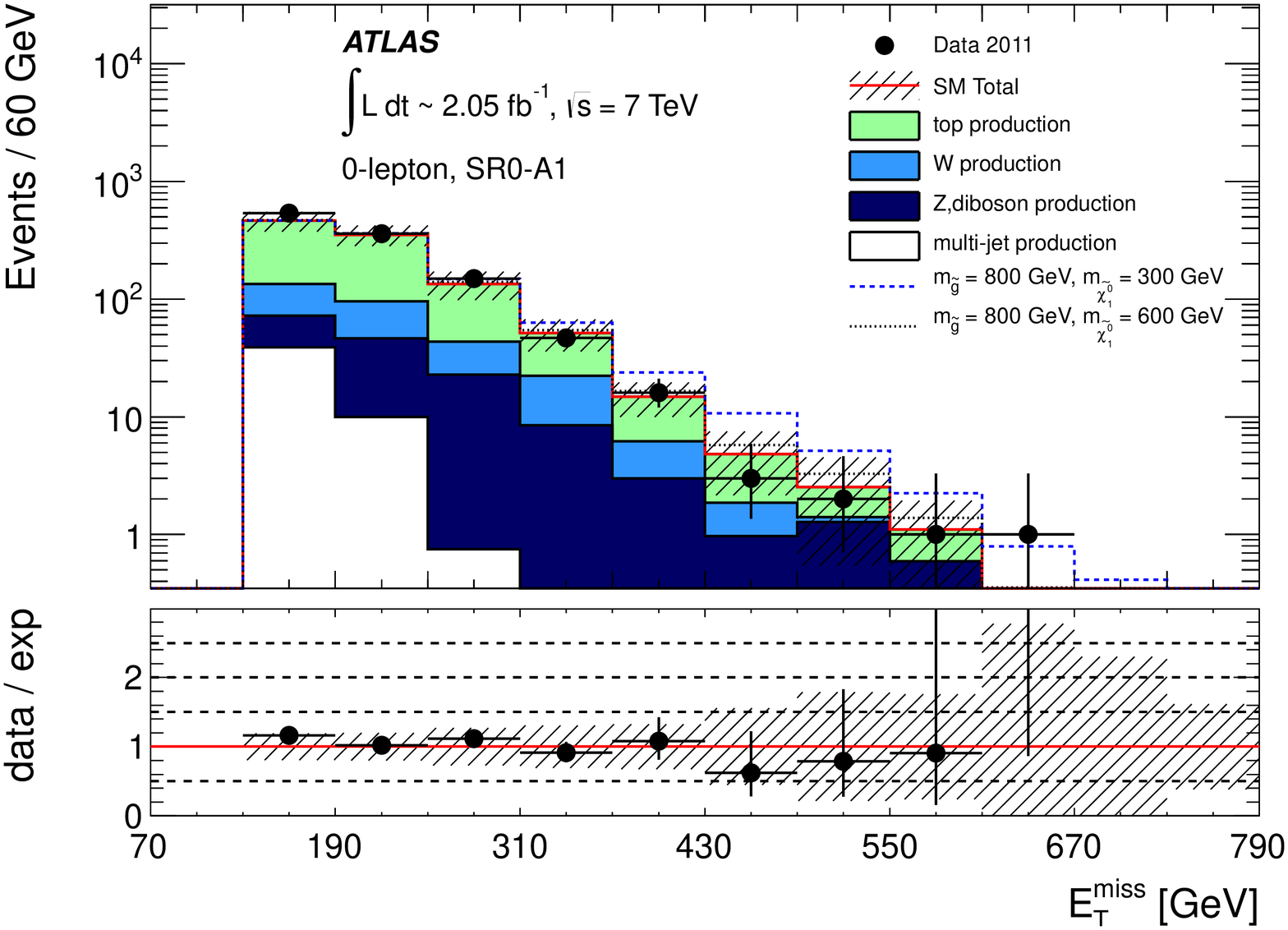}\includegraphics[width=\mywidth\columnwidth]{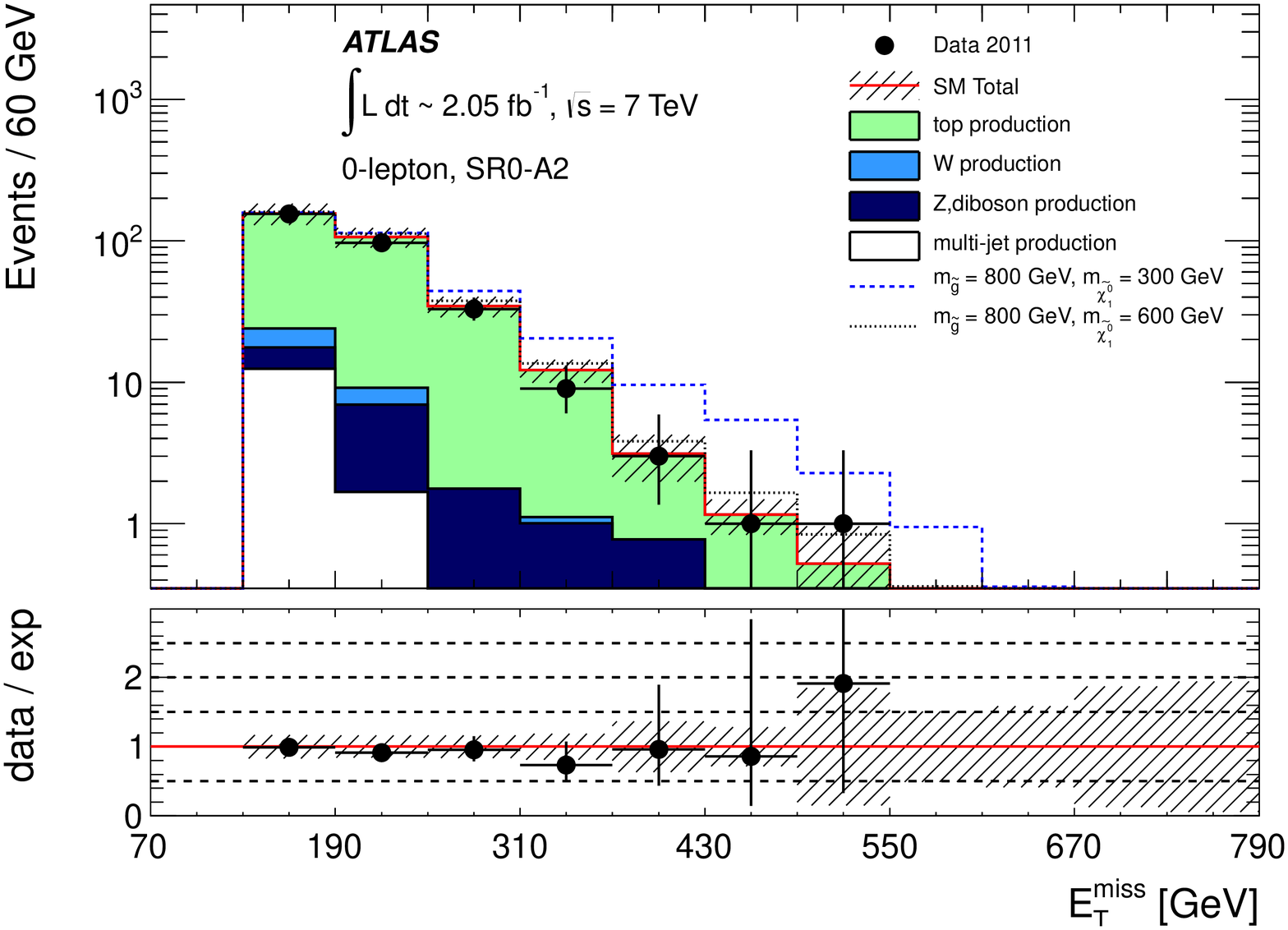} 	    
\caption{Distribution of the effective mass (top) and \etmiss (bottom) in SR0-A1 (left) and SR0-A2 (right). The hatched band shows the systematic uncertainty, which includes both experimental uncertainties (among which JES and $b$-tagging uncertainties are dominant) and theoretical uncertainties on the background normalisation and shape. The small insets show the ratio between the observed distribution and that predicted for the Standard Model background.}
\label{fig:0lep_res}
\end{figure*}

\begin{figure*}
\centering
\label{fig:1lep_meffres}\includegraphics[width=\mywidth\columnwidth]{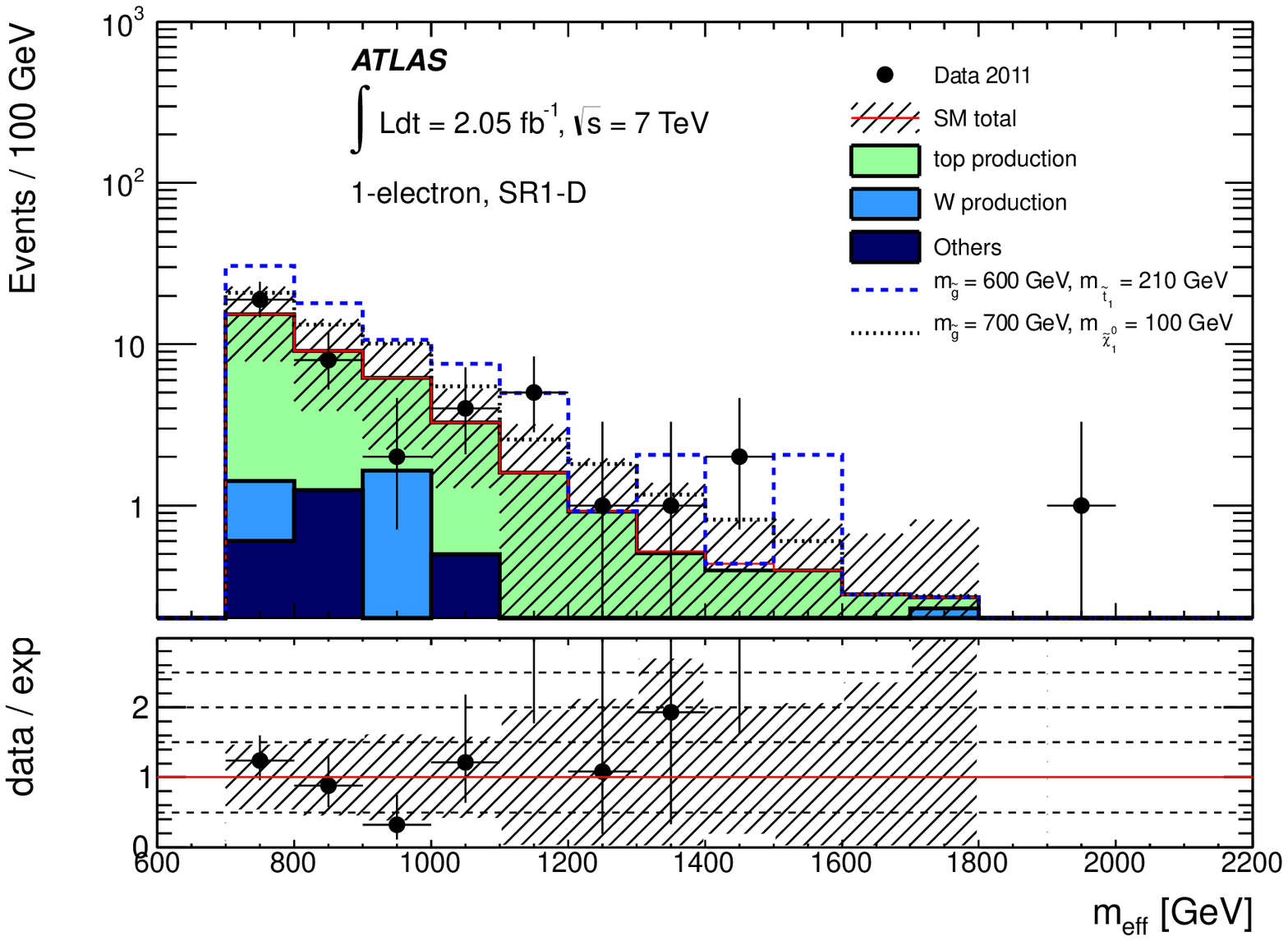}\includegraphics[width=\mywidth\columnwidth]{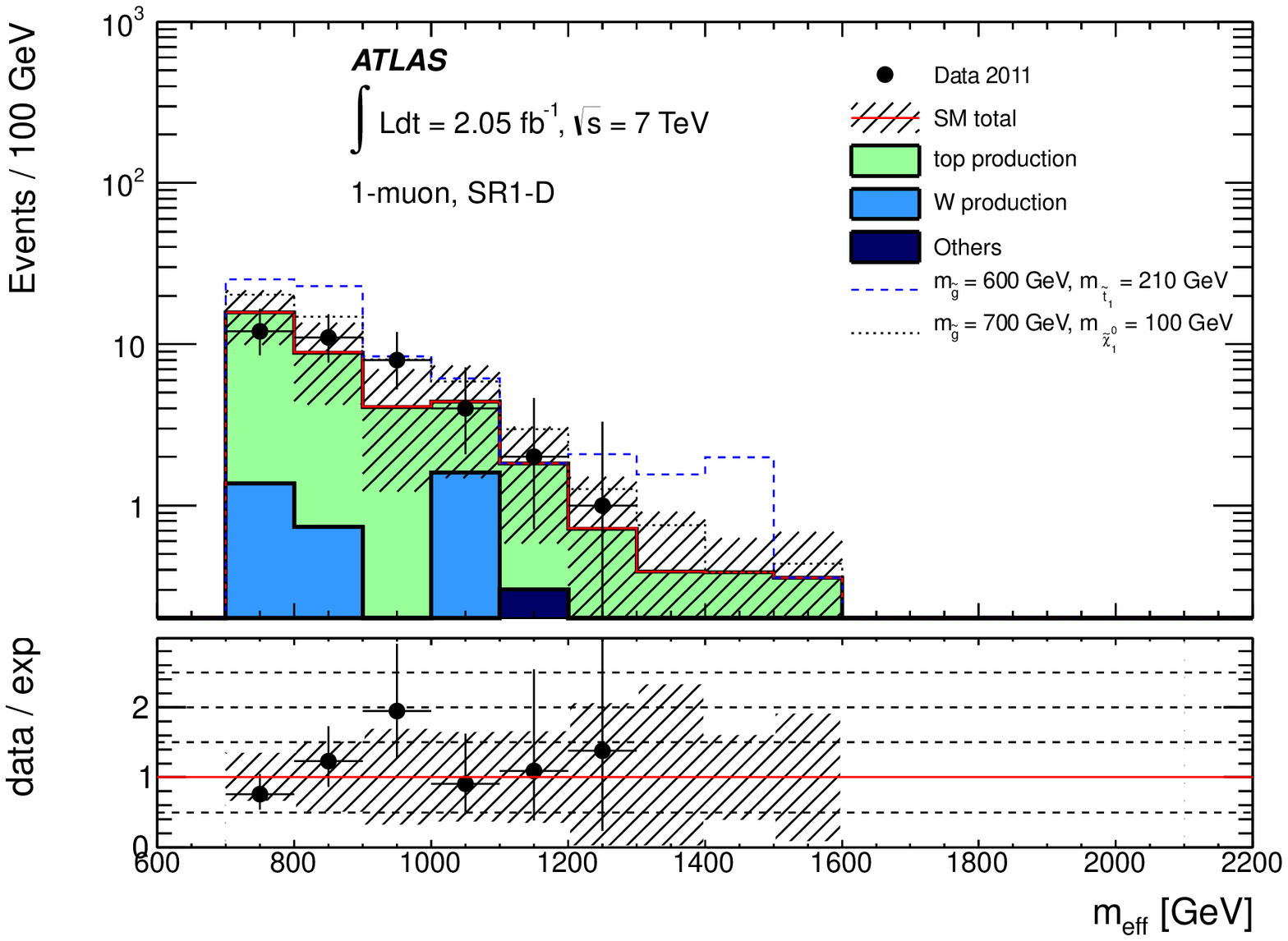} \\
\label{fig:1lep_metres}\includegraphics[width=\mywidth\columnwidth]{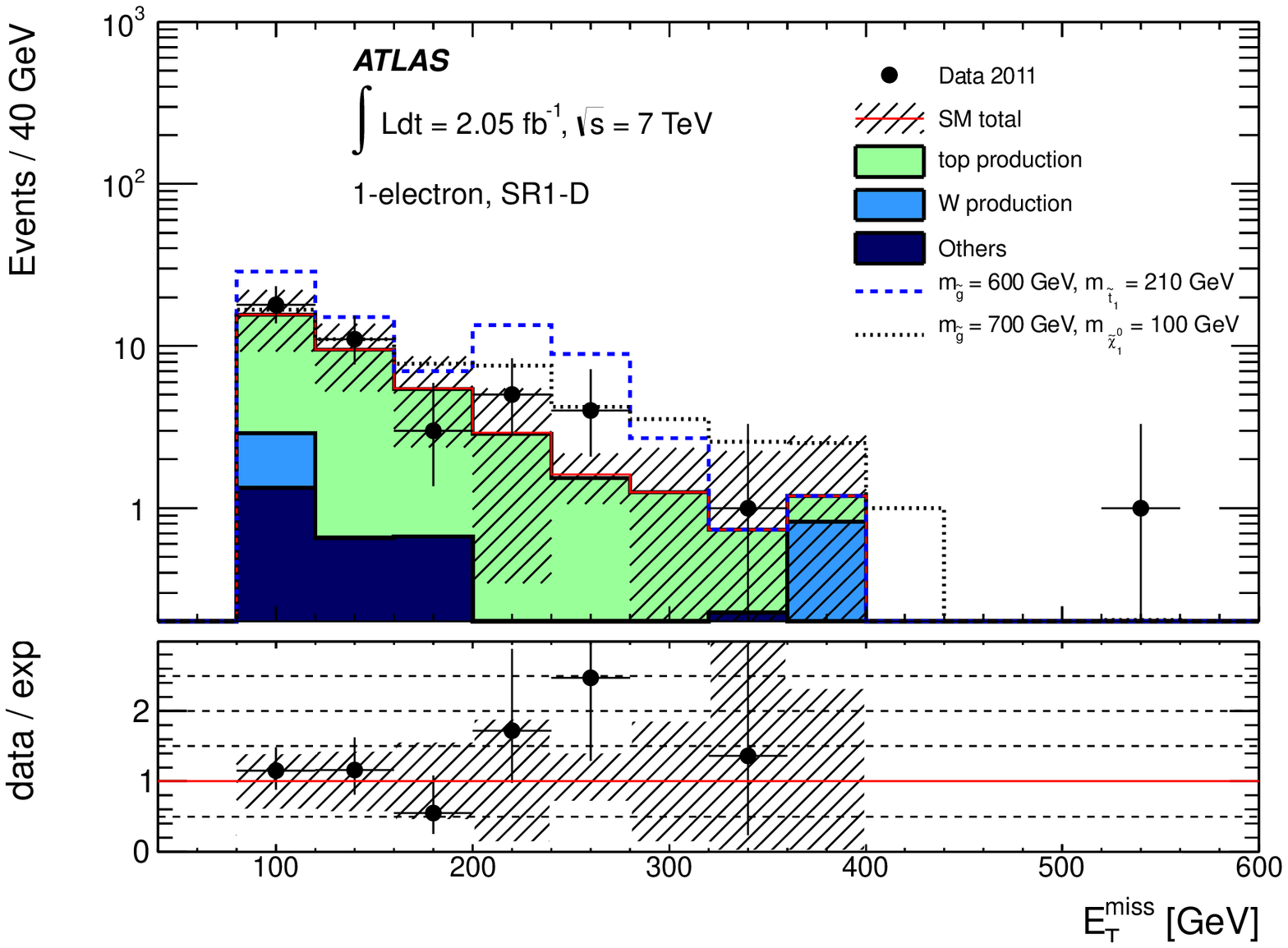}\includegraphics[width=\mywidth\columnwidth]{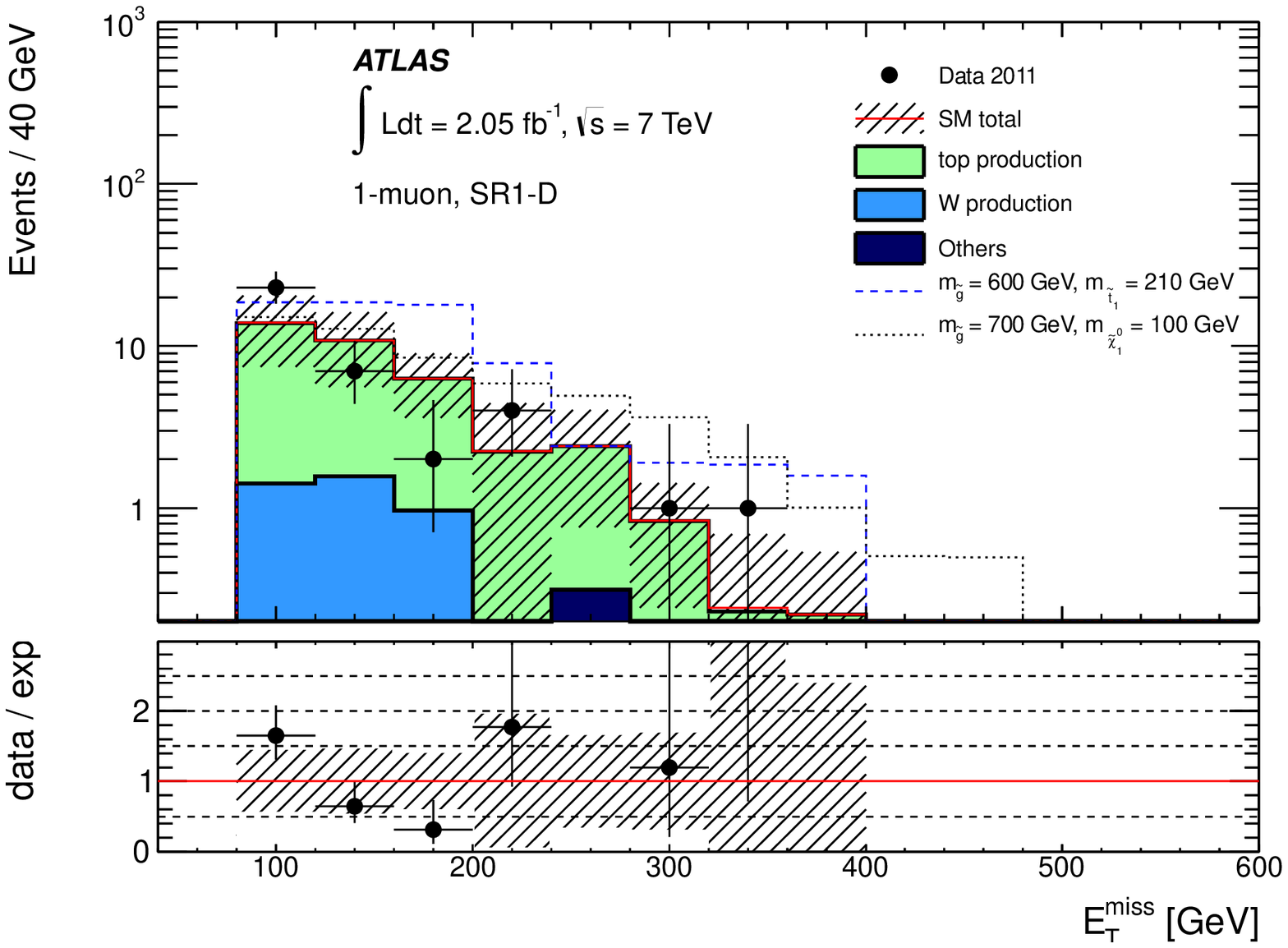} \\	    

\caption{Distribution of the effective mass (top) and \etmiss (bottom) for the 1-electron (left) and 1-muon (right) channel in SR1-D. The color labelled ``Others'' includes contributions from $Z$, di-boson and  multi-jet production processes. The hatched band shows the systematic uncertainty, which includes both experimental uncertainties (among which JES and $b$-tagging uncertainties are dominant) and theoretical uncertainties on the background normalisation and shape. The small insets show the ratio between the observed distribution and that predicted for the Standard Model background.}
\label{fig:1lep_res}
\end{figure*}

\begin{table}[htp]
 \centering
 \begin{tabular}{c|c|c|c||c||c}
   \hline
   \hline
   SR & Top & W/Z & multi-jet/ & Total  & Data\\
   &  &  & di-boson &   & \\
   \hline
  SR0-A1 &$705\pm 110$& $248 \pm 150$ & $53 \pm 21$ & $1000 \pm 180$  & 1112\\
  & (725) & & & & \\
  SR0-B1 & $119\pm 26$ & $67 \pm 42$ & $7.3 \pm 4.7$ & $190 \pm 50$  & 197\\
  & (122) & & & & \\
  SR0-C1 & $22\pm 8$ & $16 \pm 11$ & $1.5 \pm 1$ & $39 \pm 14$ & 34 \\
  & (22) & & & & \\
  SR0-A2 &$272\pm 52$  & $23 \pm 15$ & $21 \pm 12$  & $316 \pm 54$ & 299\\
  & (212) & & & & \\
  SR0-B2 &$47\pm 10$  & $4.5 \pm 3$ & $2.8 \pm 1.7$ & $54 \pm 11$ & 43\\
  & (37) & & & & \\
  SR0-C2 & $8.5\pm 3$ & $0.8 \pm 1$ & $0.5 \pm 0.4$ & $9.8\pm 3.2$ & 8\\
  & (6.6) & & & & \\
   \hline
   \hline
   \end{tabular}
   \caption{Summary of the expected and observed event yields corresponding to \lumi~\ifb\ in the six 0-lepton channel signal regions. The errors on the top contribution correspond to the total errors of Table~\ref{tab:tf_summary}. The errors quoted for all background processes include all the systematic uncertainties discussed in the text. The numbers in parentheses in the ``Top'' column are the yields predicted by the MC simulation.}
\label{tab:results}
\end{table}

\begin{table}[htp]
 \centering
 \begin{tabular}{c|c||c}
   \hline
   \hline
   SR & SM background  & Data\\
   \hline
  SR1-D ($e$)&$39\pm 12$ (39)&  43\\
  SR1-D ($\mu$)& $38\pm 14$ (37)& 38\\
  SR1-E ($e$)& $8.1\pm 3.4$ (7.9)& 11\\
  SR1-E ($\mu$)& $6.3 \pm 4.2$ (6.1)& 6\\
   \hline
   \hline
   \end{tabular}
   \caption{Summary of the expected and observed event yields corresponding to \lumi~\ifb\ in the two 1-lepton channel signal regions. The Standard Model estimation is derived with the data-driven method discussed in the text. The numbers in parentheses in the ``SM background'' column are the sum of  the yield predicted by the MC and the data-driven estimate for the multi-jet background.}
\label{tab:results_1lep}
\end{table}

The results are consistent with the Standard Model predictions, and they are therefore translated into 95\% confidence-level (CL) upper limits on 
contributions from new physics using the $CL^{}_{\rm s}$ prescription~\cite{profilelike}. 
The likelihood function used is written as 
$L({n}|s,{b},{\btheta}) = P_{\mathrm{s}} \times \, C_{\mathrm{Syst}}$;      
where ${n}$ represents the number of observed events in data, $s$ is the SUSY signal under 
consideration, ${b}$ is the background, and ${\btheta}$ represents the systematic 
uncertainties. The $P_{\mathrm{s}}$ function is a Poisson-probability distribution for event counts 
in the defined signal region and $C_{\mathrm{Syst}}$ represents the constraints on 
systematic uncertainties, which are treated as nuisance parameters with a Gaussian 
probability density function and correlated when appropriate. 

Upper limits at 95\% CL on the number of signal events in the signal   
regions are obtained independently of new physics models and assuming no signal contamination in the control regions for the 0-lepton and 1-lepton 
final states. Results for observed and expected upper limits on the number of non-SM events in 
the signal regions are shown in Table~\ref{tab:ULind}, as well as upper limits on 
the visible cross section,  $\sigma^{}_{\mathrm{vis}}$, defined as cross section times experimental acceptance and efficiency.

\begin{table}[htp]
 \centering
    \renewcommand\arraystretch{1.25}
 \begin{tabular}{c|c|c}
   \hline\hline 
   SR & \multicolumn{2}{|c}{95\% CL upper limit} \\
   \hline 
   & N events & $\sigma_{\mathrm{vis}}$(fb)  \\ 
             & obs. (exp.) & obs. (exp.) \\ \hline 
  SR0-A1 & 580 (520) & 283 (254) \\
  SR0-B1 & 133 (133) & 65 (65) \\
  SR0-C1 & 31.6 (34.6) & 15.4 (16.9) \\
  SR0-A2 & 124 (134) & 61 (66) \\
  SR0-B2 & 29.6 (31.0) & 14.4 (15.0) \\
  SR0-C2 & 8.9 (10.3) & 4.3 (5.0) \\ \hline 
  SR1-D & 45.5 (42.1) & 22.2 (20.5) \\
  SR1-E & 17.5 (15.3)& 8.5 (7.5)\\ 
   \hline\hline
  \end{tabular}
  \caption{Observed and expected 95\% CL upper limits on the non-SM contributions to all signal regions.  
   Limits are given on the number of signal events and in terms of visible cross sections. No assumptions 
are made on the possible presence of non-SM signal in the control regions. The systematic
   uncertainties on the SM background estimation are included. }  
  \label{tab:ULind}
\end{table}

\section{Interpretation in Simplified SUSY Models}

The interpretation of the results in terms of 95\% CL exclusion limits are given for several SUSY scenarios.  
The  exclusion limit contours are derived by subtracting possible signal contributions from the data yield  
in the control regions employed to estimate the SM background. The signal contamination is not negligible only for SUSY models 
leading to leptonic final states and accounts for less than 5\% of the SM predictions around the expected exclusion limit contours.  

Simplified models are characterised by well-defined SUSY particle production 
and decay modes yielding the final states under study. 
In the scenarios considered here scalar bottoms and tops are the 
only squarks to appear in the gluino decay cascade, leading to final states 
with large \bjet~multiplicity. The models listed below are addressed (in parenthesis the channel which is used for the interpretation of the result is given):    

\paragraph*{\bf Gluino-sbottom models (0-lepton):} MSSM scenarios where the $\bone$ is the lightest squark, all other 
squarks are heavier than the gluino, and $m^{}_{\gl}>m^{}_{\bone}>m^{}_{\neut}$, such that  
the branching ratio for $\gl \rightarrow \bone b$ decays is 100\%. Sbottoms  
are produced via $\gl\gl$ or by direct pair production $\bone\bone$ and are assumed to decay exclusively via 
$\bone \rightarrow b\neut$, where $m^{}_{\neut}$ is set to 
60~GeV. Exclusion limits are presented in the ($m^{}_{\gl},m^{}_{\bone}$) plane.
\paragraph*{\bf Gbb models (0-lepton):} Simplified scenarios, where $\bone$ 
is the lightest squark but $m^{}_{\gl}<m^{}_{\bone}$. 
Pair production of gluinos is the only process taken into account since 
the mass of all other sparticles apart from the $\neut$ is set above the TeV scale. 
A three-body decay via off-shell sbottom is assumed for gluino, 
such that ${\bone}{^{(*)}} \rightarrow b\neut$ (BR=100\% for $\gl \rightarrow b\bar{b}\neut$). 
Exclusion limits are presented in the ($m^{}_{\gl},m^{}_{\neut}$) plane.  
\paragraph*{\bf Gluino-stop models (1-lepton):} MSSM scenarios where the $\tone$ is the lightest squark, all other 
squarks are heavier than the gluino, and $m^{}_{\gl}>m^{}_{\tone}+m^{}_{t}$, such that  
the branching ratio for $\gl \rightarrow \tone t$ decays is 100\%. Stops   
are produced via $\gl\gl$ and $\tone\tone$ and are assumed to decay exclusively via 
$\tone \rightarrow b \chipm$. 
The neutralino mass is set to 60~GeV, the chargino mass to 120~GeV and the latter is assumed  to decay through a virtual $W$ boson 
(BR($ \chipm \to \neut l^{\pm} \nu$)=11\%). If $m_{\tone} > m^{}_{\neut} + m_t$, the decay $\tone \rightarrow t\neut$ is also kinematically allowed, with BR depending on the MSSM parameters settings. However, this mode is not considered for this interpretation, leading to conservative results, and is adopted in the $Gtt$ scenario,
 described below. Exclusion limits are presented in the 
($m^{}_{\gl},m^{}_{\tone}$) plane.
\paragraph*{\bf Gtt models (1-lepton):} Simplified scenarios,  
where $\tone$ is the lightest squark but $m^{}_{\gl}<m^{}_{\tone}$.
Pair production of gluinos is the only process taken into account since 
the mass of all other sparticles apart from the $\neut$ is set above the TeV scale. 
A three-body decay via off-shell stop is assumed for the gluino, 
such that ${\tone}{^{(*)}} \rightarrow t\neut$ (BR=100\% for $\gl \rightarrow t\bar{t}\neut$). 
Exclusion limits are presented in the ($m^{}_{\gl},m^{}_{\neut}$) plane.  
\paragraph*{\bf Gtb models (1-lepton):} Simplified scenarios, where $\bone$ and $\tone$ are the lightest squarks but 
$m^{}_{\gl}<m^{}_{\bone, \tone}$. As for the models above, pair production of gluinos is 
the only process taken into account, with gluinos decaying via virtual stops or sbottoms with a BR of 100\% 
assumed for $\tone \rightarrow b+\chipm$ and $\bone \rightarrow t+\chipm$, respectively. 
The mass difference between charginos and neutralinos is set to 2~GeV, such that  
the products of $\chipm \rightarrow \neut + ff'$ are invisible to the event selection, and 
gluino decays result in three-body final states ($b\bar{t}\neut$ or $t\bar{b}\neut$).   
Exclusion limits are presented in the ($m^{}_{\gl},m^{}_{\neut}$) plane. 
\\

The 0-lepton analysis is mostly sensitive to the SUSY scenarios where sbottom 
production dominates, whilst the 1-lepton analysis results are employed to set exclusion 
limits in models characterised by on-shell or off-shell stop production, where top-enriched 
final states are expected. 
Since several signal regions are defined for each analysis, the SR with 
the best expected sensitivity at each point in parameter space is adopted as the 
nominal result across the different planes.

The efficiency times acceptance of the selection strongly depends on the parameters of the model and the signal region considered. It varies between 5\% and 50\% in the proximity of the expected  limit for the gluino-sbottom model. For the $Gbb$ models, the efficiency times acceptance is highly dependent on the difference in mass between the gluino and the neutralino. It is about 1\% for a mass difference of about 200~GeV, and it increases up to 45\% for larger mass splitting. In the $Gtb$, gluino-stop and  $Gtt$ models, the efficiency times acceptance varies typically between 1 and 20\% in the proximity of the expected limit.

Systematic uncertainties on the signal include experimental (JES, JER, $b$-tagging) and theoretical  uncertainties. Experimental uncertainties are considered fully correlated with those obtained for the background, and they typically amount to 10-30\% depending on the signal region and  model considered. Theoretical uncertainties on the expected SUSY signal are estimated by varying the factorisation and renormalisation scales in {\tt PROSPINO} between half and twice their default values and by considering the PDF uncertainties provided by CTEQ6. Uncertainties are calculated for individual production processes and are typically 20-35\% in the vicinity of the expected limit.
 
Figure~\ref{fig:sb_gl_obs} shows the observed and expected exclusion regions 
in the ($m^{}_{\gl},m^{}_{\bone}$) plane for the gluino-sbottom model. The selection SR0-C2 provides the best 
sensitivity in most cases. If $m_{\gl}-m_{\bone}<100$~GeV, signal regions 
with one \btag are preferred, due to the lower number of expected \bjets above \pt 
thresholds.  
Gluino masses below 920~GeV are excluded for sbottom masses up to about 800~GeV. 
The exclusion is less stringent in the region with low $m_{\gl}-m_{\bone}$, 
where low \met\ is expected. 
This search extends the previous ATLAS exclusion limit in the same scenario by 
about 200~GeV, and it is complementary to direct searches for sbottom pair 
production published by the ATLAS collaboration~\cite{sbottom} using the same data-set. 
The limits do not strongly depend on the neutralino mass assumption 
as long as $m^{}_{\gl}-m^{}_{\tilde{\chi}_0}$ is larger than 300~GeV, due to the 
harsh kinematic cuts. 

The interpretation of the results in the \emph{Gbb} models, defined in the 
($m^{}_{\gl},m^{}_{\neut}$) plane at sbottom mass larger than 1 TeV, can be considered 
complementary to the previous one, defined in $m^{}_{\gl},m^{}_{\bone}$ 
at fixed $\neut$ mass. Figure~\ref{fig:GbbMax_OR} shows the expected and 
observed exclusion limit contours and the maximum 95\% upper cross section limit 
for each model. Gluino masses below 900~GeV are excluded for neutralino masses 
up to about 300~GeV.

\begin{figure}[h!]
  \begin{center}
      \includegraphics[width=0.97\columnwidth]{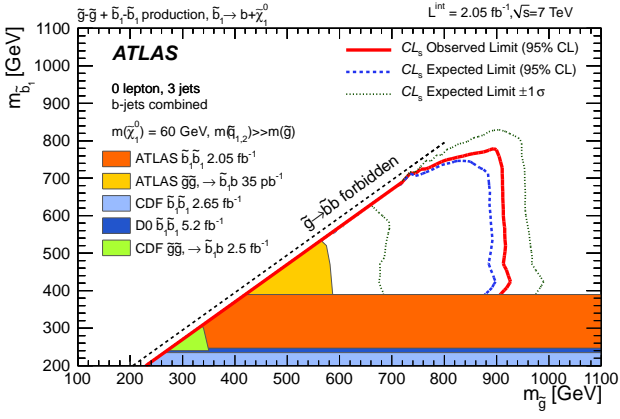}
    \end{center}
  \caption{Observed and expected 95\% CL exclusion limits in the ($m^{}_{\gl},m^{}_{\bone}$) plane (gluino-sbottom models). 
  For each scenario, the signal region providing 
  the best expected limit is chosen. The neutralino mass is assumed to be 60~GeV and the NLO cross sections are 
  calculated using {\tt PROSPINO}. 
  The result is compared to previous results from ATLAS~\cite{Aad:2011ks} and CDF~\cite{gluinosbottomCDF} searches which assume the same gluino-sbottom 
  decays hypotheses. Exclusion limits from the CDF~\cite{sbottomCDF}, D0~\cite{sbottomD0} and ATLAS~\cite{sbottom} 
  experiments on direct sbottom pair production are also shown.}
  \label{fig:sb_gl_obs}
\end{figure}
\begin{figure}[h!]
\begin{center}
\includegraphics[width=0.97\columnwidth]{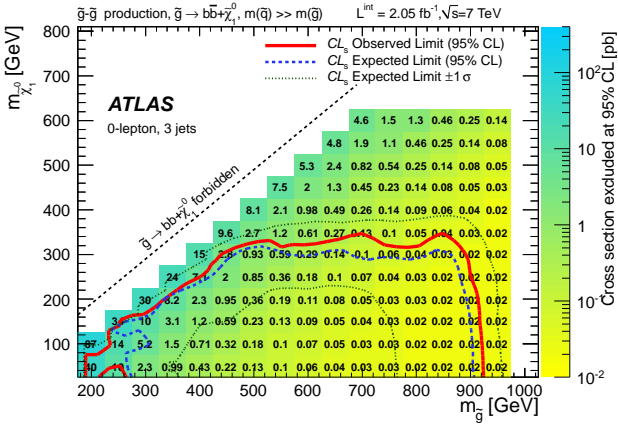}  
\end{center}
\caption{Observed and expected 95\% CL exclusion limits in the ($m^{}_{\gl},m^{}_{\neut}$) plane (\emph{Gbb} models).
  For each scenario, the signal region selection providing  the best expected limit is chosen. }
\label{fig:GbbMax_OR}
\end{figure}

Figures~\ref{fig:limit_1lepton_stop_combined} to~\ref{fig:limit_1lepton_Gtb_combined} 
report the interpretations of the 1-lepton analysis results in different scenarios. 
As for the 0-lepton results, the selection yielding the best expected limit for a 
given parameter point is used. 

Figure~\ref{fig:limit_1lepton_stop_combined} shows upper limits in the 
($m_{\tilde{g}},m_{\tilde{t}_1}$) plane for the gluino-stop model. Gluino masses below 620~\GeV\ are excluded 
at 95\% CL for stop masses up to 440~\GeV. 
The observed and expected upper limits at 95$\%$ CL extracted in the 
($m_{\tilde{g}},m_{\tilde{\chi}_1^0}$) plane for the \emph{Gtt} models are shown in 
Figure~\ref{fig:limit_1lepton_Gt_combined}. The upper cross section 
limits at 95$\%$ CL are also reported for each MSSM scenario. In this case, 
gluino masses below 750~\GeV\ are excluded at 95\% CL for $m_{\neut} = 50$~GeV while 
neutralino masses  below 160~\GeV\ are excluded at 95\% CL for $m_{\tilde{g}} = 700$~GeV. 

Figure~\ref{fig:limit_1lepton_Gtb_combined} shows upper limits at 95$\%$ CL 
for the \emph{Gtb} models. Only scenarios 
with chargino masses above the experimental limits from LEP experiments 
are considered, and gluino masses below 720~\GeV\ are excluded at 
95\% CL for $m_{\neut} = 100$~GeV while neutralino masses  below 200~\GeV\ are 
excluded at 95\% CL for $m_{\tilde{g}} = 600$~GeV. 
The contribution of the 0-lepton channel signal regions to the significance has been also evaluated for this scenario and found to be lower than that of the 1-lepton channel.

\begin{figure}
\centering
\includegraphics[width=0.97\columnwidth]{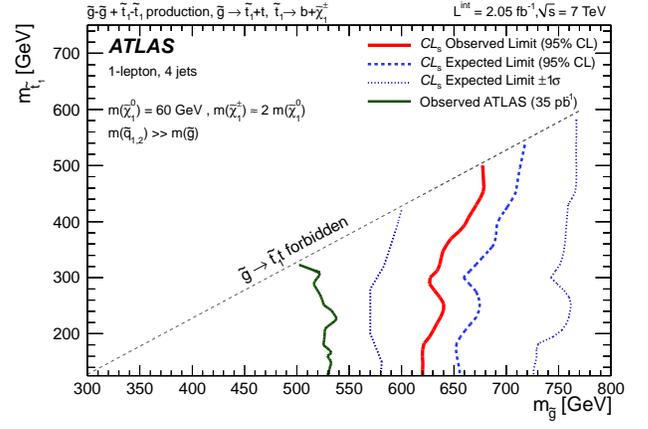}
\caption{
The observed and expected 95\% CL exclusion limits in the ($m^{}_{\gl},m^{}_{\tone}$) plane (gluino-stop models)
using the best expected limit between  SR1-D and SR1-E for each signal point. The result is compared to previous 
results from ATLAS~\cite{Aad:2011ks} searches which assume the same gluino-stop decays hypotheses.}
\label{fig:limit_1lepton_stop_combined}
\end{figure}

\begin{figure}
\centering
\includegraphics[width=0.97\columnwidth]{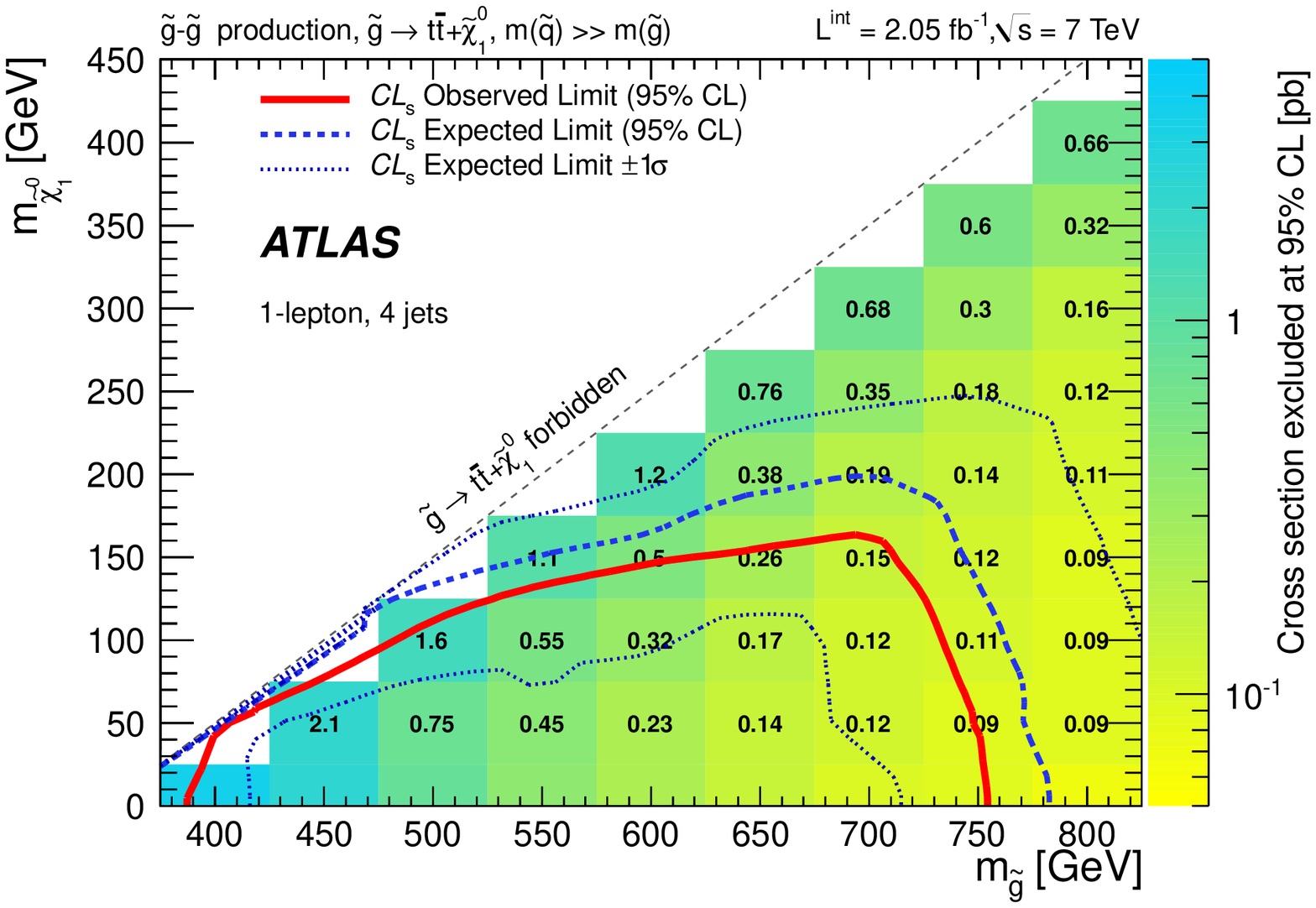}
\caption{The observed and expected 95\% CL exclusion limits in the ($m_{\tilde{g}},m_{\tilde{\chi}_1^0}$) 
plane (\emph{Gtt}) using the best expected limit between 
 SR1-D and SR1-E for each signal point.}
\label{fig:limit_1lepton_Gt_combined}
\centering
\includegraphics[width=0.97\columnwidth]{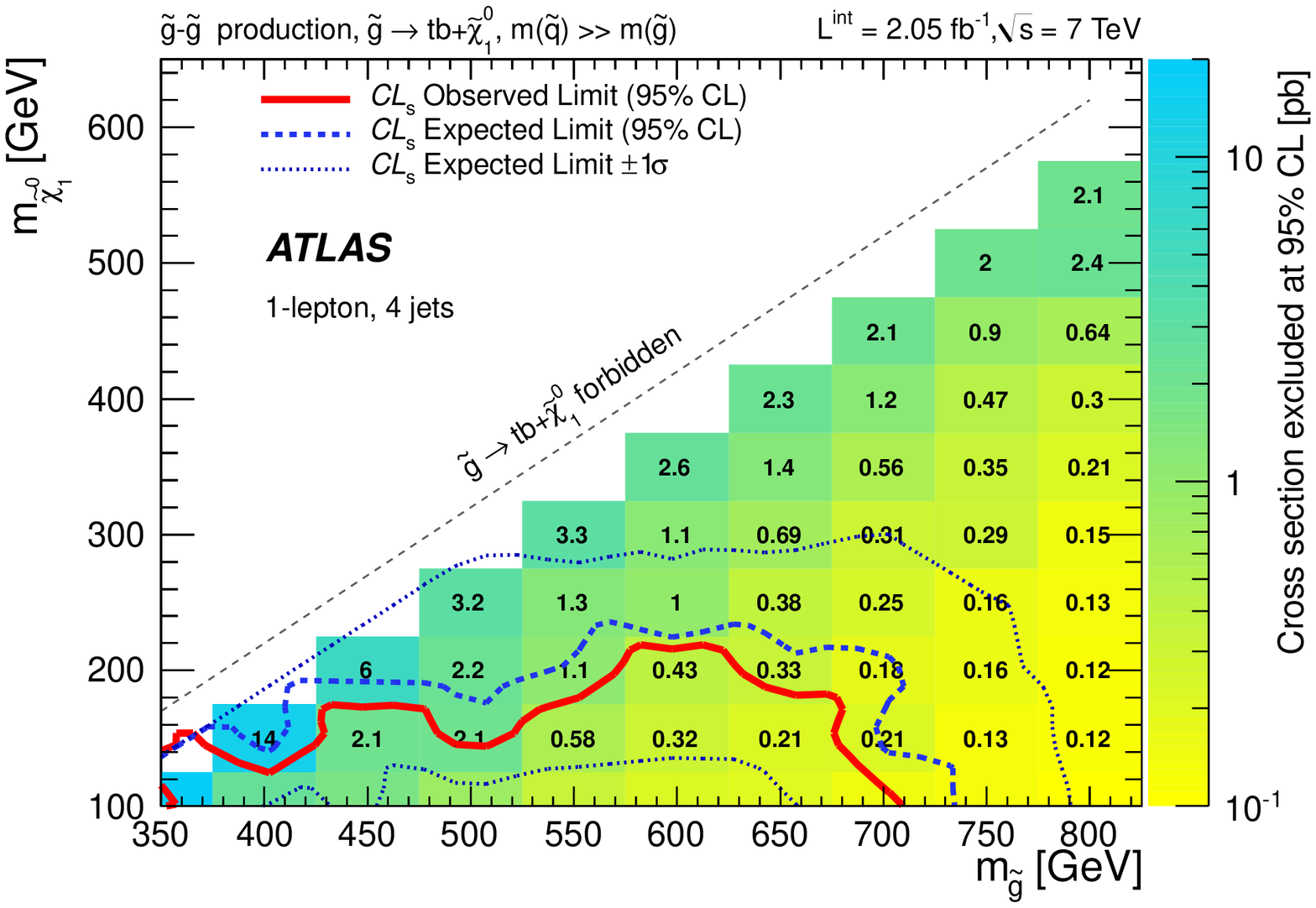}
\caption{The observed and expected 95\% CL exclusion limits in the ($m_{\tilde{g}},m_{\tilde{\chi}_1^0}$) 
 plane (\emph{Gtb} models) using the best expected limit between 
 SR1-D and SR1-E for each signal point.}
\label{fig:limit_1lepton_Gtb_combined}
\end{figure}

\section{Interpretation in SO(10) Models}

In addition to the simplified model interpretation, 
results are interpreted in the context of two SO(10) 
 models~\cite{SO10HSDR3} with $t-b-\tau$ Yukawa coupling unification : the D-term splitting model, DR3, and the Higgs splitting model, HS.     
For both models the SUSY particle mass spectrum is characterised by the low masses 
of the gluinos (300-600~GeV), charginos (100-180~GeV) and neutralinos (50-90~GeV), 
whereas all scalar particles have masses beyond the TeV scale. Depending 
on the sparticle masses, chargino-neutralino or gluino-pair production dominates. 
At low gluino masses, the three-body gluino decays $\gl \rightarrow b\bar{b}\tilde{\chi}^{0}_{1}$ and 
$\gl \rightarrow b\bar{b}\tilde{\chi}^{0}_{2}$ dominate in the DR3 and the HS model, respectively. 
Final states with high $b$-jet multiplicities are then expected in both models with a harder \met\ 
spectrum in the DR3 scenario due to the direct gluino decay into $\chi^0_1$ and with 
a higher lepton content in the HS scenario due to the subsequent decay 
$\tilde{\chi}^{0}_{2} \rightarrow  \ell \bar{\ell} \tilde{\chi}^{0}_{1} $. 
For heavy gluinos, the gluino decay modes $\gl \rightarrow bt \tilde{\chi}^{\pm}_{1}$ 
and $\gl \rightarrow t\bar{t} \tilde{\chi}^{0}_{1}$  become more relevant, enhancing 
final states with leptons in both scenarios.

Results of both 0-lepton and 1-lepton analyses have been employed to extract exclusion limits 
at 95\% CL on the gluino mass  in the two SO(10) scenarios, DR3 and HS.
The 0-lepton analysis has the best sensitivity at low gluino masses 
while the lepton-based selection is more sensitive to heavy gluinos. 
For each gluino mass, the signal region leading to the best expected significance 
is used to extract the 95\% CL exclusion limits.
Figure~\ref{fig:limit_1lepton_SO10} shows the {\tt PROSPINO} NLO cross section and the 
observed and expected  upper limit
at 95$\%$ CL for the DR3 (Figure \ref{fig:limit_1lepton_SO10}\subref{D3}) and HS (Figure \ref{fig:limit_1lepton_SO10}\subref{HS}) models  
as a function of the gluino mass. At the nominal NLO cross section, 
gluino masses below 650~GeV and 620~GeV are excluded at 95\% CL for the 
DR3 and HS models respectively. 

These limits on the gluino masses can be interpreted in terms of Yukawa coupling unification in the third generation. The degree of Yukawa 
unification is quantified by:

\begin{equation}
R = \max(f_t,f_b,f_{\tau})/\min(f_t,f_b,f_{\tau})
\end{equation}

\noindent where $f_t$,$f_b$,$f_{\tau}$ are the $t$, $b$ and $\tau$ Yukawa couplings evaluated at the scale $Q = M_{\mathrm{GUT}}$. 
In both DR3 and HS model lines, the degree of Yukawa unification increases together with the gluino mass and 
 Yukawa coupling unification occurs at a few percent level only for $m_{\tilde{g}} \leqslant$~500~GeV. 
Consequently, the most favored range of gluino masses is excluded for the two SO(10) model lines considered  as 
the degree of Yukawa unification should be further loosen up to pull the gluino mass to higher values. 
However, Yukawa coupling unification can still be realized at a few percent level for heavier gluino masses in different model lines \cite{SO10_yukawa}.

\begin{figure*}
\centering
\subfloat[]{\label{D3}\includegraphics[width=\mywidth\columnwidth]{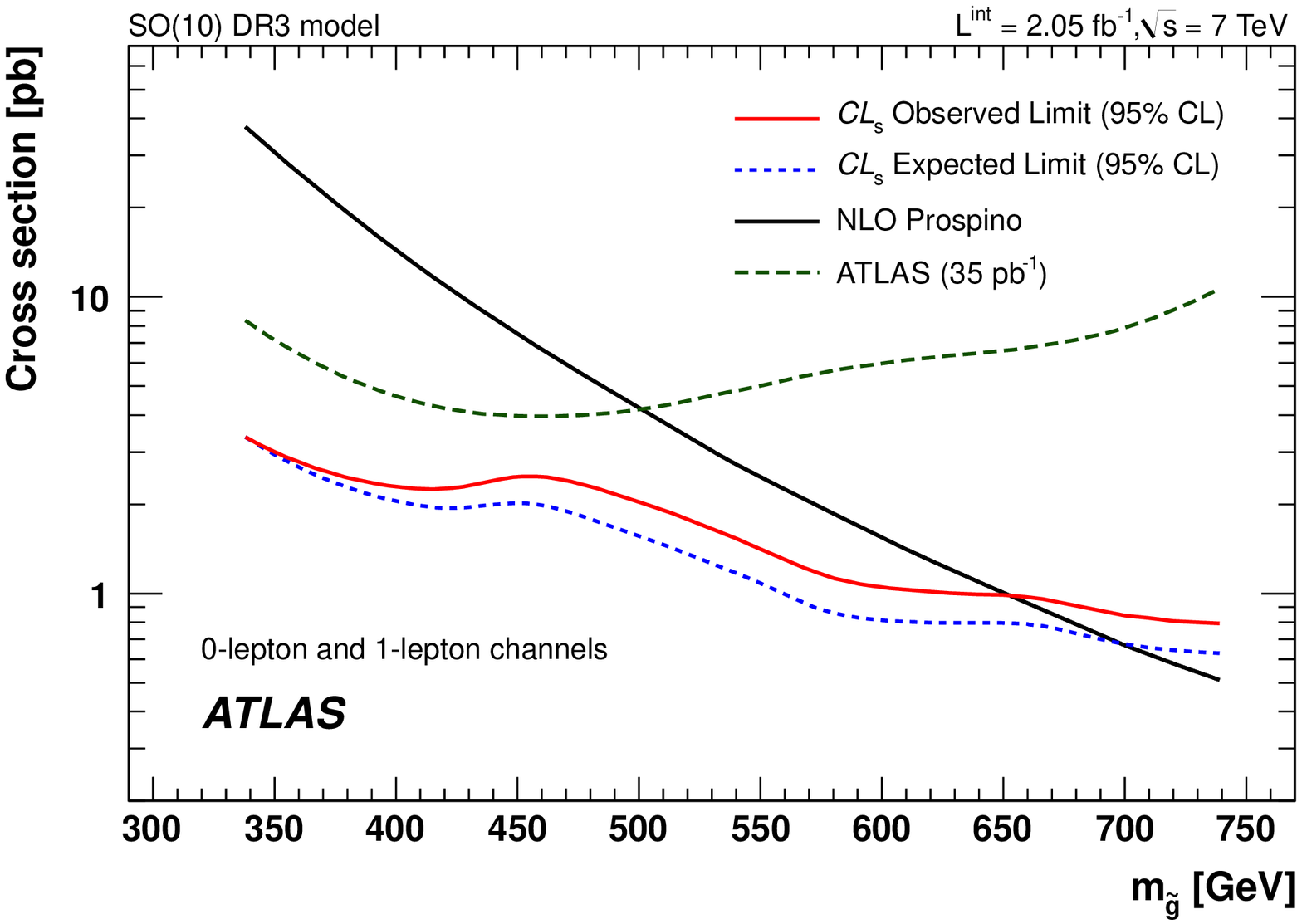}} 
\subfloat[]{\label{HS}\includegraphics[width=\mywidth\columnwidth]{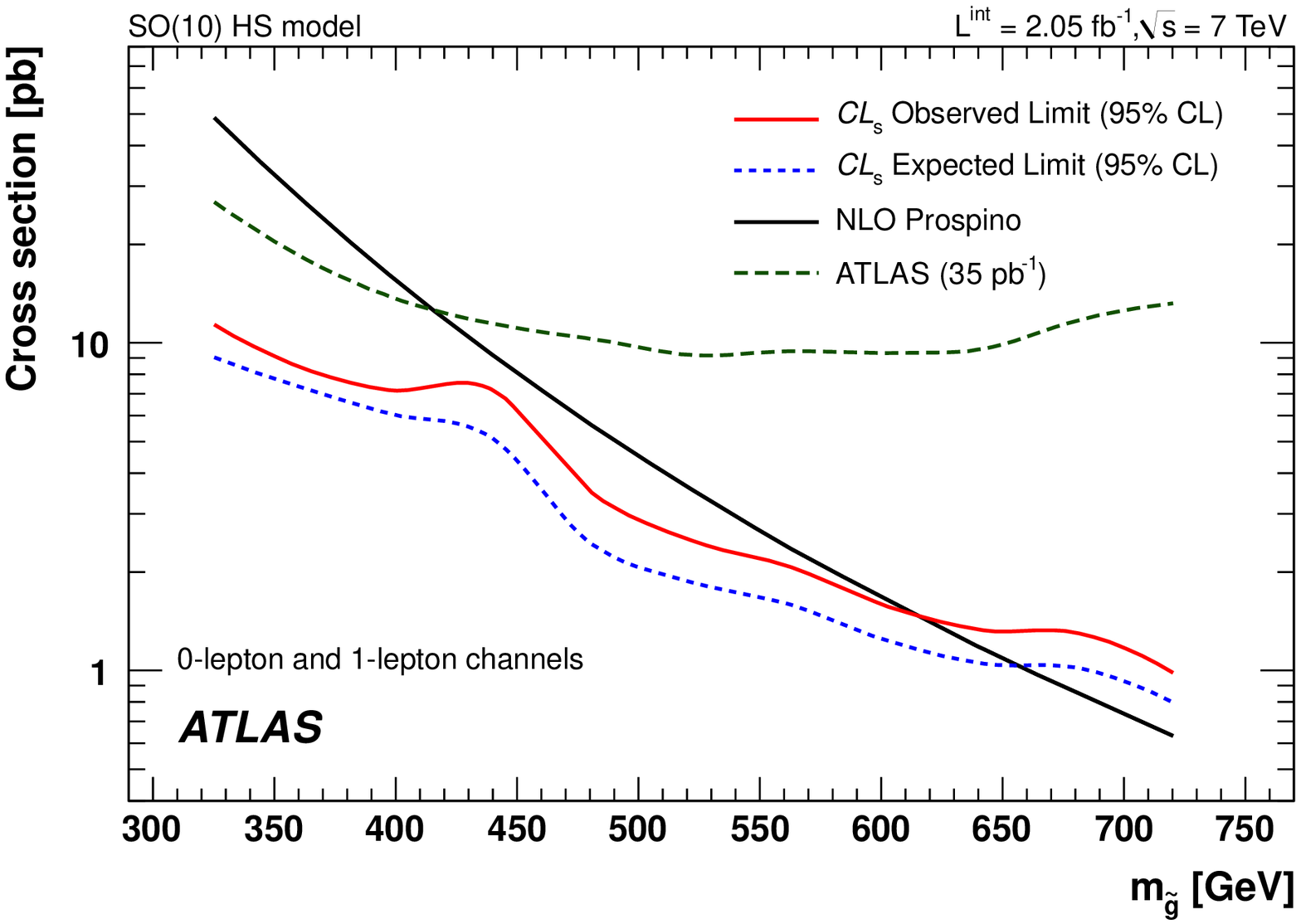}}
\caption{Cross sections as a function of the gluino mass for DR3 \protect\subref{D3} and HS
 \protect\subref{HS} models. 
The observed and expected 95\% CL exclusion limits 
are shown respectively in red and in blue. For each gluino mass, the signal region yielding the best expected limit is used.  
The NLO theoretical cross section from PROSPINO 
is shown in black. Previous limits obtained by ATLAS~\cite{Aad:2011ks} with $\mathcal{L}$ = 35~pb$^{-1}$ are superimposed for reference.}
\label{fig:limit_1lepton_SO10}
\end{figure*}

\section{Conclusions}

An updated search for supersymmetry in final states with missing transverse momentum and at least one or two $b$-jets in proton-proton collisions at 7~TeV is presented. The results are based on data corresponding to an integrated luminosity of \lumi~\ifb\ collected by ATLAS at the Large Hadron Collider during 2011. The search is sensitive mainly  to gluino-mediated production of sbottoms and stops, the supersymmetric partners of the third generation quarks, which, due to mixing effects, might be the lightest squarks. No excess above the expectations from Standard Model processes was found and the results are used to exclude parameter regions in various $R$-parity conserving SUSY models.

Gluino masses up to 800-900~GeV are excluded at 95\% CL in simplified models where the squark $\bone$ is produced either on- or off-shell and decays in 100\% of the cases into $b\neut$. 
In scenarios where the squark $\tone$ is produced (on- or off-shell) via gluino decay, gluino masses up to 620-750~GeV (depending on the specific model considered) are excluded at 95\% CL. 
In models where gluinos decay via an off-shell stop or sbottom ($bt \neut$ final states), gluino masses are excluded up to about 720~GeV for a neutralino mass of 100~GeV.

In specific models based on the gauge group SO(10), gluinos with masses below 650~GeV and 620~GeV are excluded for the DR3 and HS models, respectively. 
This analysis significantly extends the previous published limits on the same subject by the ATLAS and CMS collaborations.

\section*{Acknowledgements}

We thank CERN for the very successful operation of the LHC, as well as the
support staff from our institutions without whom ATLAS could not be
operated efficiently.

We acknowledge the support of ANPCyT, Argentina; YerPhI, Armenia; ARC,
Australia; BMWF, Austria; ANAS, Azerbaijan; SSTC, Belarus; CNPq and FAPESP,
Brazil; NSERC, NRC and CFI, Canada; CERN; CONICYT, Chile; CAS, MOST and NSFC,
China; COLCIENCIAS, Colombia; MSMT CR, MPO CR and VSC CR, Czech Republic;
DNRF, DNSRC and Lundbeck Foundation, Denmark; EPLANET and ERC, European Union;
IN2P3-CNRS, CEA-DSM/IRFU, France; GNAS, Georgia; BMBF, DFG, HGF, MPG and AvH
Foundation, Germany; GSRT, Greece; ISF, MINERVA, GIF, DIP and Benoziyo Center,
Israel; INFN, Italy; MEXT and JSPS, Japan; CNRST, Morocco; FOM and NWO,
Netherlands; RCN, Norway; MNiSW, Poland; GRICES and FCT, Portugal; MERYS
(MECTS), Romania; MES of Russia and ROSATOM, Russian Federation; JINR; MSTD,
Serbia; MSSR, Slovakia; ARRS and MVZT, Slovenia; DST/NRF, South Africa;
MICINN, Spain; SRC and Wallenberg Foundation, Sweden; SER, SNSF and Cantons of
Bern and Geneva, Switzerland; NSC, Taiwan; TAEK, Turkey; STFC, the Royal
Society and Leverhulme Trust, United Kingdom; DOE and NSF, United States of
America.

The crucial computing support from all WLCG partners is acknowledged
gratefully, in particular from CERN and the ATLAS Tier-1 facilities at
TRIUMF (Canada), NDGF (Denmark, Norway, Sweden), CC-IN2P3 (France),
KIT/GridKA (Germany), INFN-CNAF (Italy), NL-T1 (Netherlands), PIC (Spain),
ASGC (Taiwan), RAL (UK) and BNL (USA) and in the Tier-2 facilities
worldwide.

\bibliography{PUB_glmed}

\begin{thebibliography}{71}%
\makeatletter
\providecommand \@ifxundefined [1]{%
 \@ifx{#1\undefined}
}%
\providecommand \@ifnum [1]{%
 \ifnum #1\expandafter \@firstoftwo
 \else \expandafter \@secondoftwo
 \fi
}%
\providecommand \@ifx [1]{%
 \ifx #1\expandafter \@firstoftwo
 \else \expandafter \@secondoftwo
 \fi
}%
\providecommand \natexlab [1]{#1}%
\providecommand \enquote  [1]{``#1''}%
\providecommand \bibnamefont  [1]{#1}%
\providecommand \bibfnamefont [1]{#1}%
\providecommand \citenamefont [1]{#1}%
\providecommand \href@noop [0]{\@secondoftwo}%
\providecommand \href [0]{\begingroup \@sanitize@url \@href}%
\providecommand \@href[1]{\@@startlink{#1}\@@href}%
\providecommand \@@href[1]{\endgroup#1\@@endlink}%
\providecommand \@sanitize@url [0]{\catcode `\\12\catcode `\$12\catcode
  `\&12\catcode `\#12\catcode `\^12\catcode `\_12\catcode `\%12\relax}%
\providecommand \@@startlink[1]{}%
\providecommand \@@endlink[0]{}%
\providecommand \url  [0]{\begingroup\@sanitize@url \@url }%
\providecommand \@url [1]{\endgroup\@href {#1}{\urlprefix }}%
\providecommand \urlprefix  [0]{URL }%
\providecommand \Eprint [0]{\href }%
\providecommand \doibase [0]{http://dx.doi.org/}%
\providecommand \selectlanguage [0]{\@gobble}%
\providecommand \bibinfo  [0]{\@secondoftwo}%
\providecommand \bibfield  [0]{\@secondoftwo}%
\providecommand \translation [1]{[#1]}%
\providecommand \BibitemOpen [0]{}%
\providecommand \bibitemStop [0]{}%
\providecommand \bibitemNoStop [0]{.\EOS\space}%
\providecommand \EOS [0]{\spacefactor3000\relax}%
\providecommand \BibitemShut  [1]{\csname bibitem#1\endcsname}%
\let\auto@bib@innerbib\@empty
\bibitem [{\citenamefont {Miyazawa}(1966)}]{Miyazawa:1966}%
  \BibitemOpen
  \bibfield  {author} {\bibinfo {author} {\bibfnamefont {H.}~\bibnamefont
  {Miyazawa}},\ }\href {\doibase 10.1143/PTP.36.1266} {\bibfield  {journal}
  {\bibinfo  {journal} {Prog. Theor. Phys.}\ }\textbf {\bibinfo {volume} {36
  (6)}},\ \bibinfo {pages} {1266} (\bibinfo {year} {1966})}\BibitemShut
  {NoStop}%
\bibitem [{\citenamefont {Ramond}(1971)}]{Ramond:1971gb}%
  \BibitemOpen
  \bibfield  {author} {\bibinfo {author} {\bibfnamefont {P.}~\bibnamefont
  {Ramond}},\ }\href {\doibase 10.1103/PhysRevD.3.2415} {\bibfield  {journal}
  {\bibinfo  {journal} {Phys. Rev.}\ }\textbf {\bibinfo {volume} {D3}},\
  \bibinfo {pages} {2415} (\bibinfo {year} {1971})}\BibitemShut {NoStop}%
\bibitem [{\citenamefont {Golfand}\ and\ \citenamefont
  {Likhtman}(1971)}]{Golfand:1971iw}%
  \BibitemOpen
  \bibfield  {author} {\bibinfo {author} {\bibfnamefont {Y.}~\bibnamefont
  {Golfand}}\ and\ \bibinfo {author} {\bibfnamefont {E.}~\bibnamefont
  {Likhtman}},\ }\href@noop {} {\bibfield  {journal} {\bibinfo  {journal} {JETP
  Lett.}\ }\textbf {\bibinfo {volume} {13}},\ \bibinfo {pages} {323} (\bibinfo
  {year} {1971})}\BibitemShut {NoStop}%
\bibitem [{\citenamefont {Neveu}\ and\ \citenamefont
  {Schwarz}(1971{\natexlab{a}})}]{Neveu:1971rx}%
  \BibitemOpen
  \bibfield  {author} {\bibinfo {author} {\bibfnamefont {A.}~\bibnamefont
  {Neveu}}\ and\ \bibinfo {author} {\bibfnamefont {J.~H.}\ \bibnamefont
  {Schwarz}},\ }\href {\doibase 10.1016/0550-3213(71)90448-2} {\bibfield
  {journal} {\bibinfo  {journal} {Nucl. Phys.}\ }\textbf {\bibinfo {volume}
  {B31}},\ \bibinfo {pages} {86} (\bibinfo {year}
  {1971}{\natexlab{a}})}\BibitemShut {NoStop}%
\bibitem [{\citenamefont {Neveu}\ and\ \citenamefont
  {Schwarz}(1971{\natexlab{b}})}]{Neveu:1971iv}%
  \BibitemOpen
  \bibfield  {author} {\bibinfo {author} {\bibfnamefont {A.}~\bibnamefont
  {Neveu}}\ and\ \bibinfo {author} {\bibfnamefont {J.~H.}\ \bibnamefont
  {Schwarz}},\ }\href {\doibase 10.1103/PhysRevD.4.1109} {\bibfield  {journal}
  {\bibinfo  {journal} {Phys. Rev.}\ }\textbf {\bibinfo {volume} {D4}},\
  \bibinfo {pages} {1109} (\bibinfo {year} {1971}{\natexlab{b}})}\BibitemShut
  {NoStop}%
\bibitem [{\citenamefont {Gervais}\ and\ \citenamefont
  {Sakita}(1971)}]{Gervais:1971ji}%
  \BibitemOpen
  \bibfield  {author} {\bibinfo {author} {\bibfnamefont {J.}~\bibnamefont
  {Gervais}}\ and\ \bibinfo {author} {\bibfnamefont {B.}~\bibnamefont
  {Sakita}},\ }\href {\doibase 10.1016/0550-3213(71)90351-8} {\bibfield
  {journal} {\bibinfo  {journal} {Nucl. Phys.}\ }\textbf {\bibinfo {volume}
  {B34}},\ \bibinfo {pages} {632} (\bibinfo {year} {1971})}\BibitemShut
  {NoStop}%
\bibitem [{\citenamefont {Volkov}\ and\ \citenamefont
  {Akulov}(1973)}]{Volkov:1973ix}%
  \BibitemOpen
  \bibfield  {author} {\bibinfo {author} {\bibfnamefont {D.}~\bibnamefont
  {Volkov}}\ and\ \bibinfo {author} {\bibfnamefont {V.}~\bibnamefont
  {Akulov}},\ }\href {\doibase 10.1016/0370-2693(73)90490-5} {\bibfield
  {journal} {\bibinfo  {journal} {Phys. Lett.}\ }\textbf {\bibinfo {volume}
  {B46}},\ \bibinfo {pages} {109} (\bibinfo {year} {1973})}\BibitemShut
  {NoStop}%
\bibitem [{\citenamefont {Wess}\ and\ \citenamefont
  {Zumino}(1974{\natexlab{a}})}]{Wess:1973kz}%
  \BibitemOpen
  \bibfield  {author} {\bibinfo {author} {\bibfnamefont {J.}~\bibnamefont
  {Wess}}\ and\ \bibinfo {author} {\bibfnamefont {B.}~\bibnamefont {Zumino}},\
  }\href {\doibase 10.1016/0370-2693(74)90578-4} {\bibfield  {journal}
  {\bibinfo  {journal} {Phys. Lett.}\ }\textbf {\bibinfo {volume} {B49}},\
  \bibinfo {pages} {52} (\bibinfo {year} {1974}{\natexlab{a}})}\BibitemShut
  {NoStop}%
\bibitem [{\citenamefont {Wess}\ and\ \citenamefont
  {Zumino}(1974{\natexlab{b}})}]{Wess:1974tw}%
  \BibitemOpen
  \bibfield  {author} {\bibinfo {author} {\bibfnamefont {J.}~\bibnamefont
  {Wess}}\ and\ \bibinfo {author} {\bibfnamefont {B.}~\bibnamefont {Zumino}},\
  }\href {\doibase 10.1016/0550-3213(74)90355-1} {\bibfield  {journal}
  {\bibinfo  {journal} {Nucl. Phys.}\ }\textbf {\bibinfo {volume} {B70}},\
  \bibinfo {pages} {39} (\bibinfo {year} {1974}{\natexlab{b}})}\BibitemShut
  {NoStop}%
\bibitem [{\citenamefont {Weinberg}(1976)}]{Weinberg:1975gm}%
  \BibitemOpen
  \bibfield  {author} {\bibinfo {author} {\bibfnamefont {S.}~\bibnamefont
  {Weinberg}},\ }\href {\doibase 10.1103/PhysRevD.13.974} {\bibfield  {journal}
  {\bibinfo  {journal} {Phys. Rev.}\ }\textbf {\bibinfo {volume} {D13}},\
  \bibinfo {pages} {974} (\bibinfo {year} {1976})}\BibitemShut {NoStop}%
\bibitem [{\citenamefont {Gildener}(1976)}]{Gildener:1976ai}%
  \BibitemOpen
  \bibfield  {author} {\bibinfo {author} {\bibfnamefont {E.}~\bibnamefont
  {Gildener}},\ }\href {\doibase 10.1103/PhysRevD.14.1667} {\bibfield
  {journal} {\bibinfo  {journal} {Phys. Rev.}\ }\textbf {\bibinfo {volume}
  {D14}},\ \bibinfo {pages} {1667} (\bibinfo {year} {1976})}\BibitemShut
  {NoStop}%
\bibitem [{\citenamefont {Weinberg}(1979)}]{Weinberg:1979bn}%
  \BibitemOpen
  \bibfield  {author} {\bibinfo {author} {\bibfnamefont {S.}~\bibnamefont
  {Weinberg}},\ }\href {\doibase 10.1103/PhysRevD.19.1277} {\bibfield
  {journal} {\bibinfo  {journal} {Phys. Rev.}\ }\textbf {\bibinfo {volume}
  {D19}},\ \bibinfo {pages} {1277} (\bibinfo {year} {1979})}\BibitemShut
  {NoStop}%
\bibitem [{\citenamefont {Susskind}(1979)}]{Susskind:1978ms}%
  \BibitemOpen
  \bibfield  {author} {\bibinfo {author} {\bibfnamefont {L.}~\bibnamefont
  {Susskind}},\ }\href {\doibase 10.1103/PhysRevD.20.2619} {\bibfield
  {journal} {\bibinfo  {journal} {Phys. Rev.}\ }\textbf {\bibinfo {volume}
  {D20}},\ \bibinfo {pages} {2619} (\bibinfo {year} {1979})}\BibitemShut
  {NoStop}%
\bibitem [{\citenamefont {Fayet}(1976)}]{Fayet:1976et}%
  \BibitemOpen
  \bibfield  {author} {\bibinfo {author} {\bibfnamefont {P.}~\bibnamefont
  {Fayet}},\ }\href {\doibase 10.1016/0370-2693(76)90319-1} {\bibfield
  {journal} {\bibinfo  {journal} {Phys. Lett.}\ }\textbf {\bibinfo {volume}
  {B64}},\ \bibinfo {pages} {159} (\bibinfo {year} {1976})}\BibitemShut
  {NoStop}%
\bibitem [{\citenamefont {Fayet}(1977)}]{Fayet:1977yc}%
  \BibitemOpen
  \bibfield  {author} {\bibinfo {author} {\bibfnamefont {P.}~\bibnamefont
  {Fayet}},\ }\href {\doibase 10.1016/0370-2693(77)90852-8} {\bibfield
  {journal} {\bibinfo  {journal} {Phys. Lett.}\ }\textbf {\bibinfo {volume}
  {B69}},\ \bibinfo {pages} {489} (\bibinfo {year} {1977})}\BibitemShut
  {NoStop}%
\bibitem [{\citenamefont {Farrar}\ and\ \citenamefont
  {Fayet}(1978)}]{Farrar:1978xj}%
  \BibitemOpen
  \bibfield  {author} {\bibinfo {author} {\bibfnamefont {G.~R.}\ \bibnamefont
  {Farrar}}\ and\ \bibinfo {author} {\bibfnamefont {P.}~\bibnamefont {Fayet}},\
  }\href {\doibase 10.1016/0370-2693(78)90858-4} {\bibfield  {journal}
  {\bibinfo  {journal} {Phys. Lett.}\ }\textbf {\bibinfo {volume} {B76}},\
  \bibinfo {pages} {575} (\bibinfo {year} {1978})}\BibitemShut {NoStop}%
\bibitem [{\citenamefont {Fayet}(1979)}]{Fayet:1979sa}%
  \BibitemOpen
  \bibfield  {author} {\bibinfo {author} {\bibfnamefont {P.}~\bibnamefont
  {Fayet}},\ }\href {\doibase 10.1016/0370-2693(79)91229-2} {\bibfield
  {journal} {\bibinfo  {journal} {Phys. Lett.}\ }\textbf {\bibinfo {volume}
  {B84}},\ \bibinfo {pages} {416} (\bibinfo {year} {1979})}\BibitemShut
  {NoStop}%
\bibitem [{\citenamefont {Dimopoulos}\ and\ \citenamefont
  {Georgi}(1981)}]{Dimopoulos:1981zb}%
  \BibitemOpen
  \bibfield  {author} {\bibinfo {author} {\bibfnamefont {S.}~\bibnamefont
  {Dimopoulos}}\ and\ \bibinfo {author} {\bibfnamefont {H.}~\bibnamefont
  {Georgi}},\ }\href {\doibase 10.1016/0550-3213(81)90522-8} {\bibfield
  {journal} {\bibinfo  {journal} {Nucl. Phys.}\ }\textbf {\bibinfo {volume}
  {B193}},\ \bibinfo {pages} {150} (\bibinfo {year} {1981})}\BibitemShut
  {NoStop}%
\bibitem [{\citenamefont {Fritzsch}\ and\ \citenamefont
  {Minkowski}(1975)}]{SO10_2}%
  \BibitemOpen
  \bibfield  {author} {\bibinfo {author} {\bibfnamefont {H.}~\bibnamefont
  {Fritzsch}}\ and\ \bibinfo {author} {\bibfnamefont {P.}~\bibnamefont
  {Minkowski}},\ }\href@noop {} {\bibfield  {journal} {\bibinfo  {journal}
  {Annals of Physics}\ }\textbf {\bibinfo {volume} {93}},\ \bibinfo {pages}
  {193} (\bibinfo {year} {1975})}\BibitemShut {NoStop}%
\bibitem [{\citenamefont {Gell-Mann}\ \emph {et~al.}(1978)\citenamefont
  {Gell-Mann}, \citenamefont {Ramond},\ and\ \citenamefont {Slansky}}]{SO10}%
  \BibitemOpen
  \bibfield  {author} {\bibinfo {author} {\bibfnamefont {M.}~\bibnamefont
  {Gell-Mann}}, \bibinfo {author} {\bibfnamefont {P.}~\bibnamefont {Ramond}}, \
  and\ \bibinfo {author} {\bibfnamefont {R.}~\bibnamefont {Slansky}},\
  }\href@noop {} {\bibfield  {journal} {\bibinfo  {journal} {Rev. Mod. Phys.}\
  }\textbf {\bibinfo {volume} {50}},\ \bibinfo {pages} {721} (\bibinfo {year}
  {1978})}\BibitemShut {NoStop}%
\bibitem [{\citenamefont {{ATLAS
  Collaboration}}(2011{\natexlab{a}})}]{Aad:2011ks}%
  \BibitemOpen
  \bibfield  {author} {\bibinfo {author} {\bibnamefont {{ATLAS
  Collaboration}}},\ }\href {\doibase 10.1016/j.physletb.2011.06.015}
  {\bibfield  {journal} {\bibinfo  {journal} {Phys. Lett.}\ }\textbf {\bibinfo
  {volume} {B701}},\ \bibinfo {pages} {398} (\bibinfo {year}
  {2011}{\natexlab{a}})}\BibitemShut {NoStop}%
\bibitem [{\citenamefont {{CMS Collaboration}}(2011)}]{cms_bjet}%
  \BibitemOpen
  \bibfield  {author} {\bibinfo {author} {\bibnamefont {{CMS Collaboration}}},\
  }\href {\doibase 10.1007/JHEP08(2011)155} {\bibfield  {journal} {\bibinfo
  {journal} {JHEP}\ }\textbf {\bibinfo {volume} {07}},\ \bibinfo {pages} {113}
  (\bibinfo {year} {2011})}\BibitemShut {NoStop}%
\bibitem [{sbo()}]{sbottom}%
  \BibitemOpen
  \href@noop {} {}\bibinfo {howpublished} {{ATLAS Collaboration,
  arXiv:1112.3832 [hep-ex] (2012). Accepted by \PR Lett}}\BibitemShut {NoStop}%
\bibitem [{\citenamefont {{ATLAS Collaboration}}(2008)}]{DetectorPaper:2008}%
  \BibitemOpen
  \bibfield  {author} {\bibinfo {author} {\bibnamefont {{ATLAS
  Collaboration}}},\ }\href {\doibase 10.1088/1748-0221/3/08/S08003} {\bibfield
   {journal} {\bibinfo  {journal} {JINST}\ }\textbf {\bibinfo {volume} {3}},\
  \bibinfo {pages} {S08003} (\bibinfo {year} {2008})}\BibitemShut {NoStop}%
\bibitem [{atl()}]{atlascoordinates}%
  \BibitemOpen
  \href@noop {} {}\bibinfo {howpublished} {{ATLAS uses a right-handed
  coordinate system, with the $z$ axis along the beam line. The azimuthal angle
  $\phi$ is measured around the beam axis and the polar angle $\theta$ is the
  angle from the beam axis. The pseudorapidity is defined as $\eta = -\ln
  \tan(\theta/2)$. The distance~$\Delta R$ in the $\eta-\phi$ space is defined
  as $\Delta R = \sqrt{(\Delta\eta)^2+(\Delta\phi)^2}$.}}\BibitemShut {Stop}%
\bibitem [{mca()}]{mcatnlo}%
  \BibitemOpen
  \href@noop {} {}\bibinfo {howpublished} {{S. Frixione and B. Webber,
  hep-ph/0601192 (2006)}}\BibitemShut {NoStop}%
\bibitem [{\citenamefont {Nadolsky}\ \emph {et~al.}(2008)\citenamefont
  {Nadolsky} \emph {et~al.}}]{cteq66}%
  \BibitemOpen
  \bibfield  {author} {\bibinfo {author} {\bibfnamefont {P.~M.}\ \bibnamefont
  {Nadolsky}} \emph {et~al.},\ }\href@noop {} {\bibfield  {journal} {\bibinfo
  {journal} {Phys. Rev.}\ }\textbf {\bibinfo {volume} {D78}},\ \bibinfo {pages}
  {013004} (\bibinfo {year} {2008})}\BibitemShut {NoStop}%
\bibitem [{\citenamefont {Frixione}\ \emph {et~al.}(2007)\citenamefont
  {Frixione} \emph {et~al.}}]{powheg}%
  \BibitemOpen
  \bibfield  {author} {\bibinfo {author} {\bibfnamefont {S.}~\bibnamefont
  {Frixione}} \emph {et~al.},\ }\href@noop {} {\bibfield  {journal} {\bibinfo
  {journal} {JHEP}\ }\textbf {\bibinfo {volume} {11}},\ \bibinfo {pages} {070}
  (\bibinfo {year} {2007})}\BibitemShut {NoStop}%
\bibitem [{Ker(2004)}]{Kersevan:2004yg}%
  \BibitemOpen
  \href@noop {} {}\bibinfo {howpublished} {{B. Kersevan and E. Richter-Was,
  hep-ph/0405247}} (\bibinfo {year} {2004})\BibitemShut {NoStop}%
\bibitem [{\citenamefont {Mangano}\ \emph {et~al.}(2003)\citenamefont {Mangano}
  \emph {et~al.}}]{alpgen}%
  \BibitemOpen
  \bibfield  {author} {\bibinfo {author} {\bibfnamefont {M.}~\bibnamefont
  {Mangano}} \emph {et~al.},\ }\href@noop {} {\bibfield  {journal} {\bibinfo
  {journal} {JHEP}\ }\textbf {\bibinfo {volume} {07}},\ \bibinfo {pages} {001}
  (\bibinfo {year} {2003})}\BibitemShut {NoStop}%
\bibitem [{\citenamefont {Pumplin}\ \emph {et~al.}(2002)\citenamefont {Pumplin}
  \emph {et~al.}}]{cteq6l1}%
  \BibitemOpen
  \bibfield  {author} {\bibinfo {author} {\bibfnamefont {J.}~\bibnamefont
  {Pumplin}} \emph {et~al.},\ }\href@noop {} {\bibfield  {journal} {\bibinfo
  {journal} {JHEP}\ }\textbf {\bibinfo {volume} {07}},\ \bibinfo {pages} {012}
  (\bibinfo {year} {2002})}\BibitemShut {NoStop}%
\bibitem [{\citenamefont {Corcella}\ \emph {et~al.}(2001)\citenamefont
  {Corcella} \emph {et~al.}}]{herwig}%
  \BibitemOpen
  \bibfield  {author} {\bibinfo {author} {\bibfnamefont {G.}~\bibnamefont
  {Corcella}} \emph {et~al.},\ }\href@noop {} {\bibfield  {journal} {\bibinfo
  {journal} {JHEP}\ }\textbf {\bibinfo {volume} {01}},\ \bibinfo {pages} {010}
  (\bibinfo {year} {2001})}\BibitemShut {NoStop}%
\bibitem [{\citenamefont {Butterworth}\ \emph {et~al.}(1996)\citenamefont
  {Butterworth}, \citenamefont {Forshaw},\ and\ \citenamefont
  {Seymour}}]{jimmy}%
  \BibitemOpen
  \bibfield  {author} {\bibinfo {author} {\bibfnamefont {J.}~\bibnamefont
  {Butterworth}}, \bibinfo {author} {\bibfnamefont {J.~R.}\ \bibnamefont
  {Forshaw}}, \ and\ \bibinfo {author} {\bibfnamefont {M.}~\bibnamefont
  {Seymour}},\ }\href {\doibase 10.1007/s002880050286} {\bibfield  {journal}
  {\bibinfo  {journal} {\ZfP}\ }\textbf {\bibinfo {volume} {C72}},\ \bibinfo
  {pages} {637} (\bibinfo {year} {1996})}\BibitemShut {NoStop}%
\bibitem [{\citenamefont {Alwall}\ \emph {et~al.}(2011)\citenamefont {Alwall},
  \citenamefont {Herquet}, \citenamefont {Maltoni}, \citenamefont {Mattelaer},\
  and\ \citenamefont {Stelzer}}]{madgraph}%
  \BibitemOpen
  \bibfield  {author} {\bibinfo {author} {\bibfnamefont {J.}~\bibnamefont
  {Alwall}}, \bibinfo {author} {\bibfnamefont {M.}~\bibnamefont {Herquet}},
  \bibinfo {author} {\bibfnamefont {F.}~\bibnamefont {Maltoni}}, \bibinfo
  {author} {\bibfnamefont {O.}~\bibnamefont {Mattelaer}}, \ and\ \bibinfo
  {author} {\bibfnamefont {T.}~\bibnamefont {Stelzer}},\ }\href {\doibase
  10.1007/JHEP06(2011)128} {\bibfield  {journal} {\bibinfo  {journal} {JHEP}\
  }\textbf {\bibinfo {volume} {06}},\ \bibinfo {pages} {128} (\bibinfo {year}
  {2011})}\BibitemShut {NoStop}%
\bibitem [{\citenamefont {Sj{\"{o}}strand}\ \emph {et~al.}(2006)\citenamefont
  {Sj{\"{o}}strand}, \citenamefont {Mrenna},\ and\ \citenamefont
  {Skands}}]{pythia}%
  \BibitemOpen
  \bibfield  {author} {\bibinfo {author} {\bibfnamefont {T.}~\bibnamefont
  {Sj{\"{o}}strand}}, \bibinfo {author} {\bibfnamefont {S.}~\bibnamefont
  {Mrenna}}, \ and\ \bibinfo {author} {\bibfnamefont {P.}~\bibnamefont
  {Skands}},\ }\href@noop {} {\bibfield  {journal} {\bibinfo  {journal} {JHEP}\
  }\textbf {\bibinfo {volume} {0605}},\ \bibinfo {pages} {026} (\bibinfo {year}
  {2006})}\BibitemShut {NoStop}%
\bibitem [{\citenamefont {Bahr}\ \emph {et~al.}(2008)\citenamefont {Bahr} \emph
  {et~al.}}]{Bahr:2008pv}%
  \BibitemOpen
  \bibfield  {author} {\bibinfo {author} {\bibfnamefont {M.}~\bibnamefont
  {Bahr}} \emph {et~al.},\ }\href {\doibase 10.1140/epjc/s10052-008-0798-9}
  {\bibfield  {journal} {\bibinfo  {journal} {Eur. Phys. J.}\ }\textbf
  {\bibinfo {volume} {C58}},\ \bibinfo {pages} {639} (\bibinfo {year}
  {2008})}\BibitemShut {NoStop}%
\bibitem [{\citenamefont {Beenakker}\ \emph {et~al.}(1998)\citenamefont
  {Beenakker}, \citenamefont {Kramer}, \citenamefont {Plehn}, \citenamefont
  {Spira},\ and\ \citenamefont {Zerwas}}]{Beenakker:1997ut}%
  \BibitemOpen
  \bibfield  {author} {\bibinfo {author} {\bibfnamefont {W.}~\bibnamefont
  {Beenakker}}, \bibinfo {author} {\bibfnamefont {M.}~\bibnamefont {Kramer}},
  \bibinfo {author} {\bibfnamefont {T.}~\bibnamefont {Plehn}}, \bibinfo
  {author} {\bibfnamefont {M.}~\bibnamefont {Spira}}, \ and\ \bibinfo {author}
  {\bibfnamefont {P.}~\bibnamefont {Zerwas}},\ }\href {\doibase
  10.1016/S0550-3213(98)00014-5} {\bibfield  {journal} {\bibinfo  {journal}
  {Nucl. Phys.}\ }\textbf {\bibinfo {volume} {B515}},\ \bibinfo {pages} {3}
  (\bibinfo {year} {1998})}\BibitemShut {NoStop}%
\bibitem [{\citenamefont {Stump}\ \emph {et~al.}(2003)\citenamefont {Stump}
  \emph {et~al.}}]{cteq6m}%
  \BibitemOpen
  \bibfield  {author} {\bibinfo {author} {\bibfnamefont {D.}~\bibnamefont
  {Stump}} \emph {et~al.},\ }\href@noop {} {\bibfield  {journal} {\bibinfo
  {journal} {JHEP}\ }\textbf {\bibinfo {volume} {10}},\ \bibinfo {pages} {046}
  (\bibinfo {year} {2003})}\BibitemShut {NoStop}%
\bibitem [{atl(2010{\natexlab{a}})}]{atlastune1}%
  \BibitemOpen
  \href@noop {} {}\bibinfo {howpublished} {{ATLAS Collaboration,
  \href{http://cdsweb.cern.ch/record/1303025/}{ATL-PHYS-PUB-2010-014}.
  http://cdsweb.cern.ch/record/1303025}} (\bibinfo {year}
  {2010}{\natexlab{a}})\BibitemShut {NoStop}%
\bibitem [{atl(2010{\natexlab{b}})}]{atlastune2}%
  \BibitemOpen
  \href@noop {} {}\bibinfo {howpublished} {{ATLAS Collaboration,
  \href{http://cdsweb.cern.ch/record/1277665/}{ATLAS-CONF-2010-031}.
  http://cdsweb.cern.ch/record/1277665 }} (\bibinfo {year}
  {2010}{\natexlab{b}})\BibitemShut {NoStop}%
\bibitem [{\citenamefont {{ATLAS
  Collaboration}}(2010{\natexlab{a}})}]{atlassimulation}%
  \BibitemOpen
  \bibfield  {author} {\bibinfo {author} {\bibnamefont {{ATLAS
  Collaboration}}},\ }\href {\doibase 10.1140/epjc/s10052-010-1429-9}
  {\bibfield  {journal} {\bibinfo  {journal} {\EPJ}\ }\textbf {\bibinfo
  {volume} {C70}},\ \bibinfo {pages} {823} (\bibinfo {year}
  {2010}{\natexlab{a}})}\BibitemShut {NoStop}%
\bibitem [{\citenamefont {Agostinelli}\ \emph {et~al.}(2003)\citenamefont
  {Agostinelli} \emph {et~al.}}]{geant4}%
  \BibitemOpen
  \bibfield  {author} {\bibinfo {author} {\bibfnamefont {S.}~\bibnamefont
  {Agostinelli}} \emph {et~al.} (\bibinfo {collaboration} {GEANT4}),\
  }\href@noop {} {\bibfield  {journal} {\bibinfo  {journal} {Nucl. Instrum.
  Meth.}\ }\textbf {\bibinfo {volume} {A506}},\ \bibinfo {pages} {250}
  (\bibinfo {year} {2003})}\BibitemShut {NoStop}%
\bibitem [{\citenamefont {Campbell}\ and\ \citenamefont {Ellis}(1999)}]{mcfm}%
  \BibitemOpen
  \bibfield  {author} {\bibinfo {author} {\bibfnamefont {J.}~\bibnamefont
  {Campbell}}\ and\ \bibinfo {author} {\bibfnamefont {R.}~\bibnamefont
  {Ellis}},\ }\href@noop {} {\bibfield  {journal} {\bibinfo  {journal} {\PR}\
  }\textbf {\bibinfo {volume} {D60}},\ \bibinfo {pages} {113006} (\bibinfo
  {year} {1999})}\BibitemShut {NoStop}%
\bibitem [{\citenamefont {Campbell}\ \emph {et~al.}(2011)\citenamefont
  {Campbell}, \citenamefont {Ellis},\ and\ \citenamefont {Williams}}]{mcfm2}%
  \BibitemOpen
  \bibfield  {author} {\bibinfo {author} {\bibfnamefont {J.~M.}\ \bibnamefont
  {Campbell}}, \bibinfo {author} {\bibfnamefont {R.~K.}\ \bibnamefont {Ellis}},
  \ and\ \bibinfo {author} {\bibfnamefont {C.}~\bibnamefont {Williams}},\
  }\href {\doibase 10.1007/JHEP07(2011)018} {\bibfield  {journal} {\bibinfo
  {journal} {JHEP}\ }\textbf {\bibinfo {volume} {07}},\ \bibinfo {pages} {018}
  (\bibinfo {year} {2011})}\BibitemShut {NoStop}%
\bibitem [{\citenamefont {Hamberg}\ \emph {et~al.}(1991)\citenamefont
  {Hamberg}, \citenamefont {van Neerven},\ and\ \citenamefont
  {Matsuura}}]{Hamberg:1990np}%
  \BibitemOpen
  \bibfield  {author} {\bibinfo {author} {\bibfnamefont {R.}~\bibnamefont
  {Hamberg}}, \bibinfo {author} {\bibfnamefont {W.~L.}\ \bibnamefont {van
  Neerven}}, \ and\ \bibinfo {author} {\bibfnamefont {T.}~\bibnamefont
  {Matsuura}},\ }\href@noop {} {\bibfield  {journal} {\bibinfo  {journal}
  {Nucl. Phys.}\ }\textbf {\bibinfo {volume} {B359}},\ \bibinfo {pages} {343}
  (\bibinfo {year} {1991})},\ \bibinfo {note} {erratum-ibid. {\bf
  B644}:403-404,2002}\BibitemShut {NoStop}%
\bibitem [{\citenamefont {Melnikov}\ and\ \citenamefont
  {Petriello}(2006{\natexlab{a}})}]{Melnikov:2006kv}%
  \BibitemOpen
  \bibfield  {author} {\bibinfo {author} {\bibfnamefont {K.}~\bibnamefont
  {Melnikov}}\ and\ \bibinfo {author} {\bibfnamefont {F.}~\bibnamefont
  {Petriello}},\ }\href {\doibase 10.1103/PhysRevD.74.114017} {\bibfield
  {journal} {\bibinfo  {journal} {Phys. Rev.}\ }\textbf {\bibinfo {volume}
  {D74}},\ \bibinfo {pages} {114017} (\bibinfo {year}
  {2006}{\natexlab{a}})}\BibitemShut {NoStop}%
\bibitem [{\citenamefont {Melnikov}\ and\ \citenamefont
  {Petriello}(2006{\natexlab{b}})}]{Melnikov:2006di}%
  \BibitemOpen
  \bibfield  {author} {\bibinfo {author} {\bibfnamefont {K.}~\bibnamefont
  {Melnikov}}\ and\ \bibinfo {author} {\bibfnamefont {F.}~\bibnamefont
  {Petriello}},\ }\href {\doibase 10.1103/PhysRevLett.96.231803} {\bibfield
  {journal} {\bibinfo  {journal} {\PRL}\ }\textbf {\bibinfo {volume} {96}},\
  \bibinfo {pages} {231803} (\bibinfo {year} {2006}{\natexlab{b}})}\BibitemShut
  {NoStop}%
\bibitem [{\citenamefont {Bonciani}\ \emph {et~al.}(1998)\citenamefont
  {Bonciani}, \citenamefont {Catani}, \citenamefont {Mangano},\ and\
  \citenamefont {Nason}}]{Bonciani:1998vc}%
  \BibitemOpen
  \bibfield  {author} {\bibinfo {author} {\bibfnamefont {R.}~\bibnamefont
  {Bonciani}}, \bibinfo {author} {\bibfnamefont {S.}~\bibnamefont {Catani}},
  \bibinfo {author} {\bibfnamefont {M.~L.}\ \bibnamefont {Mangano}}, \ and\
  \bibinfo {author} {\bibfnamefont {P.}~\bibnamefont {Nason}},\ }\href@noop {}
  {\bibfield  {journal} {\bibinfo  {journal} {Nucl. Phys.}\ }\textbf {\bibinfo
  {volume} {B529}},\ \bibinfo {pages} {424} (\bibinfo {year}
  {1998})}\BibitemShut {NoStop}%
\bibitem [{\citenamefont {Moch}\ and\ \citenamefont
  {Uwer}(2008)}]{Moch:2008qy}%
  \BibitemOpen
  \bibfield  {author} {\bibinfo {author} {\bibfnamefont {S.}~\bibnamefont
  {Moch}}\ and\ \bibinfo {author} {\bibfnamefont {P.}~\bibnamefont {Uwer}},\
  }\href {\doibase 10.1103/PhysRevD.78.034003} {\bibfield  {journal} {\bibinfo
  {journal} {Phys. Rev.}\ }\textbf {\bibinfo {volume} {D78}},\ \bibinfo {pages}
  {034003} (\bibinfo {year} {2008})}\BibitemShut {NoStop}%
\bibitem [{\citenamefont {Beneke}\ \emph {et~al.}(2010)\citenamefont {Beneke},
  \citenamefont {Czakon}, \citenamefont {Falgari}, \citenamefont {Mitov},\ and\
  \citenamefont {Schwinn}}]{Beneke:2009ye}%
  \BibitemOpen
  \bibfield  {author} {\bibinfo {author} {\bibfnamefont {M.}~\bibnamefont
  {Beneke}}, \bibinfo {author} {\bibfnamefont {M.}~\bibnamefont {Czakon}},
  \bibinfo {author} {\bibfnamefont {P.}~\bibnamefont {Falgari}}, \bibinfo
  {author} {\bibfnamefont {A.}~\bibnamefont {Mitov}}, \ and\ \bibinfo {author}
  {\bibfnamefont {C.}~\bibnamefont {Schwinn}},\ }\href {\doibase
  10.1016/j.physletb.2010.05.038} {\bibfield  {journal} {\bibinfo  {journal}
  {\PL}\ }\textbf {\bibinfo {volume} {B690}},\ \bibinfo {pages} {483} (\bibinfo
  {year} {2010})}\BibitemShut {NoStop}%
\bibitem [{201()}]{2011mk}%
  \BibitemOpen
  \href@noop {} {}\bibinfo {howpublished} {{ATLAS Collaboration,
  arXiv:1110.3174 [hep-ex] (2011). Accepted by \EPJ C}}\BibitemShut {NoStop}%
\bibitem [{ATL()}]{ATLASPV}%
  \BibitemOpen
  \href@noop {} {}\bibinfo {note} {In case of multiple vertices, the primary
  vertex is taken to be the one for which the sum of the square of momenta of
  the associated tracks $\pt^2$ is the largest.}\BibitemShut {Stop}%
\bibitem [{\citenamefont {Cacciari}\ \emph {et~al.}(2008)\citenamefont
  {Cacciari}, \citenamefont {Salam},\ and\ \citenamefont
  {Soyez}}]{Cacciari:2008gp}%
  \BibitemOpen
  \bibfield  {author} {\bibinfo {author} {\bibfnamefont {M.}~\bibnamefont
  {Cacciari}}, \bibinfo {author} {\bibfnamefont {G.}~\bibnamefont {Salam}}, \
  and\ \bibinfo {author} {\bibfnamefont {G.}~\bibnamefont {Soyez}},\
  }\href@noop {} {\bibfield  {journal} {\bibinfo  {journal} {JHEP}\ }\textbf
  {\bibinfo {volume} {04}},\ \bibinfo {pages} {063} (\bibinfo {year}
  {2008})}\BibitemShut {NoStop}%
\bibitem [{\citenamefont {Cacciari}\ and\ \citenamefont
  {Salam}(2006)}]{Cacciari200657}%
  \BibitemOpen
  \bibfield  {author} {\bibinfo {author} {\bibfnamefont {M.}~\bibnamefont
  {Cacciari}}\ and\ \bibinfo {author} {\bibfnamefont {G.}~\bibnamefont
  {Salam}},\ }\href {\doibase DOI: 10.1016/j.physletb.2006.08.037} {\bibfield
  {journal} {\bibinfo  {journal} {\PL B}\ }\textbf {\bibinfo {volume} {641}},\
  \bibinfo {pages} {57 } (\bibinfo {year} {2006})}\BibitemShut {NoStop}%
\bibitem [{JES()}]{JES}%
  \BibitemOpen
  \href@noop {} {}\bibinfo {howpublished} {{ATLAS Collaboration,
  arXiv:1112.6426 [hep-ex] (2011). Submitted to \EPJ C}}\BibitemShut {NoStop}%
\bibitem [{\citenamefont {{ATLAS Collaboration}}(2011{\natexlab{b}})}]{btag}%
  \BibitemOpen
  \bibfield  {author} {\bibinfo {author} {\bibnamefont {{ATLAS
  Collaboration}}},\ }\href@noop {} {}\bibinfo {howpublished}
  {{\href{http://cdsweb.cern.ch/record/1369219}{ATLAS-CONF-2011-102}.
  http://cdsweb.cern.ch/record/1369219}} (\bibinfo {year}
  {2011}{\natexlab{b}})\BibitemShut {NoStop}%
\bibitem [{\citenamefont {{ATLAS
  Collaboration}}(2012{\natexlab{a}})}]{Aad:2012re}%
  \BibitemOpen
  \bibfield  {author} {\bibinfo {author} {\bibnamefont {{ATLAS
  Collaboration}}},\ }\href {\doibase 10.1140/epjc/s10052-011-1844-6}
  {\bibfield  {journal} {\bibinfo  {journal} {Eur. Phys. J.}\ }\textbf
  {\bibinfo {volume} {C72}},\ \bibinfo {pages} {1844} (\bibinfo {year}
  {2012}{\natexlab{a}})}\BibitemShut {NoStop}%
\bibitem [{\citenamefont {{ATLAS
  Collaboration}}(2011{\natexlab{c}})}]{Aad:2011dr}%
  \BibitemOpen
  \bibfield  {author} {\bibinfo {author} {\bibnamefont {{ATLAS
  Collaboration}}},\ }\href@noop {} {\bibfield  {journal} {\bibinfo  {journal}
  {\EPJ}\ }\textbf {\bibinfo {volume} {C71}},\ \bibinfo {pages} {1630}
  (\bibinfo {year} {2011}{\natexlab{c}})}\BibitemShut {NoStop}%
\bibitem [{\citenamefont {{ATLAS
  Collaboration}}(2011{\natexlab{d}})}]{lumiconf}%
  \BibitemOpen
  \bibfield  {author} {\bibinfo {author} {\bibnamefont {{ATLAS
  Collaboration}}},\ }\href@noop {} {}\bibinfo {howpublished}
  {{\href{http://cdsweb.cern.ch/record/1376384}{ATLAS-CONF-2011-116}.
  http://cdsweb.cern.ch/record/1376384}} (\bibinfo {year}
  {2011}{\natexlab{d}})\BibitemShut {NoStop}%
\bibitem [{\citenamefont {{ATLAS Collaboration}}(2012{\natexlab{b}})}]{Tapas}%
  \BibitemOpen
  \bibfield  {author} {\bibinfo {author} {\bibnamefont {{ATLAS
  Collaboration}}},\ }\href@noop {} {}\bibinfo {howpublished}
  {{\href{http://cdsweb.cern.ch/record/1423602/}{CERN-PH-EP-2012-061}.
  http://cdsweb.cern.ch/record/1423602/}} (\bibinfo {year}
  {2012}{\natexlab{b}})\BibitemShut {NoStop}%
\bibitem [{aad()}]{aad:2011ib}%
  \BibitemOpen
  \href@noop {} {}\bibinfo {howpublished} {{ATLAS Collaboration,
  arXiv:1109.6572 [hep-ex] (2011). Accepted by \PL B}}\BibitemShut {NoStop}%
\bibitem [{\citenamefont {{ATLAS
  Collaboration}}(2010{\natexlab{b}})}]{w-paper}%
  \BibitemOpen
  \bibfield  {author} {\bibinfo {author} {\bibnamefont {{ATLAS
  Collaboration}}},\ }\href {\doibase 10.1007/JHEP12(2010)060} {\bibfield
  {journal} {\bibinfo  {journal} {JHEP}\ }\textbf {\bibinfo {volume} {12}},\
  \bibinfo {pages} {060} (\bibinfo {year} {2010}{\natexlab{b}})}\BibitemShut
  {NoStop}%
\bibitem [{\citenamefont {{ATLAS Collaboration}}(2010{\natexlab{c}})}]{JER}%
  \BibitemOpen
  \bibfield  {author} {\bibinfo {author} {\bibnamefont {{ATLAS
  Collaboration}}},\ }\href@noop {} {}\bibinfo {howpublished}
  {{\href{http://cdsweb.cern.ch/record/1281311/}{ATLAS-CONF-2010-054}.
  http://cdsweb.cern.ch/record/1281311/}} (\bibinfo {year}
  {2010}{\natexlab{c}})\BibitemShut {NoStop}%
\bibitem [{\citenamefont {{ATLAS
  Collaboration}}(2011{\natexlab{e}})}]{toppaper}%
  \BibitemOpen
  \bibfield  {author} {\bibinfo {author} {\bibnamefont {{ATLAS
  Collaboration}}},\ }\href@noop {} {\bibfield  {journal} {\bibinfo  {journal}
  {\EPJ}\ }\textbf {\bibinfo {volume} {C71}},\ \bibinfo {pages} {1577}
  (\bibinfo {year} {2011}{\natexlab{e}})}\BibitemShut {NoStop}%
\bibitem [{\citenamefont {{ATLAS
  Collaboration}}(2011{\natexlab{f}})}]{Aad:2011kp}%
  \BibitemOpen
  \bibfield  {author} {\bibinfo {author} {\bibnamefont {{ATLAS
  Collaboration}}},\ }\href@noop {} {\bibfield  {journal} {\bibinfo  {journal}
  {Phys. Lett.}\ }\textbf {\bibinfo {volume} {B707}},\ \bibinfo {pages} {418}
  (\bibinfo {year} {2011}{\natexlab{f}})}\BibitemShut {NoStop}%
\bibitem [{\citenamefont {Cowan}\ \emph {et~al.}(2011)\citenamefont {Cowan},
  \citenamefont {Cranmer}, \citenamefont {Gross},\ and\ \citenamefont
  {Vitells}}]{profilelike}%
  \BibitemOpen
  \bibfield  {author} {\bibinfo {author} {\bibfnamefont {G.}~\bibnamefont
  {Cowan}}, \bibinfo {author} {\bibfnamefont {K.}~\bibnamefont {Cranmer}},
  \bibinfo {author} {\bibfnamefont {E.}~\bibnamefont {Gross}}, \ and\ \bibinfo
  {author} {\bibfnamefont {O.}~\bibnamefont {Vitells}},\ }\href {\doibase
  10.1140/epjc/s10052-011-1554-0} {\bibfield  {journal} {\bibinfo  {journal}
  {\EPJ}\ }\textbf {\bibinfo {volume} {C71}},\ \bibinfo {pages} {1554}
  (\bibinfo {year} {2011})}\BibitemShut {NoStop}%
\bibitem [{\citenamefont {{CDF Collaboration, T. Aaltonen \emph{ et
  al.}}}(2009)}]{gluinosbottomCDF}%
  \BibitemOpen
  \bibfield  {author} {\bibinfo {author} {\bibnamefont {{CDF Collaboration, T.
  Aaltonen \emph{ et al.}}}},\ }\href {\doibase 10.1103/PhysRevLett.102.221801}
  {\bibfield  {journal} {\bibinfo  {journal} {\PRL}\ }\textbf {\bibinfo
  {volume} {102}},\ \bibinfo {pages} {221801} (\bibinfo {year}
  {2009})}\BibitemShut {NoStop}%
\bibitem [{\citenamefont {{CDF Collaboration, T. Aaltonen \emph{et al.}
  }}(2010)}]{sbottomCDF}%
  \BibitemOpen
  \bibfield  {author} {\bibinfo {author} {\bibnamefont {{CDF Collaboration, T.
  Aaltonen \emph{et al.} }}},\ }\href {\doibase 10.1103/PhysRevLett.105.081802}
  {\bibfield  {journal} {\bibinfo  {journal} {Phys. Rev. Lett.}\ }\textbf
  {\bibinfo {volume} {105}},\ \bibinfo {pages} {081802} (\bibinfo {year}
  {2010})}\BibitemShut {NoStop}%
\bibitem [{\citenamefont {{D0 Collaboration, V.M. Abazov \emph{et
  al.}}}(2010)}]{sbottomD0}%
  \BibitemOpen
  \bibfield  {author} {\bibinfo {author} {\bibnamefont {{D0 Collaboration, V.M.
  Abazov \emph{et al.}}}},\ }\href@noop {} {\bibfield  {journal} {\bibinfo
  {journal} {Phys. Lett.}\ }\textbf {\bibinfo {volume} {B693}},\ \bibinfo
  {pages} {95} (\bibinfo {year} {2010})}\BibitemShut {NoStop}%
\bibitem [{\citenamefont {Baer}\ \emph {et~al.}(2010)\citenamefont {Baer},
  \citenamefont {Kraml}, \citenamefont {Lessa},\ and\ \citenamefont
  {Sekmen}}]{SO10HSDR3}%
  \BibitemOpen
  \bibfield  {author} {\bibinfo {author} {\bibfnamefont {H.}~\bibnamefont
  {Baer}}, \bibinfo {author} {\bibfnamefont {S.}~\bibnamefont {Kraml}},
  \bibinfo {author} {\bibfnamefont {A.}~\bibnamefont {Lessa}}, \ and\ \bibinfo
  {author} {\bibfnamefont {S.}~\bibnamefont {Sekmen}},\ }\href {\doibase
  10.1007/JHEP02(2010)055} {\bibfield  {journal} {\bibinfo  {journal} {JHEP}\
  }\textbf {\bibinfo {volume} {1002}},\ \bibinfo {pages} {055} (\bibinfo {year}
  {2010})}\BibitemShut {NoStop}%
\bibitem [{\citenamefont {Baer}\ \emph {et~al.}()\citenamefont {Baer},
  \citenamefont {Raza},\ and\ \citenamefont {Shafi}}]{SO10_yukawa}%
  \BibitemOpen
  \bibfield  {author} {\bibinfo {author} {\bibfnamefont {H.}~\bibnamefont
  {Baer}}, \bibinfo {author} {\bibfnamefont {H.}~\bibnamefont {Raza}}, \ and\
  \bibinfo {author} {\bibfnamefont {Q.}~\bibnamefont {Shafi}},\ }\href@noop {}
  {}\bibinfo {howpublished} {{arXiv:1201.5668 [hep-ph] (2012)}}\BibitemShut
  {NoStop}%
\end{thebibliography}%
\onecolumngrid
\clearpage
\begin{flushleft}
{\Large The ATLAS Collaboration}

\bigskip

G.~Aad$^{\rm 48}$,
B.~Abbott$^{\rm 110}$,
J.~Abdallah$^{\rm 11}$,
S.~Abdel~Khalek$^{\rm 114}$,
A.A.~Abdelalim$^{\rm 49}$,
A.~Abdesselam$^{\rm 117}$,
O.~Abdinov$^{\rm 10}$,
B.~Abi$^{\rm 111}$,
M.~Abolins$^{\rm 87}$,
O.S.~AbouZeid$^{\rm 157}$
% ...
\bigskip

% ...
$^{10}$ Institute of Physics, Azerbaijan Academy of Sciences, Baku, Azerbaijan\\
$^{11}$ Institut de F\'isica d'Altes Energies and Departament de F\'isica de la Universitat Aut\`onoma  de Barcelona and ICREA, Barcelona, Spain\\
% ...
$^{48}$ Fakult\"{a}t f\"{u}r Mathematik und Physik, Albert-Ludwigs-Universit\"{a}t, Freiburg i.Br., Germany\\
$^{49}$ Section de Physique, Universit\'e de Gen\`eve, Geneva, Switzerland\\
% ...
$^{87}$ Department of Physics and Astronomy, Michigan State University, East Lansing MI, United States of America\\
% ...
$^{110}$ Homer L. Dodge Department of Physics and Astronomy, University of Oklahoma, Norman OK, United States of America\\
$^{111}$ Department of Physics, Oklahoma State University, Stillwater OK, United States of America\\
% ...
$^{114}$ LAL, Univ. Paris-Sud and CNRS/IN2P3, Orsay, France\\
% ...
$^{117}$ Department of Physics, Oxford University, Oxford, United Kingdom\\
% ...
$^{157}$ Department of Physics, University of Toronto, Toronto ON, Canada\\
% ...
\end{flushleft}


\end{document}